\documentclass[11pt]{article}
\usepackage{savesym}
\usepackage{amsmath}
\savesymbol{iint}
\usepackage{txfonts}
\restoresymbol{TXF}{iint}
\usepackage{graphicx}
\usepackage{latexsym}
\usepackage{indentfirst}
\usepackage{rotating}
\usepackage{enumerate}
\usepackage{multirow}
\renewcommand{\Pr}{\mathsf{P}}

\makeatletter
\@addtoreset{equation}{section}\makeatother

\newtheorem{Theorem}{Theorem}[section]

\newtheorem{Proposition}[Theorem]{Proposition}
\newtheorem{Assumption}{Assumption}
\newtheorem{lemma}{\indent \bf Lemma}
\newtheorem{Remark}{Remark}[section]

\makeatletter
\addcontentsline{toc}{section}{References}
\@addtoreset{equation}{section}\makeatother

\date{}

\begin{document}

\title{Bayesian Ultrahigh-Dimensional Screening via MCMC}
\author{Zuofeng Shang and Ping Li\\
Department of Statistical Science\\
Cornell University\\
Ithaca, NY 14850}
\maketitle
\vspace{5mm}
\begin{abstract}
We \footnote{This paper was submitted in 2012.} 
explore the theoretical and numerical property of a fully Bayesian model selection method in sparse ultrahigh-dimensional settings, i.e., $p\gg n$,
where $p$ is the number of covariates and $n$ is the sample size.
Our method consists of (1) a hierarchical Bayesian model with a novel prior placed over the model space which includes a hyperparameter $t_n$ controlling the model size,
and (2) an efficient MCMC algorithm for automatic and stochastic search of the models. Our theory shows that, when specifying $t_n$ correctly,
the proposed method yields selection consistency, i.e., the posterior probability of the true model asymptotically approaches one; when $t_n$ is misspecified,
the selected model is still asymptotically nested in the true model. The theory also reveals insensitivity of the selection result with respect to the choice of $t_n$.
In implementations, a reasonable prior is further assumed on $t_n$ which allows us to draw its samples stochastically. Our approach conducts selection, estimation and even inference in a unified framework.
No additional prescreening or dimension reduction step is needed. Two novel $g$-priors are proposed to make our approach more flexible.
A simulation study is given to display the numerical advantage of our method.
\end{abstract}

\newpage
\section{Introduction}

Suppose the $n$-dimensional response vector
$\textbf{Y}=(y_1,\ldots,y_n)^T$ and the $n$ by $p$ covariate matrix
$\textbf{X}=(\textbf{X}_1,\ldots,\textbf{X}_p)$ are linked by the linear model
\begin{equation}\label{true:model}
\textbf{Y}=\textbf{X}\boldsymbol{\beta}+\boldsymbol{\epsilon},
\end{equation}
where the $\textbf{X}_j$s, $j=1,\ldots,p$, are $n$-vectors, $\boldsymbol{\beta}=(\beta_1,\ldots,\beta_p)^T$
is an unknown $p$-vector of regression coefficients, and $\boldsymbol{\epsilon}=(\epsilon_1,\ldots,\epsilon_n)^T$ is an $n$-vector of random errors.
The true parameter vector $\beta$ contains $s_n$ nonzero components and $p-s_n$ zeros.
Here we assume $p\gg n$, i.e., $p/n\rightarrow\infty$ as $n\rightarrow\infty$, but ideally restrict $s_n=o(n)$,
i.e., the true model is sparse. Our goal is to explore an automatic fully Bayesian procedure for
selecting and estimating the nonzero $\beta_j$s in (\ref{true:model}), in the ``large-$p$-small-$n$" scenario.

In frequentist settings, there is a vast amount of literature about variable selection in sparse ultrahigh-dimensional models.
We only list a few representative ones. Based on LASSO,
\cite{ZY06, MB06, VD08, ZH08, MY09} obtained selection consistency when $p$ is growing exponentially with $n$,
i.e., $\log{p}=O(n^a)$ for some $a>0$. Selection consistency here means, as $n$ goes to infinity, with probability approaching one the selected model is the true model.
\cite{HHM08} considered bridge regression, a link between the LASSO
and ridge regression, and obtained selection consistency. \cite{LF09} proposed a unified approach based on regularized
least squares with a class of concave penalties. \cite{FL08, FS10} proposed sure independence screening (SIS) based on correlations.
\cite{WLL09} proved selection consistency using BIC criteria.
\cite{WR09} examined several multi-stage selection approaches. \cite{SPZ12} applied a regularized likelihood approach based on
nonconvex constraints and proved selection and estimation consistency. \cite{BKM10} proposed a new method for variable selection
without using penalty. There are many other frequentist approaches handling this research area; see \cite{FL10} for an insightful review.

In Bayesian framework, selection consistency is somewhat different from the one in frequentist setting.
Unlike the frequentist setting which treats the true model as fixed a priori, Bayesian approaches assume
the model as a random element which has $2^p$ possible choices. Under proper Bayesian hierarchical models,
it is possible to derive the posterior distribution of the model. In other words, the posterior probabilities of all the $2^p$
models are achievable. We say such procedure is \textit{posterior consistent}
if the posterior probability of the true model converges to one. A nice property of the Bayesian approach
is that it can evaluate all the possible models based on the posterior probabilities and provide a stochastic search,
though an MCMC procedure might be needed.
Besides, it can simultaneously conduct estimation and inference over the selected coefficients
through the posterior samples.

Posterior consistency has been theoretically established when $p$ is fixed
(see \cite{FLS01,MG05,LPMCB08,CGMM09}).
\cite{LPMCB08} obtained posterior consistency in the setting of mixture of $g$-priors for fixed $p$. \cite{SC11} extended these results
to the growing $p$ situations. Their results cover both $p\le n$ and $p\gg n$. For $p\gg n$, they examined a two-step procedure. Explicitly,
in step I a dimension reduction (or prescreening) procedure such as  SIS proposed by \cite{FL08} is
performed to obtain a reduced model space, and in step II the Bayesian selection
procedure is performed over the reduced model space. However, the two-step scheme has several drawbacks.
According to \cite{FL08}, to yield better selection accuracy, the data has to be divided into two subsamples with one
for SIS and the other for Bayesian model selection. This additional prescreening step introduces additional complexity in applications,
and very often one has to determine the sizes of both subsamples, though a default choice may be an equal separation.
Furthermore, in many high-dimensional problems, the number of predictors $p$ can be much larger than the sample size $n$,
so the sizes of both subsamples
become even smaller. Usual Bayesian selection procedures based on
a smaller subset of the data may cause selection inaccuracy. Motivated by these considerations, an automatic one-step Bayesian
method, which does not involve any prescreening or dimension reduction procedure,
is highly needed and useful in both theoretical and applied aspects.
Related theoretical results on Bayesian model selection include \cite{BGM03,MGC10,GMCM10}
who proved consistency of Bayes factors when $p=O(n)$.
\cite{JR12} placed a set of novel non-local priors over the model coefficients
and proved posterior consistency for $p\le n$.
Recently, \cite{BR12} proposed a non-fully Bayesian selection method which works under $p\gg n$
but requires thresholding the marginal posterior means of $\boldsymbol{\beta}$.

In this paper, we explore the theoretical and numerical property of a fully
Bayesian model selection procedure in sparse ultrahigh-dimensional situations where $p$ is allowed to grow exponentially with $n$.
In our approach, stochastic model search, parameter estimation and even inference
can be simultaneously conducted in a unified framework,
though an MCMC procedure is employed for these goals. No additional steps such as dimension reduction or thresholding are needed.
Our model includes a hyperparameter controlling the size of the target models, namely, the size-control parameter.
A set of mild sufficient conditions are provided under which posterior consistency holds when this size-control parameter is correctly specified,
i.e., it is greater than the size of the true model. We also examine
the selection performance when the size-control parameter is misspecified. To the best of our knowledge, our work is the first one establishing posterior consistency
of the fully Bayesian model selection method in ultrahigh-dimensional settings, and theoretically examining the effect of a misspecified size-control parameter on model selection result.
To make the model more flexible,
we propose two new types of $g$-priors extending those in \cite{ZS80,LPMCB08} to ultrahigh-dimensional settings.
Posterior consistency under these priors is established. A prior over the size-control parameter
is considered which largely avoids misspecification, and induces a nontrivial extension of the traditional sampling scheme.
The simulation study reveals that the proposed method is computationally accurate and convenient.

The rest of this paper is organized as follows.
In Section \ref{sec:model}, a Bayesian hierarchical model involving suitable priors is explicitly given.
Section \ref{sec:main:results} contains the theoretical results which justify posterior consistency and evaluate the effect
of misspecifying the hyperparameter controlling the model size in various situations
including the $g$-prior. New types of $g$-priors are constructed in this section. We also briefly discuss the credible interval construction over the selected coefficients.
Section \ref{sec:computation} presents the computational details involving a constrained blockwise sampling procedure.
In Section \ref{simulation:sec}, a simulation study is given to demonstrate the performance.
All the technical proofs are given in the appendix.

\section{A hierarchical model with a size-control prior on model space}\label{sec:model}

Before formally describing our models, we first introduce some notation that are used frequently throughout this paper.
Define $\gamma_j=I(\beta_j\neq 0)$, i.e., the 0-1 variable indicating the exclusion
or inclusion of $\beta_j$, and define $\boldsymbol{\gamma}=(\gamma_1,\ldots,\gamma_p)^T$. Throughout we use $|\boldsymbol{\gamma}|$
to denote the number of ones in $\boldsymbol{\gamma}$. Clearly, each $\boldsymbol{\gamma}$ corresponds to
a candidate model $\textbf{Y}=\textbf{X}_{\boldsymbol{\gamma}}\boldsymbol{\beta}_{\boldsymbol{\gamma}}+\boldsymbol{\epsilon}$, where $\textbf{X}_{\boldsymbol{\gamma}}$ is an
$n\times |\gamma|$ submatrix of $\textbf{X}$, and $\boldsymbol{\beta}_{\boldsymbol{\gamma}}$ is the subvector
(with size $|\boldsymbol{\gamma}|$) of $\boldsymbol{\beta}$, whose columns and elements are indexed by the nonzero components of
$\boldsymbol{\gamma}$, respectively. The $2^p$ possible $\boldsymbol{\gamma}$s correspond to the $2^p$ different
models, which form the entire model space. For any $\boldsymbol{\gamma}$ and $\boldsymbol{\gamma}'$, let
$(\boldsymbol{\gamma}\backslash\boldsymbol{\gamma}')_j=I(\gamma_j=1,\gamma'_j=0)$, and $(\boldsymbol{\gamma}\cap\boldsymbol{\gamma}')_j=I(\gamma_j=1,\gamma'_j=1)$.
Thus, $\boldsymbol{\gamma}\backslash\boldsymbol{\gamma}'$ is the 0-1 vector indicating the variables
present in $\boldsymbol{\gamma}$ but absent in $\boldsymbol{\gamma}'$, and $\boldsymbol{\gamma}\cap\boldsymbol{\gamma}'$ is the 0-1 vector indicating the variables
present in both $\boldsymbol{\gamma}$ and $\boldsymbol{\gamma}'$. We say that $\boldsymbol{\gamma}$ is
nested in $\boldsymbol{\gamma}'$ (denoted by $\boldsymbol{\gamma}\subset\boldsymbol{\gamma}'$) if
$\boldsymbol{\gamma}\backslash\boldsymbol{\gamma}'$ is zero. Denote the true model coefficient vector by
$\boldsymbol{\beta}^0$ and the corresponding 0-1 vector by $\boldsymbol{\gamma}^0$, and let
$s_n=|\boldsymbol{\gamma}^0|$ denote the size of the true model.

We adopt a normal linear model between the response and covariates, i.e.,
\begin{equation}\label{fbm:1}
\textbf{Y}|\boldsymbol{\beta},\sigma^2\sim N(\textbf{X}\boldsymbol{\beta},\sigma^2 \textbf{I}_n).
\end{equation}
Suitable prior distributions are required for the parameters $\boldsymbol{\beta}$ and $\sigma^2$. We adopt the ``spike-and-slab" prior
for $\beta_j$s, i.e
\begin{equation}\label{fbm:2}
\beta_j|\gamma_j,\sigma^2\sim (1-\gamma_j)\delta_0+\gamma_j N(0,c_j\sigma^2),
\end{equation}
where $\delta_0(\cdot)$ is the point mass measure concentrating on zero, and $c_j$'s are temporarily assumed to be fixed. Note that $c_j$'s
are used to control the variance of the nonzero coefficients, and therefore are called the variance-control parameters.
In next sections we will treat the mixture of $g$-prior setup, i.e., assuming priors on $c_j$'s.
The ``spike-and-slab" prior has been explored in various applied aspects by \cite{SK96,CPV98,CG00,WGN04,LS12}.

We place an inverse $\chi^2$ prior on $\sigma^2$, i.e.,
\begin{equation}\label{fmb:3}
1/\sigma^2\sim \chi_\nu^2,
\end{equation}
where $\nu$ is a fixed hyperparameter. Other choices such as the noninformative priors or inverse Gamma priors can also be applied.
The theoretical results derived in this paper can be extended without further difficulty to these situations.

A prior probability, namely, $p(\gamma)$, should be assigned to each candidate model $\gamma$, i.e.,
\begin{equation}\label{fmb:4}
\boldsymbol{\gamma}\sim p(\boldsymbol{\gamma}).
\end{equation}
A popular choice of $p(\boldsymbol{\gamma})$ is the so-called independent Bernoulli prior used \cite{GM93,GM97,GF00,BFV01,BVF02,NK05,LS09,SC12},
or the Bernoulli-Beta prior used by \cite{SB10,BR12,LS12}.
The independent Bernoulli prior assumes each covariate to be included in the model with
 probability $\theta_j$, i.e., $p(\boldsymbol{\gamma})=\prod\limits_{j=1}^p \theta_j^{\gamma_j}(1-\theta_j)^{1-\gamma_j}$
with $\theta_j$s being fixed. The Bernoulli-Beta prior assumes further a Beta prior over $\theta_j$s.

In many practical applications, such as genewise selection, only a small amount of covariates should be included in the model,
which can be treated as, in Bayesian terminology, prior information.
Thus, most of the candidate models, especially those with large model sizes, should be assigned a tiny or even zero prior probability.
In Bernoulli prior, this can be achieved by assuming a very small but positive $\theta_j$. Due to the huge number of candidate models,
of which most are ``incorrect", even though each ``incorrect" model is assigned a very small prior probability, the aggregated prior probability
over all the ``incorrect" models can still be large. This will severely affect the accuracy of the Bayesian model selection procedure when $p\gg n$.
Here we propose a novel prior that
only assigns positive weights to the models with smaller sizes, i.e, a size-control prior on model space. Namely,
\begin{equation}\label{fbm:5}
p(\boldsymbol{\gamma})=\left\{\begin{array}{cc} \pi_{\boldsymbol{\gamma}},&\,\,\textrm{if $|\boldsymbol{\gamma}|\le t_n$,}\\
                                    0, &\textrm{otherwise,}\end{array}\right.
\end{equation}
where $\pi_{\boldsymbol{\gamma}}$ for $|\boldsymbol{\gamma}|\le t_n$ are fixed positive numbers, and $t_n\in (0,n)$ is an integer-valued hyperparameter
controlling the sizes of the candidate models.
Clearly, (\ref{fbm:5}) is more powerful than Bernoulli or Bernoulli-Beta prior to screen out the models with larger sizes.
When the number of nonzeros in $\boldsymbol{\beta}^0$, i.e., $s_n$, is small so that $t_n>s_n$,
this implies (\ref{fbm:5}) is powerful to screen out the ``incorrect" models with greater sizes.

Based on the above Bayesian hierarchical model (\ref{fbm:1})-(\ref{fbm:5}), the joint posterior distribution for $(\boldsymbol{\beta},\boldsymbol{\gamma},\sigma^2)$ can be derived.
For simplicity, denote $\textbf{Z}=(\textbf{Y},\textbf{X})$ to be the full data variable. The joint posterior distribution is then
\begin{eqnarray}\label{post:dist}
p(\boldsymbol{\beta},\boldsymbol{\gamma},\sigma^2|\textbf{Z})&\propto& p(\textbf{Z}|\boldsymbol{\beta},\sigma^2)p(\boldsymbol{\beta}|\sigma^2,\boldsymbol{\gamma})p(\boldsymbol{\gamma})p(\sigma^2)\nonumber\\
&\propto&\sigma^{-(n+\nu+2)}\exp\left(-\frac{\|\textbf{Y}-\textbf{X}\boldsymbol{\beta}\|^2+1}{2\sigma^2}\right) p(\boldsymbol{\gamma})
\prod\limits_{j\in \boldsymbol{\gamma}}\left[\frac{1}{\sqrt{c_j}\sigma}\phi\left(\frac{\beta_j}{\sqrt{c_j}\sigma}\right)\right]\prod\limits_{j\in-\boldsymbol{\gamma}}\delta_0(\beta_j),
\end{eqnarray}
where $\phi(\cdot)$ is the density function of the standard normal random variable, $j\in\boldsymbol{\gamma}$ means the index $j\in \{1,\ldots,p\}$ satisfies $\gamma_j=1$
and $j\in -\boldsymbol{\gamma}$ means $\gamma_j=0$, $p(\boldsymbol{\gamma})$ is the prior defined as in (\ref{fbm:5}).
Integrating out $\boldsymbol{\beta}$ and $\sigma^2$ in (\ref{post:dist}) one obtains
\begin{equation}\label{p:gamma}
p(\boldsymbol{\gamma}|\textbf{Z})\propto \det(\textbf{W}_{\boldsymbol\gamma})^{-1/2}p(\boldsymbol{\gamma})\left(1+\textbf{Y}^T(\textbf{I}_n-\textbf{X}_{\boldsymbol{\gamma}} \textbf{U}_{\boldsymbol{\gamma}}^{-1} \textbf{X}_{\boldsymbol{\gamma}}^T)\textbf{Y}\right)^{-(n+\nu)/2},
\end{equation}
where $\textbf{W}_{\boldsymbol{\gamma}}=\boldsymbol{\Sigma}_{\boldsymbol{\gamma}}^{1/2}\textbf{U}_{\boldsymbol{\gamma}}\Sigma_\gamma^{1/2}$,
$\textbf{U}_{\boldsymbol{\gamma}}=\boldsymbol{\Sigma}_{\boldsymbol{\gamma}}^{-1}+ \textbf{X}_{\boldsymbol{\gamma}}^T\textbf{X}_{\boldsymbol{\gamma}}$,
and $\boldsymbol{\Sigma}_{\boldsymbol{\gamma}}$ denotes the principle submatrix of $\boldsymbol{\Sigma}=\textrm{diag}(c_1,\ldots,c_p)$ indexed by $\boldsymbol{\gamma}$.
Here we adopt the convention that $\textbf{X}_{\emptyset}=0$ and
$\boldsymbol{\Sigma}_\emptyset=\textbf{U}_\emptyset=\textbf{W}_\emptyset=1$, where $\emptyset$ means the null model,
i.e., the vector $\boldsymbol{\gamma}$ with all elements being zero.

The optimal model $\widehat{\boldsymbol{\gamma}}$ is chosen to maximize (\ref{p:gamma}), i.e.,
\begin{equation}\label{gamma:hat}
\widehat{\boldsymbol{\gamma}}=\arg\max\limits_{\boldsymbol{\gamma}} p(\boldsymbol{\gamma}|\textbf{Z}).
\end{equation}
In other words, $\widehat{\boldsymbol{\gamma}}$ achieves the highest posterior probability among all the possible models. When $|\boldsymbol{\gamma}|>t_n$, $p(\boldsymbol{\gamma}|\textbf{Z})=0$.
So maximizing (\ref{gamma:hat}) is actually performed over a smaller model space named as the target model space.
We name the model selection procedure (\ref{gamma:hat})
as Bayesian ultrahigh-dimensional screening. Ideally we hope to show that the selected model $\widehat{\boldsymbol{\gamma}}$ is asymptotically exactly the true model $\boldsymbol{\gamma}^0$.
This is equivalent to showing that $p(\boldsymbol{\gamma}^0|\textbf{Z})$ is asymptotically greater than $p(\boldsymbol{\gamma}|\textbf{Z})$ for any $\boldsymbol{\gamma}\neq \boldsymbol{\gamma}^0$, which holds
if $p(\boldsymbol{\gamma}^0|\textbf{Z})$ converges to one in certain mode.

\section{Main results}\label{sec:main:results}

In this section, we present our main results on posterior consistency.
Throughout we suppose $\boldsymbol{\gamma}^0\neq \emptyset$, that is, the true model is not empty.
Our first result shows that when properly choosing $t_n\ge s_n$,
under certain mild conditions, $p(\boldsymbol{\gamma}^0|\textbf{Z})$ converges in probability to one, where convergence holds uniformly for $c_j$'s lying within certain ranges.
Since typically $s_n$ is unknown, one may face a risk of misspecifying $t_n$ so that $t_n$ is actually smaller than $s_n$. Theoretical results are thus needed to
examine this situation. Our second result shows that when $0<t_n<s_n$, with probability approaching one, the selected $\widehat{\boldsymbol{\gamma}}$ is nonnull and is nested to the true model,
implying that all the selected variables are significant although there are other significant variables not selected.

Throughout this whole section, we define $\textbf{P}_{\boldsymbol{\gamma}}=
\textbf{X}_{\boldsymbol{\gamma}}(\textbf{X}_{\boldsymbol{\gamma}}^T\textbf{X}_{\boldsymbol{\gamma}})^{-1}\textbf{X}_{\boldsymbol{\gamma}}^T$, i.e., the
projection matrix based on $\textbf{X}_{\boldsymbol{\gamma}}$. We adopt the convention that $\textbf{P}_\emptyset=0$.
Let $\lambda_{-}(\textbf{A})$ and $\lambda_{+}(\textbf{A})$ be the minimal and maximal eigenvalues of the square matrix $\textbf{A}$. Suppose there exist positive sequences $\underline{\phi}_n$
and $\bar{\phi}_n$ such that $\underline{\phi}_n\le c_j\le \bar{\phi}_n$ for $j=1,\ldots,p$.
Denote $k_n=\|\boldsymbol{\beta}^0_{\boldsymbol{\gamma}^0}\|^2$ and $\psi_n=\min\limits_{j\in \boldsymbol{\gamma}^0}|\beta^0_j|$,
where $\beta^0_j$ denotes the $j$th element of $\boldsymbol{\beta}^0$ and $\|\cdot\|$ denotes the $\ell_2$-norm.

\subsection{When $t_n\ge s_n$}\label{large:tn}
\renewcommand{\theAssumption}{A.\arabic{Assumption}}
\setcounter{Assumption}{0}

We first consider the case $t_n\ge s_n$, that is, the size-control parameter $t_n$ is correctly specified
as being greater than or equal to the size of the true model. In this case,
the true model $\boldsymbol{\gamma}^0$ has positive posterior probability, and thus, is among our target model space.

To prove $p(\boldsymbol{\gamma}^0|\textbf{Z})$ asymptotically approaches one,
we introduce some useful notation and technical assumptions.
Define $S_1(t_n)=\{\boldsymbol{\gamma}|\boldsymbol{\gamma}^0\subset\boldsymbol{\gamma}, \boldsymbol{\gamma}\neq\boldsymbol{\gamma}^0, |\boldsymbol{\gamma}|\le t_n\}$ and
$S_2(t_n)=\{\boldsymbol{\gamma}|\boldsymbol{\gamma}^0\,\, \textrm{is not nested in}\,\, \boldsymbol{\gamma}, |\boldsymbol{\gamma}|\le t_n\}$. It
is clear that $S_1(t_n)$ and $S_2(t_n)$ are disjoint, and $S(t_n)$ defined by $S(t_n)=S_1(t_n)\bigcup S_2(t_n) \bigcup\{\boldsymbol{\gamma}^0\}$ is
the class of all models with size not exceeding $t_n$.
To insure a flexible choice of $t_n$, we assume $t_n\in [s_n,r_n]$ for some integer $r_n>s_n$.
Our result in this section shows that when properly fixing the upper bound $r_n$,
any choice of $t_n\in [s_n,r_n]$ will guarantee that the true model is selected. This says that
the selection result is somewhat insensitive to the choice of $t_n$ within certain range.

\begin{Assumption}\label{A1} There exists a positive constant $c_0$ such that, as $n\rightarrow\infty$, with probability approaching one, for any $t_n\in [s_n,r_n]$,
\[
1/c_0\le \min\limits_{\boldsymbol{\gamma}\in S_2(t_n)}\lambda_{-}\left(\frac{1}{n}\textbf{X}_{\boldsymbol{\gamma}^0\backslash\boldsymbol{\gamma}}^T
(\textbf{I}_n-\textbf{P}_{\boldsymbol{\gamma}})\textbf{X}_{\boldsymbol{\gamma}^0\backslash\boldsymbol{\gamma}}\right)\le \max\limits_{\boldsymbol{\gamma}\in S_2(t_n)}\lambda_{+}
\left(\frac{1}{n}\textbf{X}_{\boldsymbol{\gamma}^0\backslash\boldsymbol{\gamma}}^T\textbf{X}_{\boldsymbol{\gamma}^0\backslash\boldsymbol{\gamma}}\right)\le c_0,
\]
and
\[
\min\limits_{\boldsymbol{\gamma}\in S_1(t_n)}\lambda_-\left(\frac{1}{n}\textbf{X}_{\boldsymbol{\gamma}\backslash\boldsymbol{\gamma}^0}^T
(\textbf{I}_n-\textbf{P}_{\boldsymbol{\gamma}^0})\textbf{X}_{\boldsymbol{\gamma}\backslash\boldsymbol{\gamma}^0}^T\right)\ge 1/c_0.
\]
\end{Assumption}

\begin{Assumption}\label{A2}
$\sup\limits_n\max\limits_{\substack{\boldsymbol{\gamma}\in S(t_n),\\ t_n\in[s_n,r_n]}}\frac{p(\boldsymbol{\gamma})}{p(\boldsymbol{\gamma}^0)}<\infty$.
\end{Assumption}

\begin{Assumption}\label{A3} The sequences $s_n$, $r_n$, $\bar{\phi}_n$, $\underline{\phi}_n$, $k_n$, and
$\psi_n$ satisfy, as $n\rightarrow\infty$,
\begin{enumerate}[(i).]
\item $s_n=o(n)$;
\item $n\psi_n^2\rightarrow\infty$;
\item $s_n< r_n\le n/2$ and $r_n\log{p}=o(n\log(1+\min\{1,\psi_n^2\}))$;
\item $s_n\log(1+c_0 n\bar{\phi}_n)=o(n\log(1+\min\{1,\psi_n^2\}))$;
\item $\log{p}=o(\log{\underline{\phi}_n})$ and $k_n=O(\underline{\phi}_n)$.
\end{enumerate}
\end{Assumption}

\begin{Remark}\label{rem:1}

We briefly discuss the validity of Assumptions \ref{A1} to \ref{A3}. We first have the following result
showing that Assumption \ref{A1} holds under a very broad range of situations. Its proof is similar to that of Proposition 2.1 in \cite{SC11},
and thus is omitted.

\begin{Proposition}\label{matrix:class}
Assumption \ref{A1} is satisfied if there exists $c_0>0$ such that
\begin{equation}\label{eq:matrix:class}
c_0^{-1}\le\min\limits_{|\boldsymbol{\gamma}|\le 2r_n}\lambda_{-}\left(\frac{1}{n}\textbf{X}^T_{\boldsymbol{\gamma}}
\textbf{X}_{\boldsymbol{\gamma}}\right)\le\max\limits_{|\boldsymbol{\gamma}|\le 2r_n}\lambda_+\left(\frac{1}{n}\textbf{X}^T_{\boldsymbol{\gamma}}\textbf{X}_{\boldsymbol{\gamma}}\right)\le c_0.
\end{equation}
\end{Proposition}
(\ref{eq:matrix:class}) is called the \emph{sparse Riesz condition}, a standard condition in the study of high-dimensional problems; see \cite{ZH08,MY09}
for applications in LASSO. Proposition \ref{matrix:class} confirms that the \emph{sparse Riesz condition} is even stronger than our Assumption \ref{A1}.
Assumption \ref{A2} holds if we place indifference prior over $\boldsymbol{\gamma}$ with $|\boldsymbol{\gamma}|\le t_n$, which implies $\frac{p(\boldsymbol{\gamma})}{p(\boldsymbol{\gamma}^0)}=1$.

To see when Assumption \ref{A3} holds, let us consider a simple scenario. Suppose $\psi_n=n^{-k_1}$, $s_n=n^{k_2}$, $r_n=n^{k_3}$ and $\log{p}=n^{k_4}$,
where $k_4>0$, $k_1,k_2,k_3$ are nonnegative satisfying $k_2<k_3$ and $2k_1+k_3+k_4<1$. Furthermore, $\log{k_n}=O(\log{n})$ which is a weaker assumption than \cite{J07}.
Then it can be shown directly that $\bar{\phi}_n$ and $\underline{\phi}_n$ with $\log{\bar{\phi_n}}=o(n^{1-2k_1-k_2})$ and $n^{k_4}=o(\log{\underline{\phi}_n})$
satisfy Assumption \ref{A3}. In this simple situation, both $\bar{\phi}_n$ and $\underline{\phi}_n$ are growing exponentially with $n$. In other words,
they have to be large enough to support the high-dimensional selection.
Here we want to emphasize that the upper bound for $\bar{\phi}_n$ and the lower bound for $\underline{\phi}_n$ are both necessary for selecting the true model;
see \cite{SK96} for heuristic explanations in a lower-dimensional situation.

\end{Remark}

\begin{Theorem}\label{main:thm1}
Under Assumptions \ref{A1} through \ref{A3}, as $n\rightarrow\infty$,
\[
\min\limits_{s_n\le t_n\le r_n}\inf\limits_{\underline{\phi}_n\le c_1,\ldots,c_p\le \bar{\phi}_n} p(\boldsymbol{\gamma}^0|\textbf{Z})\rightarrow 1,\,\,\textrm{in probability}.
\]
\end{Theorem}

The proof of Theorem \ref{main:thm1} is given in the appendix.
Theorem \ref{main:thm1} provides a set of sufficient conditions under which, uniformly for $c_j$s $\in [\underline{\phi}_n,\bar{\phi}_n]$
and $t_n\in [s_n,r_n]$, posterior consistency holds. In other words,
selection accuracy is not sensitive to the values of these hyperparameters when they are in a proper range.
These conditions are satisfied when $p=O(\exp(n^{k_4}))$ for some $k_4\in (0,1)$ (see Remark \ref{rem:1}),
thus, Theorem \ref{main:thm1} holds in ultrahigh-dimensional settings.
The proof of Theorem \ref{main:thm1} relies on finding the sharp upper bounds of the Bayes factors between models including $t_n$ as a component.
It is shown that uniformly for $t_n\in [s_n,r_n]$ with $s_n$ and $r_n$ growing at certain rates, all these upper bounds can be well
managed so that the posterior probability of the true model converges to one.
In next section, we further examine the performance of our Bayesian selection method when $t_n$ is misspecified, i.e., $t_n<s_n$.

In computations (Section \ref{sec:computation}), to enhance flexibility, we further assume a prior $p(t_n)$ over $t_n$.
Concretely, in simulation study (Section \ref{simulation:sec}) we chose the improper prior $p(t_n)=I(t_n\le m_n)$
with some given $m_n>0$. Here $m_n$ represents our prior belief on the range of $s_n$, the number of true nonzeros.
To be conservative, we set $m_n=n/2$, a commonly accepted upper bound in sparse high-dimensional problems (see \cite{CT05}),
but still find satisfactory selection accuracy.

Here we want to compare Theorem \ref{main:thm1} with literature. There are two major types of Bayesian model selection procedures explored in literature, i.e.,
the Bayes factor and the fully Bayesian approach based on hierarchical models like (\ref{fbm:1})--(\ref{fbm:5}).
Bayes factor is a useful tool for pairwise model comparison and is equivalent to the fully Bayesian model selection when $p$ is fixed (see \cite{BP96,LPMCB08}).
When $p\le n$ is increasing with $n$, these two types of selection methods are not equivalent (see \cite{SC11}).
In this case, \cite{BGM03,MGC10,GMCM10} proved consistency for Bayes factors which holds even for $p=O(n)$.

In contrast, the fully Bayesian approach evaluates all the $2^p$ models and selects the model with the highest posterior probability,
and thus, is essentially different from Bayes factor in the setting of growing $p$. Important literature includes \cite{FLS01,MG05,LPMCB08,CGMM09}
who showed selection consistency for fixed $p$. Later on these results were generalized to increasing $p$ with $p\le n$
in a range of hierarchical models; see \cite{SC11,JR12}.
To the best of our knowledge, Theorem \ref{main:thm1} is the first result establishing posterior consistency for a fully Bayesian method
in ultrahigh-dimensional settings.
\cite{S11} also describes a two-step procedure so that selection consistency holds for $p\gg n$.
Of course this procedure is not fully Bayesian since a preliminary step such as SIS is performed before formal selection.
Instead, the selection method introduced in this paper is performed by directly fitting the hierarchical model (\ref{fbm:1})--(\ref{fbm:5}). No additional steps such as SIS
or posterior mean thresholding considered by \cite{BR12} are needed. The key is the application of the prior (\ref{fbm:5}).
We believe when adopting this prior, other existing results valid for $p\le n$ can also be extended to $p\gg n$.


\subsection{When $0<t_n<s_n$}\label{small:tn}
\renewcommand{\theAssumption}{B.\arabic{Assumption}}
\setcounter{Assumption}{0}

Now we turn to the case of misspecifying the hyperparameter $t_n$ so that actually $0<t_n<s_n$.
In this case, the true model $\boldsymbol{\gamma}^0$, which has posterior probability zero,
is outside our target model space and thus is impossible to be selected out.
We will show that even in this false setting the selected model $\widehat{\boldsymbol{\gamma}}$ is asymptotically nested in the true model.
In other words, all the selected variables are significant ones which ought to be included in the model.

Define $T_0(t_n)=\{\boldsymbol{\gamma}| 0\le |\boldsymbol{\gamma}|\le t_n, \boldsymbol{\gamma}\subset\boldsymbol{\gamma}^0\}$,
$T_1(t_n)=\{\boldsymbol{\gamma}| 0<|\boldsymbol{\gamma}|\le t_n,\boldsymbol{\gamma}\cap\boldsymbol{\gamma}^0\neq\emptyset,
\,\,\textrm{$\boldsymbol{\gamma}$ is not nested in $\boldsymbol{\gamma}^0$}\}$, and $T_2(t_n)=\{\boldsymbol{\gamma}| 0<|\boldsymbol{\gamma}|\le t_n, \boldsymbol{\gamma}\cap\boldsymbol{\gamma}^0=\emptyset\}$.
It is easy to see that $T_0(t_n), T_1(t_n), T_2(t_n)$ are disjoint and $T(t_n)=T_0(t_n)\cup T_1(t_n)\cup T_2(t_n)$ is exactly the class of $\boldsymbol{\gamma}$ with $|\boldsymbol{\gamma}|\le t_n$. Throughout this section,
we make the following assumptions.

\begin{Assumption}\label{B1}
There exist a positive constant $d_0$ and a positive sequence $\rho_n$ such that, when $n\rightarrow\infty$, with probability approaching one,
\begin{equation}\label{B1:eq}
d_0^{-1}\le \min\limits_{\substack{|\boldsymbol{\gamma}|\le s_n\\ \boldsymbol{\gamma}\neq\emptyset}}\lambda_{-}\left(\frac{1}{n}\textbf{X}_{\boldsymbol{\gamma}}^T \textbf{X}_{\boldsymbol{\gamma}}\right)\le
\max\limits_{\substack{|\boldsymbol{\gamma}|\le s_n\\ \boldsymbol{\gamma}\neq\emptyset}}\lambda_{+}\left(\frac{1}{n}\textbf{X}_{\boldsymbol{\gamma}}^T \textbf{X}_{\boldsymbol{\gamma}}\right)\le d_0,\,\,\textrm{and}\,\,
\end{equation}
\begin{equation}\label{B1:eq2}
\max\limits_{\substack{\boldsymbol{\gamma}\in T(t_n),\\ 0<t_n<s_n}}\lambda_{+}\left(\textbf{X}_{\boldsymbol{\gamma}^0\backslash\boldsymbol{\gamma}}^T
\textbf{P}_{\boldsymbol{\gamma}} \textbf{X}_{\boldsymbol{\gamma}^0\backslash\boldsymbol{\gamma}}\right)\le \rho_n.
\end{equation}
\end{Assumption}

\begin{Assumption}\label{B2}
$\sup\limits_n\max\limits_{\substack{\boldsymbol{\gamma},\boldsymbol{\gamma}'\in T(t_n),\\ 0<t_n<s_n}}\frac{p(\boldsymbol{\gamma})}{p(\boldsymbol{\gamma}')}<\infty$.
\end{Assumption}

\begin{Assumption}\label{B3}
The sequences $s_n$, $\underline{\phi}_n$, $k_n$,
$\psi_n$ and $\rho_n$ satisfy, as $n\rightarrow\infty$,
\begin{enumerate}[(i).]
\item $s_n=o(n)$;
\item $n\psi_n^2\rightarrow\infty$;
\item $s_n=o(n\psi_n^2)$;
\item $k_n=O(\underline{\phi}_n)$;
\item $\max\{\rho_n,s_n^2\log{p}\}=o(\min\{n,\log(\underline{\phi}_n)\})$.
\end{enumerate}
\end{Assumption}

\begin{Remark}\label{rem:2}
Before stating our main theorems in this section, let us examine the validity of the Assumptions \ref{B1}--\ref{B3}.
Assumption \ref{B2} holds if we adopt the indifference prior, i.e., $p(\boldsymbol{\gamma})$ is positive constant for all $\boldsymbol{\gamma}\in T(t_n)$.
The following result demonstrates the validity of Assumption \ref{B1} in a special situation, though we believe
this condition may still hold in more general cases.

\begin{Proposition}\label{valid:B1}
Suppose the rows of $\textbf{X}$ are \textit{iid} copies of $(\xi_1,\ldots,\xi_p)$ which is a zero-mean Gaussian vector with $E\{\xi_j^2\}=1$ for $1\le j\le p$.
The vector $(\xi_1,\ldots,\xi_p)$ is a subvector of the infinite population sequence $\{\xi_j, j=1,2,\ldots\}$ which satisfies the Riesz condition, i.e., (4.5)
in \cite{ZH08}. Furthermore, $s_n\log{p}=o(n)$ and $\xi_j$'s are independent.
Then Assumption \ref{B1} holds for $\rho_n=\alpha s_n^2\log{p}$ with any constant $\alpha>4$.
\end{Proposition}

Proposition \ref{valid:B1} is proved in the Appendix.
In the setting of Proposition \ref{valid:B1}, we may choose $\rho_n\asymp s_n^2\log{p}$.
Suppose $\log{k_n}=O(\log{n})$ and choose $\underline{\phi}_n$ such that $\log{\underline{\phi}_n}>n$.
Let $\psi_n=n^{-k_1}$, $s_n=n^{k_2}$ and $\log{p}=n^{k_4}$,
where $k_4>0$, $k_1,k_2$ are nonnegative satisfying $2k_1+k_2+k_4<1$ and $2k_2+k_4<1$.
Then it can be easily verified that Assumption \ref{B3} holds in this particular situation.
Clearly, Assumptions \ref{B3} and \ref{A3} are not contradictive in that there exist $\{p,s_n,\psi_n,\underline{\phi}_n,\bar{\phi}_n,k_n\}$ satisfying both conditions.
The difference is that Assumption \ref{A3} also involves $r_n$, i.e., the upper bound for the hyperparameter $t_n$, while Assumption \ref{B3} does not since $t_n$ has already been assumed to be bounded by $s_n$.
The careful readers may also notice that, unlike Assumption \ref{A3} which places both upper bound for $\bar{\phi}_n$ and lower bound for $\underline{\phi}_n$,
in Assumption \ref{B3}, only lower bound for $\underline{\phi}_n$ is assumed.
The reason is, in the subsequent Theorem \ref{main:thm2}, we allow $\widehat{\boldsymbol{\gamma}}=\emptyset$, a model in $T_0(t_n)$.
This case is preferred when all the $c_j$s tend to infinity (corresponding to $\bar{\phi}_n=\infty$); see \cite{SK96}.
Thus, the upper bound for $\bar{\phi}_n$ is not necessary.
Actually, in the below Theorem \ref{main:thm3} where we show in a situation that $\widehat{\boldsymbol{\gamma}}$ is nested in the true model but $\widehat{\boldsymbol{\gamma}}\neq\emptyset$,
an upper bound for $\bar{\phi}_n$ will still be needed.
\end{Remark}

Next we state our first theorem in this section.

\begin{Theorem}\label{main:thm2}
Under Assumptions \ref{B1}--\ref{B3}, as $n\rightarrow\infty$,
\[
\max\limits_{0<t_n<s_n}\sup\limits_{\underline{\phi}_n\le c_1,\ldots, c_p\le \bar{\phi}_n}
\frac{\max\limits_{\boldsymbol{\gamma}\in T_1(t_n)\cup T_2(t_n)}p(\boldsymbol{\gamma}|\textbf{Z})}{\max\limits_{\boldsymbol{\gamma}\in T_0(t_n)}p(\boldsymbol{\gamma}|\textbf{Z})}\rightarrow0, \textrm{in probability}.
\]
\end{Theorem}

The proof of Theorem \ref{main:thm2} can be found in the Appendix.
Theorem \ref{main:thm2} examines the situation of misspecifying the hyperparameter $t_n$, i.e., $t_n<s_n$. It says that in such situation, even though the selected model $\widehat{\boldsymbol{\gamma}}$
cannot be the true model since necessarily $|\widehat{\boldsymbol{\gamma}}|<s_n$, $\widehat{\boldsymbol{\gamma}}$ can still be nested to the true model with probability approaching one.
Furthermore, convergence holds uniformly for $0<t_n\le s_n$ and $c_j$s within certain range.

Theorem \ref{main:thm2} allows $\widehat{\boldsymbol{\gamma}}=\emptyset$.
However, when the true model is nonnull, we may ask further if $\widehat{\boldsymbol{\gamma}}$ can be nonnull. The following result provides a positive answer to this question.
The price we pay is an additional assumption to separate a nonnull model from the null model.

\begin{Theorem}\label{main:thm3}
Suppose we happen to choose some $t_n$ from $(0,s_n)$. Let Assumptions \ref{B1}--\ref{B3} be satisfied. If, in addition, Assumption \ref{A3} (iv) holds,
and there is $\boldsymbol{\gamma}\in T_0(t_n)\backslash\{\emptyset\}$, such that $\|\boldsymbol{\beta}^0_{\boldsymbol{\gamma}^0\backslash\boldsymbol{\gamma}}\|^2\le f_0\|\boldsymbol{\beta}^0_{\boldsymbol{\gamma}}\|^2$,
where $f_0>0$ is constant. Then as $n\rightarrow\infty$, $\sup\limits_{\underline{\phi}_n\le c_1,\ldots,c_p\le\bar{\phi}_n} \frac{p(\emptyset|\textbf{Z})}{p(\boldsymbol{\gamma}|\textbf{Z})}=o_\Pr(1)$.
In other words, $\boldsymbol{\gamma}$ is a better choice than the null model.
\end{Theorem}

The proof of Theorem \ref{main:thm3} is given in the appendix.
In Theorem \ref{main:thm3}, we make the assumption $\|\boldsymbol{\beta}^0_{\boldsymbol{\gamma}^0\backslash\boldsymbol{\gamma}}\|^2\le f_0\|\boldsymbol{\beta}^0_{\boldsymbol{\gamma}}\|^2$.
Heuristically, $\|\boldsymbol{\beta}^0_{\boldsymbol{\gamma}}\|$ represents the information of the model $\boldsymbol{\gamma}$ and $\|\boldsymbol{\beta}^0_{\boldsymbol{\gamma}^0\backslash\boldsymbol{\gamma}}\|$
represents the information of the complement model $\boldsymbol{\gamma}^0\backslash\boldsymbol{\gamma}$. This assumption simply says that much of the information of the true model
is concentrated on $\boldsymbol{\gamma}$. Theorem \ref{main:thm3} states that with this ``information" assumption and  Assumption \ref{A3} (iv), model $\boldsymbol{\gamma}$ can successfully
outperform the null model so that $\widehat{\boldsymbol{\gamma}}\neq\emptyset$ with arbitrarily large probability. Note here Assumption \ref{A3} (iv)
is necessary since otherwise with $c_j$s approaching infinity the null model will be always preferred (see \cite{SK96}).
To the best of our knowledge, Theorems \ref{main:thm2} and \ref{main:thm3} are the first theoretical results in the fully Bayesian setting
examining the selection performance with misspecified hyperparameters.

\subsection{Extensions to the $g$-prior setting}\label{sec:g-prior}

In this section, we extend the results in Sections \ref{large:tn} and \ref{small:tn} to the $g$-prior setting. For simplicity, let all the variance-control parameters be the same,
i.e., $c_j=c$ for all $j=1,\ldots,p$. Instead of using a fixed $c$, we place over $c$ a proper prior $g(c)$, i.e., $\int_0^\infty g(c)dc=1$. Here we consider a broad functional class for $g(c)$
including the variations of the Zellner-Siow prior proposed by \cite{ZS80} and the hyper $g$-prior proposed by \cite{LPMCB08}.

Assuming a random $c\in (0,\infty)$, the conditional probability of $\boldsymbol{\gamma}$ given $(c,\textbf{Z})$ is exactly
\begin{equation}
p(\boldsymbol{\gamma}|c,\textbf{Z})\propto \det(\textbf{W}_{\boldsymbol{\gamma}})^{-1/2}p(\boldsymbol{\gamma})\left(1+\textbf{Y}^T(\textbf{I}_n-\textbf{X}_{\boldsymbol{\gamma}}
\textbf{U}_{\boldsymbol{\gamma}}^{-1} \textbf{X}_{\boldsymbol{\gamma}}^T)\textbf{Y}\right)^{-(n+\nu)/2},
\end{equation}
where $\textbf{W}_{\boldsymbol{\gamma}}$ and $\textbf{U}_{\boldsymbol{\gamma}}$, both depending $c$, are defined as in (\ref{p:gamma}).
Consequently, the posterior probability of $\boldsymbol{\gamma}$, in the setting of $g$-prior, is given by
\begin{equation}\label{gp:gamma}
p_g(\boldsymbol{\gamma}|\textbf{Z})=\int_0^\infty p(\boldsymbol{\gamma}|c,\textbf{Z})g(c)dc,
\end{equation}
where the subscript $g$ represents the posterior probability in the setting of $g$-prior.

We will prove that $p_g(\boldsymbol{\gamma}|\textbf{Z})$ shares similar probabilistic properties as those in Sections \ref{large:tn} and \ref{small:tn},
though a $g$-prior setting has been considered. \cite{LPMCB08} obtained selection consistency in the $g$-prior settings
where $p$ is fixed. Their proof relies on an application of Laplace approximation of the posterior likelihood.
Here we will use a different approach which relies on the uniform convergence results that have been derived in previous sections.
Our first theorem below treats the case when $t_n\in [s_n,r_n]$ with $r_n>s_n$ being some integer.

\begin{Theorem}\label{main:gp:thm1}
Suppose Assumptions \ref{A1}--\ref{A3} are satisfied. Furthermore, $g$ is proper and satisfies, as $n\rightarrow\infty$, $\int_0^{\underline{\phi}_n}g(c)dc=o(1)$
and $\int_{\bar{\phi}_n}^\infty g(c)dc=o(1)$. Then as $n\rightarrow\infty$, $\min\limits_{s_n\le t_n\le r_n} p_g(\boldsymbol{\gamma}^0|\textbf{Z})\rightarrow 1$, in probability.
\end{Theorem}

Theorem \ref{main:gp:thm1} is proved in the appendix. It establishes model selection consistency under the $g$-prior setting.
Again, this result uniformly holds for $t_n\in [s_n,r_n]$.

Our second and third results treat the case $0<t_n<s_n$. The proofs are given in the appendix. They state that even when one misspecifies the $t_n$ such it actually lies in $(0,s_n)$,
the selected model may still be nested in the true model, and even nonnull. However, we are only able to show the desired results for those $g$s
with compact support $[\underline{\phi}_n,\bar{\phi}_n]$, though we conjecture that these results may still hold for more general $g$s.

\begin{Theorem}\label{main:gp:thm2}
Suppose Assumptions \ref{B1}--\ref{B3} are satisfied. Furthermore, $g$ is proper and supported in $[\underline{\phi}_n,\bar{\phi}_n]$, i.e.,
$g(c)=0$ if $c\notin[\underline{\phi}_n,\bar{\phi}_n]$. Then as $n\rightarrow\infty$,
$\max\limits_{0<t_n<s_n}\frac{\max\limits_{\boldsymbol{\gamma}\in T_1(t_n)\cup T_2(t_n)}p_g(\boldsymbol{\gamma}|\textbf{Z})}
{\max\limits_{\boldsymbol{\gamma}\in T_0(t_n)}p_g(\boldsymbol{\gamma}|\textbf{Z})}\rightarrow0$, in probability.
\end{Theorem}

\begin{Theorem}\label{main:gp:thm3}
Suppose we happen to choose some $t_n$ from $(0,s_n)$. Let Assumptions \ref{B1}--\ref{B3} be satisfied. If, in addition, Assumption \ref{A3} (iv) holds,
and there is $\boldsymbol{\gamma}\in T_0(t_n)\backslash\{\emptyset\}$, such that $\|\boldsymbol{\beta}^0_{\boldsymbol{\gamma}^0\backslash\boldsymbol{\gamma}}\|^2\le f_0\|\boldsymbol{\beta}^0_{\boldsymbol{\gamma}}\|^2$,
where $f_0>0$ is constant. Furthermore, $g$ is proper and supported in $[\underline{\phi}_n,\bar{\phi}_n]$. Then as $n\rightarrow\infty$,
$\frac{p_g(\emptyset|\textbf{Z})}{p_g(\boldsymbol{\gamma}|\textbf{Z})}=o_\Pr(1)$.
In other words, $\boldsymbol{\gamma}$ is a better choice than the null model in the setting of $g$-prior.
\end{Theorem}

\subsection{Generalized Zellner-Siow prior and generalized hyper-$g$ prior}\label{sec:g:example:prior}

In this section, motivated from \cite{ZS80} and \cite{LPMCB08} in fixed $p$ scenario,
two new types of $g$-priors will be proposed. The first one is a generalization of Zeller-Siow prior motivated from \cite{ZS80}.
The second one is a generalization of the hyper-$g$ prior motivated from \cite{LPMCB08}. Both variations never appeared in literature and are nontrivial.

The original form of Zellner-Siow prior is $g(c)\propto c^{-3/2}\exp(-n/(2c))$; see \cite{LPMCB08}.
However, as demonstrated in our simulation study,
the accuracy of using this prior severely decreases in high-dimensional setting. The reason is, as revealed in the discussions in Remark \ref{rem:1},
to achieve more accurate selection, one has to shift the range $[\underline{\phi}_n,\bar{\phi}_n]$ to be suitably large. A possible choice is to
make both $\underline{\phi}_n$ and $\bar{\phi}_n$ exponentially growing with $n$. To achieve selection consistency in the $g$-prior setting, one may
choose $g$ concentrated on $[\underline{\phi}_n,\bar{\phi}_n]$ (see Theorem \ref{main:gp:thm1}), implying that the mode of $g$ is, say, exponentially growing with $n$;
see the original form of Zellner-Sior prior with mode $n/3$.
This motivates us to consider the following \emph{generalized} Zellner-Siow prior
\begin{equation}\label{GZS}
g(c)= \frac{p^{ab_n}}{\Gamma(a)}c^{-a-1}\exp(-p^{b_n}/c),
\end{equation}
where $a>0, b_n>0$ are fixed hyperparameters. The prior in (\ref{GZS}) is actually $IG(a,p^{b_n})$ with mode $p^{b_n}/(a+1)$. A nice property of this prior is its conjugacy for which we can use
a Gibbs sampler step to draw the $c$ samples. A proper choice is a constant $a>0$ and $b_n\asymp \log{n}$. With direct calculations we have
$\int_0^{\underline{\phi}_n} g(c)dc=(\Gamma(a))^{-1}\int_{p^{b_n}\underline{\phi}_n^{-1}}^\infty c^{a-1}\exp(-c)dc$ and
$\int_{\bar{\phi}_n}^\infty g(c)dc=(\Gamma(a))^{-1}\int_0^{p^{b_n}\bar{\phi}_n^{-1}}c^{a-1}\exp(-c)dc$.
Thus, with $\underline{\phi}_n=p^{\sqrt{b_n}}$ and $\bar{\phi}_n=p^{b_n^2}$, both integrals are $o(1)$, i.e., $g$ satisfies the condition in Theorem \ref{main:gp:thm1}.
Note that this condition is violated for $a=0$.
Furthermore, it follows from the discussions in Remark \ref{rem:1} that such choice of $\underline{\phi}_n$ and $\bar{\phi}_n$ also fulfill Assumption \ref{A3}
for $s_n,\psi_n, p$ specified therein. This shows that the prior (\ref{GZS}) can indeed induce consistent Bayesian selection.

Next we intend to explore our second type of $g$-prior. Following \cite{LPMCB08}, the motivation of the hyper-$g$ prior is
that the shrinkage factor $c/(1+c)$ has most of the mass near $1$, for which they assume $c/(1+c)$ to have beta distribution with hyperparameters properly managed.
However, as demonstrated in our simulation study, the hyper-$g$ prior or the hyper-$g/n$ prior considered in \cite{LPMCB08},
though work well in lower-dimensional situation, does not work well in high-dimensional setting.
The reason is similar to that for the conventional Zeller-Siow prior, i.e, the mode of these $g$-priors are not large enough to support high-dimensional selection.
From this point of view, we consider $c/(1+c)\sim \textrm{Beta}(\alpha_n,b)$, leading to the following \emph{generalized} hyper-$g$ prior
\begin{equation}\label{Ghyper:gprior}
g(c)= \frac{\Gamma(\alpha_n+b)}{\Gamma(\alpha_n)\Gamma(b)}\cdot\frac{c^{\alpha_n-1}}{(1+c)^{\alpha_n+b}},
\end{equation}
where $b>0$ is constant and $\alpha_n=p^{a_n}+1$ with $a_n\asymp\log{n}$. Obviously, the mode of our generalized hyper-$g$ prior is $(\alpha_n-1)/(b+1)$.
With $\underline{\phi}_n=p^{\sqrt{a}_n}$
and $\bar{\phi}_n=p^{a_n^2}$, by direct calculations, it can be verified that
$\int_0^{\underline{\phi}_n} g(c)dc=O(\alpha_n^{b-1}\exp(-\alpha_n/(1+\underline{\phi}_n)))=o(1)$
and $\int_{\bar{\phi}_n}^\infty g(c)dc=O(\alpha_n^b/(1+\bar{\phi}_n))=o(1)$. Therefore, the proposed generalized hyper-g prior
also satisfies the assumptions in Theorem \ref{main:gp:thm1}, implying the selection consistency.

In implementations we simply choose $a=b=0$ to achieve the maximum prior modes for both generalized Zellner-Siow prior and generalized hyper-g prior,
though they may violate the limit conditions in Theorem \ref{main:gp:thm1}. Our empirical results in Section \ref{simulation:sec} demonstrate satisfactory performance
of such choice.

\subsection{Simultaneous credible intervals}\label{sec:SCI}

In many applications, model selection is just an initial step. After selecting the model, it is important to further make inference on the selected variables,
e.g., constructing simultaneous credible intervals for the nonzero features.

Suppose one has selected model $\boldsymbol{\gamma}$, and the goal is to further build credible intervals for $\beta_j$s with $\gamma_j=1$. To ease technical arguments,
we assume known $\sigma^2$ and $c_j$s, $\gamma_j=1$ for $j=1,\ldots,r$, and $\gamma_j=0$ for $j=r+1,\ldots,p$. Therefore the hierarchical model becomes
\[
\textbf{Y}|\boldsymbol{\beta}\sim N(\textbf{X}\boldsymbol{\beta},\sigma^2 \textbf{I}_n),\,\,\,\,\beta_j\sim (1-\gamma_j)\delta_0+\gamma_j N(0,c_j\sigma^2).
\]
With straightforward calculations one can show that $\boldsymbol{\beta}_{\boldsymbol{\gamma}}$ follows $N(\boldsymbol{\xi},\sigma^2 \textbf{U}_{\boldsymbol{\gamma}}^{-1})$,
where $\boldsymbol{\xi}=\textbf{U}_{\boldsymbol{\gamma}}^{-1}\textbf{X}_{\boldsymbol{\gamma}}^T \textbf{Y}$ and $\textbf{U}_{\boldsymbol{\gamma}}$ is defined as in (\ref{p:gamma}).
Thus, the marginal posterior distribution for $\beta_j$ for $j=1,\ldots,r$ is $\beta_j\sim N(\xi_j,\sigma_j^2)$,
where $\xi_j$ is the $j$th component of $\boldsymbol{\xi}$, and $\sigma_j^2$ is the $j$th diagonal element of $\sigma^2 \textbf{U}_{\boldsymbol{\gamma}}^{-1}$.
The $100\times (1-\alpha)\%$ credible interval for $\beta_j$ is thus
\begin{equation}\label{Bayes:SCI}
\textrm{CI}_j: \xi_j\pm c_{\alpha/2}\sigma_j,\,\,\,\,j=1,\ldots,r,
\end{equation}
where $c_{\alpha/2}$ is the lower $(\alpha/2)$-th quantile of the standard normal distribution.

To see the performance of the intervals $\textrm{CI}_j$s, we use the concept of Bayes false coverage rate (FCR) considered by
\cite{ZH12}. Namely, let $V$ be the number of the $\textrm{CI}_j$s which do not cover $\beta_j$.
Then $\textrm{FCR}=E\{V/r\}$. Since for any $j=1,\ldots,r$, $\Pr(\beta_j\notin\textrm{CI}_j|Z)=\alpha$,
it follows by Theorem 2 of \cite{ZH12} that $\textrm{FCR}\le\alpha$.
In other words, the Bayes FCR of the simultaneous credible intervals constructed in (\ref{Bayes:SCI}) can be controlled at arbitrary nominal level $\alpha$,
though a smaller $\alpha$ would enlarge the $\textrm{CI}_j$s simultaneously.

\section{Computational details}\label{sec:computation}

In this section we present the sampling details. In Section \ref{sec:fixc}, we fix $c_j$s and demonstrate how to use MCMC to draw samples from $\boldsymbol{\beta},\boldsymbol{\gamma},\sigma^2, t_n$.
In Section \ref{sec:aboutc}, we discuss various ways of handling the $c_j$s including using BIC or RIC in which the $c_j$s are fixed a priori,
or using an additional MCMC step to draw samples from $c_j$s in a $g$-prior setting.

\subsection{A constrained blockwise Gibbs sampler for automatic and stochastic model search}\label{sec:fixc}

In previous sections $t_n$, i.e., the size-control parameter in (\ref{fbm:5}), is a fixed integer.
Though the theory holds uniformly for $t_n$ within certain range, practically one still has to choose a proper one to facilitate computation.
To address this difficulty, we further place a prior on $t_n$. Specifically, to play simple, we let $p(t_n)=I(t_n\le m_n)$, i.e.,
a uniform prior on $[1,m_n]$ with $m_n$ being a predetermined integer, though other choices with more complicated forms can also be used,
which induces corresponding revisions in the following algorithm.
With this prior and based on the Bayesian hierarchical model (\ref{fbm:1})-(\ref{fbm:5}), the posterior distribution is
\begin{equation}\label{full:post:dist}
p(\boldsymbol{\beta},\boldsymbol{\gamma},\sigma^2,t_n|\textbf{Z})\propto p(\boldsymbol{\beta},\boldsymbol{\gamma},\sigma^2|t_n,\textbf{Z})p(t_n),
\end{equation}
where $p(\boldsymbol{\beta},\boldsymbol{\gamma},\sigma^2|t_n,\textbf{Z})$ is exactly given in (\ref{post:dist}). Temporarily all the $c_j$s are fixed hyperparameters.

We will present an efficient Gibbs sampler to draw posterior samples from (\ref{full:post:dist}).
In conventional Gibbs sampler one draws samples iteratively and separately from the full conditionals $p(\boldsymbol{\beta}|\boldsymbol{\gamma},\sigma^2,t_n,\textbf{Z})$,
$p(\boldsymbol{\gamma}|\boldsymbol{\beta},\sigma^2,t_n,\textbf{Z})$,
$p(\sigma^2|\boldsymbol{\beta},\boldsymbol{\gamma},t_n,\textbf{Z})$ and $p(t_n|\boldsymbol{\beta},\boldsymbol{\gamma},\sigma^2,\textbf{Z})$. However, for our specific Bayesian model, it can be shown that both the full conditionals for $\boldsymbol{\beta}$ and $\boldsymbol{\gamma}$
involve intensive matrix inversion computation, an extremely time-consuming step when data dimension is large or a long Markov chain needs to be sampled.
To ease the matrix inversion computation, \cite{LZ10} used a novel technique in structured high-dimensional model which reduces computing time.
Here we will adopt a different approach that fully avoids the computation of the inverse matrices.

To improve sampling speed, we propose a constrained blockwise Gibbs sampler motivated from \cite{GR98} and \cite{WGN04}.
The basic idea of the original blockwise Gibbs sampler is to treat each two-dimensional vector $\textbf{g}_j=(\beta_j,\gamma_j)$, for
$j=1,\ldots,p$, as a block. Instead of sampling $\boldsymbol{\beta}$
and $\boldsymbol{\gamma}$ separately, we draw them together through sampling the blocks $g_j$s iteratively.
A nice property of the blockwise Gibbs sampler is that it effectively avoids matrix inversion computation,
and therefore is more computationally efficient. However, our specific prior on the model space, i.e., the inclusion of the hyperparameter $t_n$ that controls the model size, induces nontrivial modifications in this method.
Specifically, during the sampling process, to fulfill the blockwise technique, the size of the sampled model from the previous iteration has to be less than $t_n$,
which is essentially a constrained version of the blockwise procedure.
In practical implementations, we further allow a stochastic draw from $t_n$, i.e., an automatic and stochastic control of the model sizes during posterior sampling,
which makes our procedure even more flexible.

From (\ref{post:dist}), the joint posterior of $\textbf{g}_1,\ldots,\textbf{g}_p$ given $\sigma^2$ and $t_n$ is
\begin{eqnarray}\label{block:post:dist}
&&p(\textbf{g}_1,\ldots,\textbf{g}_p|\sigma^2,t_n,\textbf{Z})\nonumber\\
&\propto& (2\pi\sigma^2)^{-\frac{|\boldsymbol{\gamma}|}{2}}p(\boldsymbol{\gamma})
\exp\left(-\frac{\|\textbf{Y}-\textbf{X}\boldsymbol{\beta}\|^2+\boldsymbol{\beta}_{\boldsymbol{\gamma}}^T\boldsymbol{\Sigma}_{\boldsymbol{\gamma}}^{-1}\boldsymbol{\beta}_{\boldsymbol{\gamma}}}{2\sigma^2}\right)\cdot\prod\limits_{j\in\boldsymbol{\gamma}}c_j^{-1/2}\cdot
\prod\limits_{j\in-\boldsymbol{\gamma}}\delta_0(\beta_j).
\end{eqnarray}
Denote $\textbf{g}_{-j}=\{\textbf{g}_1,\ldots,\textbf{g}_{j-1},\textbf{g}_{j+1},\ldots,\textbf{g}_p\}$. If $|\boldsymbol{\gamma}_{-j}|>t_n$, i.e.,
the number of indexes $k$ with $k\neq j$ and $\gamma_k=1$ is greater than $t_n$,
then the posterior probability in (\ref{block:post:dist}) becomes zero.
So we only consider $|\boldsymbol{\gamma}_{-j}|\le t_n$.
To ease the technical arguments, suppose for each $\textbf{g}_k=(\beta_k,\gamma_k)$ with $k\neq j$,
$\gamma_k$ and $\beta_k$ ``match" each other in the sense that $\gamma_k=1$ if $\beta_k\neq0$, and $\gamma_k=0$ if $\beta_k=0$.
It can be shown directly from (\ref{block:post:dist}) that
\begin{eqnarray}\label{block:full:conditional}
&&p(\textbf{g}_j|\textbf{g}_{-j},\sigma^2,t_n,Z)\nonumber\\
&\propto& (2\pi c_j\sigma^2)^{-\frac{\gamma_j}{2}}p(\gamma_j,\boldsymbol{\gamma}_{-j})
\exp\left(-\frac{\|\textbf{Y}-\textbf{X}\boldsymbol{\beta}\|^2+\boldsymbol{\beta}_{\boldsymbol{\gamma}}^T\boldsymbol{\Sigma}_{\boldsymbol{\gamma}}^{-1}\boldsymbol{\beta}_{\boldsymbol{\gamma}}}
{2\sigma^2}\right)\cdot\prod\limits_{j\in-\boldsymbol{\gamma}}\delta_0(\beta_j)\nonumber\\
&\propto& (2\pi c_j\sigma^2)^{-\frac{\gamma_j}{2}}p(\gamma_j,\boldsymbol{\gamma}_{-j})
\exp\left(-\frac{\textbf{X}_j^T\textbf{X}_j\beta_j^2-2u_j\beta_j+\boldsymbol{\beta}_{\boldsymbol{\gamma}}^T\boldsymbol{\Sigma}_{\boldsymbol{\gamma}}^{-1}\boldsymbol{\beta}_{\boldsymbol{\gamma}}}
{2\sigma^2}\right)\cdot\prod\limits_{j\in-\boldsymbol{\gamma}}\delta_0(\beta_j),
\end{eqnarray}
where $u_j=(\textbf{Y}-\textbf{X}_{-j}\boldsymbol{\beta}_{-j})^T\textbf{X}_j$ and $\textbf{X}_{-j}=\left(\textbf{X}_1,\ldots,\textbf{X}_{j-1},\textbf{X}_{j+1},\ldots,\textbf{X}_p\right)$ for $j=1,\ldots,p$.

We first consider $|\boldsymbol{\gamma}_{-j}|< t_n$. In this case, $p(\gamma_j,\boldsymbol{\gamma}_{-j})$ is always positive
since the size of $(\gamma_j,\boldsymbol{\gamma}_{-j})$, i.e., $\gamma_j+|\boldsymbol{\gamma}_{-j}|$, does not exceed $t_n$.
From (\ref{block:full:conditional}), it can be shown that the full conditionals of $(\gamma_j=1,\beta_j)$ and $(\gamma_j=0,\beta_j)$ are respectively
\begin{equation}\label{bfc:1}
p(\gamma_j=1,\beta_j|\textbf{g}_{-j},\sigma^2,t_n,\textbf{Z})\propto (2\pi c_j\sigma^2)^{-1/2} p(\gamma_j=1,\boldsymbol{\gamma}_{-j})\exp\left(-\frac{v_j^2\beta_j^2-2u_j\beta_j}{2\sigma^2}\right),
\end{equation}
where $v_j^2=\textbf{X}_j^T\textbf{X}_j+c_j^{-1}$, and
\begin{equation}\label{bfc:0}
p(\gamma_j=0,\beta_j|\textbf{g}_{-j},\sigma^2,t_n,\textbf{Z})\propto p(\gamma_j=0,\boldsymbol{\gamma}_{-j})\exp\left(-\frac{\textbf{X}_j^T\textbf{X}_j\beta_j^2-2u_j\beta_j}{2\sigma^2}\right)\delta_0(\beta_j).
\end{equation}
Integrating out $\beta_j$ in (\ref{bfc:1}) and (\ref{bfc:0}), one obtains the marginal distribution for $\gamma_j$ given by
\begin{equation}\label{bfc:gamma:1}
p(\gamma_j=1|\textbf{g}_{-j},\sigma^2,t_n,\textbf{Z})\propto \frac{p(\gamma_j=1,\boldsymbol{\gamma}_{-j})}{\sqrt{c_j}v_j}\exp\left(\frac{u_j^2}{2\sigma^2 v_j^2}\right),
\end{equation}
and
\begin{equation}\label{bfc:gamma:0}
p(\gamma_j=0|\textbf{g}_{-j},\sigma^2,t_n,\textbf{Z})\propto p(\gamma_j=0,\boldsymbol{\gamma}_{-j}).
\end{equation}
From (\ref{bfc:gamma:1}) and (\ref{bfc:gamma:0}), we can draw $\gamma_j$ marginally through
\begin{equation}\label{marginal:gamma}
p(\gamma_j=0|\textbf{g}_{-j},\sigma^2,t_n,\textbf{Z})=\frac{1}{1+\frac{1}{\sqrt{c_j}v_j}\cdot\frac{p(\gamma_j=1,\boldsymbol{\gamma}_{-j})}{p(\gamma_j=0,\boldsymbol{\gamma}_{-j})}\cdot\exp\left(\frac{u_j^2}{2\sigma^2 v_j^2}\right)}.
\end{equation}
Then by (\ref{bfc:1}) and (\ref{bfc:0}), we sample $\beta_j$ through the following marginal conditional distributions
\begin{equation}\label{conditional:beta:1}
\beta_j|\gamma_j=1,\textbf{g}_{-j},\sigma^2,t_n,\textbf{Z}\sim N\left(\frac{u_j}{v_j^2},\frac{\sigma^2}{v_j^2}\right),
\end{equation}
\begin{equation}\label{conditional:beta:0}
p(\beta_j=0|\gamma_j=0,\textbf{g}_{-j},\sigma^2,t_n,\textbf{Z})=1.
\end{equation}
Through (\ref{marginal:gamma})--(\ref{conditional:beta:0}), we can draw sample $\textbf{g}_j$ from $p(\textbf{g}_j|\textbf{g}_{-j},\sigma^2,t_n,\textbf{Z})$ in the setting $|\boldsymbol{\gamma}_{-j}|<t_n$, for $j=1,\ldots,p$.

When $|\boldsymbol{\gamma}_{-j}|=t_n$, it follows directly from (\ref{block:post:dist}) that
$p(\gamma_j=1,\beta_j,\textbf{g}_{-j}|\sigma^2,t_n,\textbf{Z})=0$,
implying $p(\gamma_j=0|\textbf{g}_{-j},\sigma^2,t_n,\textbf{Z})=1$. Using (\ref{block:full:conditional}),
it can be shown that $p(\beta_j=0|\gamma_j=0,\textbf{g}_{-j},\sigma^2,t_n,\textbf{Z})=1$. In other words,
we simply set $\beta_j=0$ to match its binary state $\gamma_j$, by which we can control the model sizes to be not exceeding $t_n$.
We should mention that this additional ``size-control" step does not appear in conventional lower-dimensional Bayesian model selection;
see \cite{GR98} or \cite{WGN04} for comparison. Here we need it to address the ultrahigh dimensionality.

From (\ref{post:dist}), it can be verified that the full conditional of $\sigma^2$ is given by
\begin{equation}\label{sigma:full:conditional}
p(\sigma^2|\boldsymbol{\beta},\boldsymbol{\gamma},t_n,\textbf{Z})\propto (\sigma^2)^{-\frac{n+|\boldsymbol{\gamma}|+\nu}{2}-1}
\exp\left(-\frac{\|\textbf{Y}-\textbf{X}\boldsymbol{\beta}\|^2+\boldsymbol{\beta}_{\boldsymbol{\gamma}}^T\boldsymbol{\Sigma}_{\boldsymbol{\gamma}}^{-1}\boldsymbol{\beta}_{\boldsymbol{\gamma}}+1}{2\sigma^2}\right),
\end{equation}
that is, $\sigma^2|\boldsymbol{\beta},\boldsymbol{\gamma},t_n,\textbf{Z}\sim IG\left(\frac{n+|\boldsymbol{\gamma}|+\nu}{2},
\frac{\|\textbf{Y}-\textbf{X}\boldsymbol{\beta}\|^2+\boldsymbol{\beta}_{\boldsymbol{\gamma}}^T\boldsymbol{\Sigma}_{\boldsymbol{\gamma}}^{-1}\boldsymbol{\beta}_{\boldsymbol{\gamma}}+1}{2}\right)$,
where $IG(a,b)$ denotes the inverse-gamma distribution with density $\pi(x)\propto x^{-a-1}\exp\left(-b/x\right)$.
Finally, given $\boldsymbol{\beta},\boldsymbol{\gamma},\sigma$, it is easy to see that $t_n$ is uniform in $[|\boldsymbol{\gamma}|,m_n]$.

To conclude, we summarize our Gibbs sampler in a fashion that can be applied directly in programming.
Set the initial stage $\gamma^{(0)}_j=0$, $\beta^{(0)}_j=0$, for $j=1,\ldots,p$,
$\sigma_{(0)}^2$ to be a random selected positive number, and $t_n^{(0)}$ to be uniform over $[1,m_n]$.
Suppose we have sampled $(\boldsymbol{\gamma}^{(l)}, \boldsymbol{\beta}^{(l)}, \sigma_{(l)}^2, t_n^{(l)})$ from the $l$th iteration.
\begin{enumerate}[(i).]
\item
Suppose, in the $(l+1)$th iteration, we have sampled
the first $j-1$ blocks, i.e., $\textbf{g}^{(l+1)}_1=(\beta^{(l+1)}_1,\gamma^{(l+1)}_1),\ldots,\textbf{g}^{(l+1)}_{j-1}=(\beta^{(l+1)}_{j-1},\gamma^{(l+1)}_{j-1})$.
Denote $\boldsymbol{\gamma}_{-j}=(\gamma^{(l+1)}_1,\ldots,\gamma^{(l+1)}_{j-1},\gamma^{(l)}_{j+1},\ldots,\gamma^{(l)}_p)$
and $\boldsymbol{\beta}_{-j}=(\beta^{(l+1)}_1,\ldots,\beta^{(l+1)}_{j-1},\beta^{(l)}_{j+1},\ldots,\beta^{(l)}_p)^T$.
To generate $\textbf{g}^{(l+1)}_j=(\beta^{(l+1)}_j,\gamma^{(l+1)}_j)$, we use the following procedure:
\begin{enumerate}[(1).]
\item If $|\boldsymbol{\gamma}_{-j}|<t_n^{(l)}$, then set $\gamma_j^{(l+1)}=0$ with probability $\frac{1}{1+\theta_j}$,
where $\theta_j=\frac{1}{\sqrt{c_j}v_j}\cdot\frac{p(\gamma_j=1,\boldsymbol{\gamma}_{-j})}{p(\gamma_j=0,\boldsymbol{\gamma}_{-j})}
\cdot\exp\left(\frac{u_j^2}{2\sigma^2 v_j^2}\right)$,
$u_j=(\textbf{Y}-\textbf{X}_{-j}\boldsymbol{\beta}_{-j})^T\textbf{X}_j$ and $v_j^2=\textbf{X}_j^T\textbf{X}_j+c_j^{-1}$.

If $\gamma^{(l+1)}_j=1$, then draw $\beta_j^{(l+1)}$ from $N\left(\frac{u_j}{v_j^2},\frac{\sigma^2}{v_j^2}\right)$.
Else, if $\gamma^{(l+1)}_j=0$, then set $\beta_j^{(l+1)}=0$.

\item If $|\boldsymbol{\gamma}_{-j}|=t_n^{(l)}$, then set $\gamma_j^{(l+1)}=0$ and $\beta_j^{(l+1)}=0$.
\end{enumerate}

\item After finishing (i) for all $j=1,\ldots,p$, denote $\boldsymbol{\gamma}^{(l+1)}$ and $\boldsymbol{\beta}^{(l+1)}$
to be the current update of $\boldsymbol{\gamma}$ and $\boldsymbol{\beta}$. Draw $\sigma^2_{(l+1)}$ from
\[
IG\left(\frac{n+|\boldsymbol{\gamma}^{(l+1)}|+\nu}{2},
\frac{\|\textbf{Y}-\textbf{X}\boldsymbol{\beta}^{(l+1)}\|^2+(\boldsymbol{\beta}_{\boldsymbol{\gamma}^{(l+1)}}^{(l+1)})^T
\boldsymbol{\Sigma}_{\boldsymbol{\gamma}^{(l+1)}}^{-1}\boldsymbol{\beta}_{\boldsymbol{\gamma}^{(l+1)}}^{(l+1)}+1}{2}\right).
\]

\item Draw $t_n^{(l+1)}$ uniformly over $[|\boldsymbol{\gamma}^{(l+1)}|,m_n]$.
\end{enumerate}

\subsection{About the $c_j$s}\label{sec:aboutc}

The choice of $c_j$s plays an important role in practical implementation of our method,
and therefore they must be well addressed. In our numerical study, we chose $c_j=c$, a constant hyperparameter for all $j=1,\ldots,p$,
though to ease the application they can be chosen as different numbers if we priorly have preferences over certain coefficients.

There are several popular ways of finding $c$ including BIC, RIC (see \cite{FG94}), and the Benchmark prior method (see \cite{FLS01}).
In these methods $c$ is fixed as $n$, $p^2$, and $\max\{n,p^2\}$ respectively.
An alternative way is to avoid finding $c$ by assuming $c$ to follow the $g$-priors such as the ones
introduced in Section \ref{sec:g:example:prior}, though an Metropolis-Hasting step might be needed to draw the $c$ samples.

Suppose $g(c)$ is a proper prior over $c$, then the full conditional of $c$ can be derived directly by
\begin{equation}\label{full:cond:c}
p(c|\boldsymbol{\beta},\boldsymbol{\gamma},\sigma^2,t_n,\textbf{Z})\propto
p(\boldsymbol{\beta}|\boldsymbol{\gamma},c,\sigma^2)g(c)\propto c^{-|\boldsymbol{\gamma}|/2}\exp\left(-\frac{\boldsymbol{\beta}_{\boldsymbol{\gamma}}^T\boldsymbol{\beta}_{\boldsymbol{\gamma}}}{2c\sigma^2}\right) g(c).
\end{equation}

When $g$ is the generalized Zellner-Siow prior specified by (\ref{GZS}), (\ref{full:cond:c}) has a closed form.
Explicitly, $c$ follows $IG\left(a+|\boldsymbol{\gamma}|/2,p^{b_n}+\|\boldsymbol{\beta}_{\boldsymbol{\gamma}}\|^2/(2\sigma^2)\right)$.

When $g$ is the generalized hyper-$g$ prior specified in (\ref{Ghyper:gprior}),
(\ref{full:cond:c}) does not have a closed form. In this case, we have to incorporate an Metropolis-Hasting step.
Technically, we reparametrize $\kappa=\log{c}$. Then the full conditional of $\kappa$ is
$p(\kappa|\boldsymbol{\beta},\boldsymbol{\gamma},\sigma^2,t_n,\textbf{Z})=p_c(\exp(\kappa)|\boldsymbol{\beta},\boldsymbol{\gamma},\sigma^2,t_n,\textbf{Z})\cdot\exp(\kappa)$,
where $p_c(\cdot|\boldsymbol{\beta},\boldsymbol{\gamma},\sigma^2,t_n,\textbf{Z})$ denotes the full conditional of $c$ specified as in (\ref{full:cond:c}).
With $\kappa_{old}$ being the current value of $\kappa$, then generate $\kappa_{new}$ from $N(\kappa_{old},\sigma_\kappa^2)$, i.e., a normal proposal,
with $\sigma_\kappa^2$ being a fixed priori. Then we accept $\kappa_{new}$ with probability
$p(\kappa_{new}|\boldsymbol{\beta},\boldsymbol{\gamma},\sigma^2,t_n,\textbf{Z})/p(\kappa_{old}|\boldsymbol{\beta},\boldsymbol{\gamma},\sigma^2,t_n,\textbf{Z})$.

\section{Simulation study}\label{simulation:sec}

In this section, a simulation study is conducted to compare the performance of different methods.
In Example \ref{exam1}, we compare our approach based on the generalized Zellner-Siow (GZS) and generalized hyper-$g$ (GHG) priors with several popular Bayesian methods. Specifically,
we examined the posterior probability of the true model using different approaches. We also looked at the FCR and length
of the simultaneous credible intervals constructed using the GZS and GHG priors. In Example \ref{exam2}, we compare our approach with SIS-SCAD and ISIS-SCAD
considered by \cite{FL08}. The median size of the selected models and median estimation error are reported.

\subsection{Example 1} \label{exam1}

For the first simulation, the data were generated from $\textbf{Y}=\textbf{X}\boldsymbol{\beta}+\boldsymbol{\epsilon}$ with $\boldsymbol{\epsilon}\sim N(\textbf{0},\sigma^2 \textbf{I}_n)$.
The entries $X_{ij}$s of $\textbf{X}$ are standard normal with the correlation between $X_{ij_1}$ and $X_{ij_2}$ being $\rho^{|j_1-j_2|}$,
i.e., the AR(1) model. To better examine the performance,
we considered a variety of situations $\sigma^2=1, 2$, $(n,p,s_n)=(100,15,2), (200,15,2), (100,1000,10), (200,1000,10)$, and $\rho=0, 0.5$.
The choice of $\rho$ represents independence and relatively higher correlation among the predictors.
Note in these situations RIC and the benchmark prior method by \cite{FLS01} coincide with each other so we only considered RIC.
The true model coefficient is $\boldsymbol{\beta}^0=(\textbf{u}_{s_n/2}^+, \textbf{u}_{s_n/2}^-,\textbf{0}_{p-s_n})^T$ for $s_n=2,10$, where $\textbf{0}_{p-s_n}$ is the $(p-s_n)$-dimensional zero vector,
$\textbf{u}_{s_n/2}^+$ ($\textbf{u}_{s_n/2}^-$) is the $(s_n/2)$-dimensional vector with components uniformly generated from $[1,5]$ ($[-5,-1]$).

We fixed $\nu=6$ in (\ref{post:dist}) somewhat arbitrarily though we found other choices also performing well.
The prior on $t_n$ was set to be $p(t_n)=I(t_n\le n/2)$,
a commonly acceptable prior sparsity assumption in many high-dimensional problems.
For GZS defined as in (\ref{GZS}), we chose $a=0$ and $b_n=d$; for GHG defined as in (\ref{Ghyper:gprior}), we chose $\alpha_n=p^d+1$ and $b=0$.
To examine sensitivity, we considered $d=2.8, 3, 3.2$, and denote the corresponding GZS and GHG priors as GZS2.8, GZS3, GZS3.2
and GHG2.8, GHG3, GHG3.2. Our study relied on $N=100$ replicated data sets $\textbf{Z}_{(v)}=(\textbf{Y}_{(v)},\textbf{X}_{(v)})$ for $v=1,\ldots,N$. Based on each data $\textbf{Z}_{(v)}$,
we generated $10000$ samples from the posterior distribution based on any of the above mentioned approaches in a variety of settings.
The first 5000 samples served as burnins, and the second half were used to conduct computation. It takes about 440.30 seconds to generate
10000 posterior samples when $p=1000$ using the parallel computing techniques on a computer with 16 CPUs and 256 GB Memory.
Convergence of the Markov chains was monitored by Gelman-Rubin's statistics; see \cite{GCSR03}.

The results contain two parts.
First, we examined the empirical posterior probability of the true model
using BIC, RIC, Zellner-Siow (ZS), hyper-$g$ (HG) and hyper-$g/n$ (HGN) priors that were considered
in \cite{LPMCB08}, and GZS, GHG
introduced in Section \ref{sec:g:example:prior}. The empirical proportion of the true model, denoted as $\widehat{p(\boldsymbol{\gamma}^0|\textbf{Z}_{(v)})}$, is an estimate of $p(\boldsymbol{\gamma}^0|\textbf{Z}_{(v)})$.
For $\eta\in (0,1)$, define $F(\eta)=\# \{1\le v\le N|  \widehat{p(\boldsymbol{\gamma}^0|\textbf{Z}_{(v)})}>\eta\}/N$.
That is, $1-F(\eta)$ is the empirical distribution function of $\widehat{p(\boldsymbol{\gamma}^0|\textbf{Z}_{(v)})}$s.
Since $\widehat{p(\boldsymbol{\gamma}^0|\textbf{Z}_{(v)})}>0.5$ implies that the true model is selected, $F(0.5)$ measures the selection accuracy.
To further examine how significantly the true model is selected, we also looked at $F(0.9)$, i.e., the empirical proportion
of $\widehat{p(\boldsymbol{\gamma}^0|\textbf{Z}_{(v)})}$ greater than $0.9$.
For each of the above mentioned situations, we examined $F(\eta)$ for $\eta=0.5, 0.9$. Obviously, the larger value of $F(\eta)$
indicates better performance.

Our empirical finding (based on the R package \textit{BAS} provided by
www.stat.duke.edu/$\sim$clyde/BAS) reveals that the value of the hyperparameter in HG and HGN recommended by \cite{LPMCB08} cannot yield
high value (close to 1) of $\widehat{p(\boldsymbol{\gamma}^0|\textbf{Z}_{(v)})}$, though correct model selection can still be achieved since
it was found to be greater than $\widehat{p(\boldsymbol{\gamma}|\textbf{Z}_{(v)})}$ for any $\boldsymbol{\gamma}\neq\boldsymbol{\gamma}^0$.
For this reason, we chose the hyperparameter in HG and HGN to be 0.1
to achieve higher value of $\widehat{p(\boldsymbol{\gamma}^0|\textbf{Z}_{(v)})}$ (see Table \ref{Table:1}). The code was written in Matlab and is available upon request.

Table \ref{Table:1} summarizes the values of $F(0.5)$ and $F(0.9)$. We found that all the approaches demonstrate satisfactory performance
when $(p,s_n)=(15,2)$. With $\sigma^2$ and $\rho$ increasing, the selection performance is slightly affected but overall is accurate enough.
When $(p,s_n)=(1000,10)$, BIC, HG, ZS and HGN cannot select the correct model, while GZS and GHG can still accurately select the true model.
The worst situation is $\sigma^2=2, \rho=0.5$, in which $F(0.5)$ all decreases to $0.80$-$0.90$. Somewhat surprisingly, RIC
can still achieve values of $F(0.5)$ up to $0.70$ when $n=100$, and even up to $0.90$ when $n=200$. This is because RIC fixes $c=p^2$, a large number to yield more accurate selection.
However, it cannot give positive values of $F(0.9)$,
indicating insignificant selection of the true model. In contrast, both GZS and GHG can give values of $F(0.9)$ over $0.80$ when $n=100$,
and even over $0.90$ when $n=200$. The results also demonstrate that selection accuracy of GZS and GHG appears to be not much sensitive to $d\in [2.8,3.2]$
in all of the situations.

\begin{table}[htp]
{\scriptsize
\begin{center}
\begin{tabular}{ccccccccccc}\hline
                       &                       &        &\multicolumn{4}{c}{$n=100$}&\multicolumn{4}{c}{$n=200$}\\ \cline{4-11}
                       &                       &        &\multicolumn{2}{c}{$(p, s_n)=(15,2)$}&\multicolumn{2}{c}{$(1000,10)$}&\multicolumn{2}{c}{$(15,2)$}&\multicolumn{2}{c}{$(1000,10)$}\\ \cline{4-11}
$\sigma^2$             &$\rho$                 & Method & $F(0.5)$ & $F(0.9)$  & $F(0.5)$ & $F(0.9)$  & $F(0.5)$ & $F(0.9)$  & $F(0.5)$ & $F(0.9)$\\ \hline
1                      &0                      & BIC    &0.94&---&---&---&1&0.31&---&---\\
                       &                       & RIC    &0.97&0.05&0.73&---&0.99&0.38&0.98&---\\
                       &                       & ZS     &0.85&---& ---&---&0.98&---&---&---\\
                       &                       & HG     &0.96&0.56& ---&---&0.96&0.71&---&---\\
                       &                       & HGN    &0.94&0.53& ---&---&0.98&0.82&---&---\\
                       &                       & GZS2.8 &0.99&0.82&0.99&0.79&1&0.99&1&0.92\\
                       &                       & GHG2.8 &0.99&0.82&0.96&0.78&1&0.97&1&0.98\\
                       &                       & GZS3   &0.99&0.90&0.99&0.89&1&0.99&1&0.97\\
                       &                       & GHG3   &0.98&0.90&1&0.94&1&0.96&1&0.97\\
                       &                       & GZS3.2 &1&0.86&1&0.93&1&0.95&1&1\\
                       &                       & GHG3.2 &0.97&0.91&0.96&0.90&1&0.93&1&0.99\\  \cline{2-11}
                       &0.5                    & BIC    &0.90&---&---&---&0.97&0.24&---&---\\
                       &                       & RIC    &0.94&0.03&0.70&---&0.97&0.31&0.96&---\\
                       &                       & ZS     &0.75&---&---&---&0.95&---&---&---\\
                       &                       & HG     &0.95&0.49&---&---&0.94&0.68&---&---\\
                       &                       & HGN    &0.92&0.56&---&---&0.98&0.84&---&---\\
                       &                       & GZS2.8 &0.98&0.80&0.95&0.70&1&0.90&1&0.91\\
                       &                       & GHG2.8 &0.98&0.82&0.96&0.78&0.98&0.89&1&0.95\\
                       &                       & GZS3   &1&0.91&0.97&0.86&1&0.95&1&0.97\\
                       &                       & GHG3   &0.99&0.89&0.94&0.87&1&0.92&1&0.97\\
                       &                       & GZS3.2 &0.98&0.87&0.95&0.91&1&0.97&1&0.99\\
                       &                       & GHG3.2 &1&0.93&0.95&0.92&1&0.98&1&0.98\\ \cline{1-11}
2                      &0                      & BIC    &0.93&---&---&---&0.98&0.23&---&---\\
                       &                       & RIC    &0.95&0.02&0.74&---&1&0.34&0.96&---\\
                       &                       & ZS     &0.83&---&---&---&0.97&---&---&---\\
                       &                       & HG     &0.92&0.36&---&---&0.97&0.57&---&---\\
                       &                       & HGN    &0.84&0.38&---&---&0.90&0.50&---&---\\
                       &                       & GZS2.8 &1&0.83&0.93&0.72&1&0.90&1&0.97\\
                       &                       & GHG2.8 &0.97&0.76&0.98&0.83&1&0.90&0.98&0.93\\
                       &                       & GZS3   &0.99&0.93&0.93&0.82&1&0.95&1&0.95\\
                       &                       & GHG3   &1&0.82&0.98&0.88&1&0.94&1&0.97\\
                       &                       & GZS3.2 &0.99&0.92&0.96&0.92&1&0.93&1&1\\
                       &                       & GHG3.2 &1&0.94&0.95&0.90&1&0.95&1&0.98\\ \cline{2-11}
                       &0.5                    & BIC    &0.90&---&---&---&0.94&0.23&---&---\\
                       &                       & RIC    &0.93&0.01&0.64&---&0.95&0.30&0.90&---\\
                       &                       & ZS     &0.80&---&---&---&0.94&---&---&---\\
                       &                       & HG     &0.83&0.37&---&---&0.95&0.51&---&---\\
                       &                       & HGN    &0.81&0.34  &---&---&0.98&0.50&---&---\\
                       &                       & GZS2.8   &0.98&0.82&0.90&0.65&0.98&0.91&0.98&0.90\\
                       &                       & GHG2.8   &0.99&0.86&0.88&0.58&1&0.90&0.97&0.92\\
                       &                       & GZS3    &0.98&0.87&0.85&0.68&1&0.91&0.93&0.91\\
                       &                       & GHG3    &0.99&0.88&0.87&0.79&1&0.90&0.99&0.91\\
                       &                       & GZS3.2   &1&0.88&0.89&0.71&1&0.94&0.93&0.90\\
                       &                       & GHG3.2   &0.97&0.92&0.84&0.74&1&0.98&0.96&0.90\\\cline{1-11}
\end{tabular}
\end{center}
\caption{\footnotesize\textit{Values of $F(\eta)$ for $\eta=0.5,0.9$ in various settings. ``---" indicates a zero-value.}}
\label{Table:1}
}
\end{table}

Second, we computed the FCR and the length of the 95\% credible intervals for the selected coefficients (based on the highest posterior probability model),
when GZS and GHG with $d=2.8,3,3.2$ were used.
The 95\% credible intervals were constructed using the formula (\ref{Bayes:SCI}).
The posterior estimates of $c$ and $\sigma^2$, obtained through posterior averages of the $c$ and $\sigma^2$ chains, were plugged in to obtain the intervals.
We should point out that the credible intervals, together with the
empirical posterior probability of the true model, were jointly obtained through the posterior samples.
In other words, model selection and credible interval construction were jointly achieved,
reflecting the ``one-step" feature of the method. Based on the 100 replicated data sets, the FCR was calculated as the mean false coverage proportions,
and the average length was recognized as the mean length of the intervals for the selected coefficients.

Table \ref{Table:2} summarizes the results. We observed that the FCRs are all controlled by 5\% except for $(\sigma^2,\rho)=(2,0.5)$.
This is consistent with the finding by \cite{ZH12} who showed that the FCR of the simultaneous credible intervals can be controlled by
the nominal level for constructing the intervals, when signal-to-noise ratio is reasonably large. When $(\sigma^2,\rho)=(2,0.5)$, FCR tends to be around 10\% reflecting the effect of
higher correlation and model error. As $n$ increases, or $\rho$ and $\sigma^2$ decrease, the average lengths of the credible intervals for the selected coefficients
become shorter. The results also reveal that using GZS and GHG with different choices of $d\in [2.8,3.2]$, the performance of the simultaneous credible intervals
appears to be not much sensitive, at least in this simulation.

\begin{table}[htp]
{\scriptsize
\begin{center}
\begin{tabular}{ccccccccccc}\hline
                       &                       &        &\multicolumn{2}{c}{$n=100$}&\multicolumn{2}{c}{$n=200$}\\ \cline{4-7}
$\sigma^2$             &$\rho$                 & Method &$(p, s_n)=(15,2)$&$(1000,10)$&$(15,2)$&$(1000,10)$\\ \hline
1                      &0                      & GZS2.8 &5.50 (38.93)&7.40 (38.58)&5.17 (27.30)&5.90 (27.29)\\
                       &                       & GHG2.8 &5.67 (38.98)&7.55 (37.82)&3.17 (27.72)&6.49 (27.44)\\
                       &                       & GZS3   &5.50 (38.94)&7.40 (38.67)&5.17 (27.30)&5.90 (27.31)\\
                       &                       & GHG3   &5.67 (39.00)&7.37 (37.97)&3.17 (27.73)&6.59 (27.45)\\
                       &                       & GZS3.2 &5.50 (38.96)&7.40 (38.72)&6.50 (27.30)&5.90 (27.29)\\
                       &                       & GHG3.2 &5.67 (39.00)&7.27 (38.07)&3.17 (27.73)&6.50 (27.46)\\ \cline{2-7}
                       &0.5                    & GZS2.8 &3.00 (44.58)&7.04 (47.62)&4.00 (32.18)&4.80 (35.10)\\
                       &                       & GHG2.8 &6.83 (44.59)&7.33 (47.10)&5.50 (31.73)&6.80 (35.01)\\
                       &                       & GZS3   &3.00 (44.60)&6.49 (47.67)&4.00 (32.18)&4.80 (35.13)\\
                       &                       & GHG3   &6.83 (44.61)&7.05 (47.14)&5.50 (31.75)&6.80 (35.02)\\
                       &                       & GZS3.2 &3.00 (44.62)&6.49 (47.79)&3.50 (32.19)&4.80 (35.17)\\
                       &                       & GHG3.2 &6.83 (44.61)&7.05 (47.29)&5.50 (31.75)&6.80 (35.02)\\ \cline{1-7}
2                      &0                      & GZS2.8 &6.50 (54.61)&6.00 (54.52)&4.00 (38.89)&6.40 (39.66)\\
                       &                       & GHG2.8 &5.17 (54.26)&7.37 (56.52)&6.50 (38.36)&5.99 (40.17)\\
                       &                       & GZS3   &6.50 (54.62)&6.22 (54.71)&4.00 (38.89)&6.50 (39.68)\\
                       &                       & GHG3   &4.83 (54.27)&7.11 (56.72)&6.50 (38.36)&5.90 (40.19)\\
                       &                       & GZS3.2 &6.50 (54.64)&6.22 (54.91)&4.00 (38.90)&6.40 (39.68)\\
                       &                       & GHG3.2 &4.83 (54.28)&7.01 (56.91)&6.50 (38.36)&5.90 (40.21)\\ \cline{2-7}
                       &0.5                    & GZS2.8 &6.00 (63.07)&8.59 (65.31)&6.83 (45.30)&5.60 (48.37)\\
                       &                       & GHG2.8 &4.50 (62.33)&8.68 (68.33)&4.50 (45.09)&6.10 (49.16)\\
                       &                       & GZS3   &6.00 (63.08)&9.09 (65.42)&6.83 (45.30)&5.50 (48.37)\\
                       &                       & GHG3   &4.00 (62.35)&9.96 (68.72)&4.50 (45.09)&5.92 (49.16)\\
                       &                       & GZS3.2 &5.50 (63.10)&9.19 (65.71)&6.83 (45.30)&5.60 (48.35)\\
                       &                       & GHG3.2 &4.50 (62.35)&9.90 (68.84)&4.50 (45.09)&5.92 (49.16)\\ \hline
\end{tabular}
\end{center}
\caption{\footnotesize\textit{100$\times$FCR (100$\times$average length) of the 95\% credible intervals for the selected coefficients constructed by GZS and GHG in various settings.}}
\label{Table:2}
}
\end{table}

\subsection{Example 2}\label{exam2}

In our second study, we adopted two simulation settings in \cite{FL08}. In Setting I,
$N=200$ data sets were generated from $\textbf{Y}=\textbf{X}\boldsymbol{\beta}+\boldsymbol{\epsilon}$ with $\boldsymbol{\epsilon}\sim N(\textbf{0},1.5^2 \textbf{I}_n)$,
where $\textbf{X}$ is $n\times p$ containing i.i.d standard Gaussian entries.
We considered $(n,p,s_n)=(200,1000,8)$ and $(800,20000,18)$, where recall $s_n$ represents the size of the true model.
In each data replication, the $s_n$ nonzero coefficients were chosen to be $(-1)^u (a+|z|)$, where $u$ was drawn from Bernoulli distribution with parameter 0.4,
$z$ was drawn from standard Gaussian distribution, and $a=4\log{n}/\sqrt{n}$ and $5\log{n}/\sqrt{n}$ corresponding to the two situations.
In \cite{FL08}, the median size of the selected models and the median of $\|\widehat{\boldsymbol{\beta}}-\boldsymbol{\beta}^0\|$ obtained from SIS-SCAD and ISIS-SCAD were reported. In Bayesian approaches,
we also looked at the median size of the selected models with the highest posterior probability, and the median of $\|\widehat{\boldsymbol{\beta}}-\boldsymbol{\beta}^0\|$,
where $\widehat{\boldsymbol{\beta}}$ was found by posterior mean of the $\boldsymbol{\beta}$ samples.
To demonstrate how stable the posterior estimate is, we also looked the standard deviations of $\|\widehat{\boldsymbol{\beta}}-\boldsymbol{\beta}^0\|$s.
We fixed $\nu=6$ in (\ref{post:dist}). The prior on $t_n$ was set to be $p(t_n)=I(t_n\le n/2)$.
Due to computational cost, we generated Markov chains with length 4000 and 1000 for $(n,p,s_n)=(200,1000,8)$ and $(800,20000,18)$ respectively.
Using Gelman-Rubin's statistics, we found that the Markov chains appear to mix well.

In Table \ref{Table:3}, we compared the median size of the selected models (MSSM) and the median of the error $\|\widehat{\boldsymbol{\beta}}-\boldsymbol{\beta}^0\|$ (ME)
obtained from SIS-SCAD, ISIS-SCAD, and the proposed Bayesian method with GZS3 and GHG3 priors, in Setting I. The performance of GZS and GHG priors with $d=2.8$ and $3.2$ is similar,
and thus, was not reported. Results based on SIS-SCAD and ISIS-SCAD were summarized from \cite{FL08}.
We observed that all the four methods yield satisfactory accuracy in coefficient estimation, and GZS3 and GHG3 perform slightly better
in yielding the correct model size. The standard deviations of $\|\widehat{\boldsymbol{\beta}}-\boldsymbol{\beta}^0\|$
using both GZS3 and GHG3 priors are around 0.08 and 0.04 (for $p=1000,20000$), reflecting the stability of
the two approaches.

\begin{table}[htp]
{\scriptsize
\begin{center}
\begin{tabular}{ccccc}\hline
$(n,p,s_n)$&SIS-SCAD&ISIS-SCAD&GZS3&GHG3\\ \hline
(200,1000,8)&15 (0.374)&13 (0.329)&8 (0.2811)&8 (0.2806)\\
            &          &          &(0.0784)&(0.0783)\\
(800,20000,18)&37 (0.288)&31 (0.246)&18 (0.2252)&18 (0.2257)\\
              &       &      &   (0.0329) & (0.0360)\\ \hline
\end{tabular}
\end{center}
\caption{\footnotesize \textit{MSSM and ME based on SIS-SCAD, ISIS-SCAD, GZS3 and GHG3 for Setting I. For SIS-SCAD and ISIS-SCAD, the numbers in the parentheses represent the MEs.
For GZS3 and GHG3, the numbers in the parentheses represent MEs (upper) and standard deviations of $\|\widehat{\boldsymbol{\beta}}-\boldsymbol{\beta}^0\|$ (lower).}}
\label{Table:3}
}
\end{table}

In Setting II, $N=200$ data sets were generated from $\textbf{Y}=\textbf{X}\boldsymbol{\beta}+\boldsymbol{\epsilon}$ with $\boldsymbol{\epsilon}\sim N(\textbf{0},\sigma^2 \textbf{I}_n)$. We considered three situations
$(n,p,s_n)=(200,1000,5), (200,1000,8), (800,20000,14)$. Correspondingly, we chose $(\sigma,a)=(1,2\log{n}/\sqrt{n}), (1.5,4\log{n}/\sqrt{n}), (2,4\log{n}/\sqrt{n})$.
The true coefficient vector $\beta^0$ was generated using the same strategy described in Setting I. The major difference in Setting II lies in generating the $X$ matrix.
Explicitly, the $s_n$ predictors $X_1,\ldots,X_{s_n}$ were generated from $N(\textbf{0},\textbf{A})$ for some positive definite covariance matrix $\textbf{A}$ with condition number $\sqrt{n}/\log{n}$.
The procedure for producing $\textbf{A}$ was described in \cite{FL08}. Then we drew $W_{s_n+1},\ldots,W_{p}$ from $N(\textbf{0},\textbf{I}_{p-s_n})$, set
$X_{j}=W_j+rX_{j-s_n}$ for $j=s_n+1,\ldots,2s_n$, and $X_j=W_j+(1-r)X_1$ for $j=2s_n+1,\ldots,p$. Here $r=1-4\log{n}/p, 1-5\log{n}/p, 1-5\log{n}/p$ for the three situations.
We still fixed $\nu=6$ in (\ref{post:dist}). The prior on $t_n$ was set to be $p(t_n)=I(t_n\le n/2)$.
The Markov chains have length 4000 and 1000 for $p=1000$ and $20000$ respectively. The chains appear to converge based on Gelman-Rubin's statistics.
In Table \ref{Table:4}, the MSSMs and the MEs of $\|\widehat{\boldsymbol{\beta}}-\boldsymbol{\beta}^0\|$ obtained from the four methods in Setting II were summarized.
Although the covariate variables now have certain dependence structure, all the four methods still perform well. In particular, GZS3 and GHG3 yield more satisfactory selection and estimation accuracy,
and produce stable results.

\begin{table}[htp]
{\scriptsize
\begin{center}
\begin{tabular}{ccccc}\hline
$(n,p,s_n)$&SIS-SCAD&ISIS-SCAD&GZS3&GHG3\\ \hline
(200,1000,5)&21 (0.331)&11 (0.223)&5 (0.1570)&5 (0.1559)\\
            &          &          &(0.0478)&(0.0477)\\
(200,1000,8)&18 (0.458)&13.5 (0.366)&8 (0.2947) &8 (0.2959)\\
            &          &          & (0.0732)&(0.0731)\\
(800,20000,14)&36 (0.367)&27 (0.315)&14 (0.2633) & 14 (0.2631) \\
              &          &           &(0.0543)& (0.0466)\\ \hline
\end{tabular}
\end{center}
\caption{\footnotesize \textit{MSSM and ME based on SIS-SCAD, ISIS-SCAD, GZS3 and GHG3 for Setting II. For SIS-SCAD and ISIS-SCAD, the numbers in the parentheses represent the MEs.
For GZS3 and GHG3, the numbers in the parentheses represent MEs (upper) and standard deviations of $\|\widehat{\boldsymbol{\beta}}-\boldsymbol{\beta}^0\|$ (lower).}}
\label{Table:4}
}
\end{table}

\section{Conclusions}

We examined posterior consistency of a fully Bayesian method in sparse high-dimensional settings.
As revealed in our main results, the prior (\ref{fbm:5}) plays an important role. This prior plays the same role as a dimension reduction step.
The difference is, unlike other methods in which dimension reduction is a separate step, using (\ref{fbm:5})
dimension reduction is fulfilled automatically and stochastically in the process of Bayesian model fitting and MCMC search,
and thus, all the statistical procedures are conducted in a unified framework. This ``one-step" fashion differs our method from the existing ones.

Tables \ref{Table:1} and \ref{Table:2} demonstrate the numerical performance of the proposed method. Overall, the performance is not much sensitive to
the choice of hypeparameter $d$ in GZS and GHG. In practice, we recommend to use $d=3$ which, at least in our simulation settings, display satisfactory accuracy.
Other choices close to it yield not much different results.

Two extensions of our method to other scopes are worth mentioning. The first one is the high-dimensional Gaussian graphical model
in which the goal is to find the associated genes through estimating the sparse precision matrix. As is well known that this problem
can be solved by Bayesian model selection approach in a completely different setting; see \cite{CS09} and the references therein.
It is possible that we can apply a prior similar to (\ref{fbm:5}) to control the size of genes during the model fitting and conduct a stochastic
search to find the associated genes.

The second direction that we intend to explore is whether our approach can be extended to generalized linear models with high-dimensionality.
Ideally, a fully Bayesian framework endowed with MCMC is possible to simplify the selection procedure, and meanwhile, conduct estimation and inference
over the selected variables. It remains open whether such computing methods can be proposed in more general modeling framework.
It is well known that in generalized linear model the posterior distribution of the model does not have closed forms.
A common method is to apply Laplace approximation; see, e.g., \cite{WG07}.
However, as pointed out by \cite{SM95} that the approximation error cannot be easily controlled in higher dimensional settings.
An alternative way might be first showing uniform convergence of the posterior probability
by fixing certain hyperparameters like Theorems \ref{main:thm1}--\ref{main:thm3},
then generalizing this to more broader situations where the posterior probability can be expressed as an intractable integral; see, e.g., Section \ref{sec:g-prior}.


\section*{\Large APPENDIX: Proofs}

To prove Theorem \ref{main:thm1}, we need the following preliminary lemma.

\begin{lemma}\label{lemma1} Suppose $\boldsymbol{\epsilon}\sim N(\textbf{0},\sigma_0^2  \textbf{I}_n)$ and recall the true model is $\textbf{Y}=\textbf{X}\boldsymbol{\beta}^0+\boldsymbol{\epsilon}$.
\begin{enumerate}[(i).]

\item Let $\boldsymbol{\nu}_{\boldsymbol{\gamma}}$ be an $n$-dimensional vector indexed by $\boldsymbol{\gamma}\in \mathcal{S}$, a subset of the model space.
Adopt the convention that $\boldsymbol{\nu}_{\boldsymbol{\gamma}}^T\boldsymbol{\epsilon}/\|\boldsymbol{\nu}_{\boldsymbol{\gamma}}\|=0$ when $\boldsymbol{\nu}_{\boldsymbol{\gamma}}=0$.
Let $\#\mathcal{S}$ denote the cardinality of $\mathcal{S}$ with $\#\mathcal{S}\ge 2$. Then
\begin{equation}\label{lemma1:eq1}
\max\limits_{\boldsymbol{\gamma}\in \mathcal{S}} \frac{|\boldsymbol{\nu}_{\boldsymbol{\gamma}}^T\boldsymbol{\epsilon}|}{\|\boldsymbol{\nu}_{\boldsymbol{\gamma}}\|}=O_\Pr\left(\sqrt{\log(\#\mathcal{S})}\right).
\end{equation}
In particular, let $\boldsymbol{\nu}_{\boldsymbol{\gamma}}=(\textbf{I}_n-\textbf{P}_{\boldsymbol{\gamma}})
\textbf{X}_{\boldsymbol{\gamma}^0\backslash\boldsymbol{\gamma}}\boldsymbol{\beta}_{\boldsymbol{\gamma}^0\backslash\boldsymbol{\gamma}}^0$ for $\boldsymbol{\gamma} \in S_2(t_n)$, we have
\begin{equation}\label{lemma1:eq2}
\max\limits_{s_n\le t_n\le r_n}\max\limits_{\boldsymbol{\gamma}\in S_2(t_n)} \frac{|\boldsymbol{\nu}_{\boldsymbol{\gamma}}^T\boldsymbol{\epsilon}|}{\|\boldsymbol{\nu}_{\boldsymbol{\gamma}}\|}=O_\Pr(\sqrt{r_n\log{p}}).
\end{equation}

\item For any fixed $\alpha>2$,
\[
\lim\limits_{n\rightarrow\infty}\Pr\left(\max\limits_{t_n\in [s_n,r_n]}\max\limits_{\boldsymbol{\gamma}\in S_1(t_n)}\boldsymbol{\epsilon}^T
(\textbf{P}_{\boldsymbol{\gamma}}-\textbf{P}_{\boldsymbol{\gamma}^0})\boldsymbol{\epsilon}/(|\boldsymbol{\gamma}|-s_n) \le \alpha\sigma_0^2\log{p}\right)=1.
\]

\item Adopt the convention that $\boldsymbol{\epsilon}^T \textbf{P}_{\boldsymbol{\gamma}}\boldsymbol{\epsilon}/|\gamma|=0$ when $\gamma$ is
null. Then for any fixed $\alpha>2$,
\[
\lim\limits_{n\rightarrow\infty}\Pr\left(\max\limits_{t_n\in [s_n,r_n]}\max\limits_{\boldsymbol{\gamma}\in S_2(t_n)}\boldsymbol{\epsilon}^T \textbf{P}_{\boldsymbol{\gamma}}\boldsymbol{\epsilon}/|\boldsymbol{\gamma}| \le \alpha\sigma_0^2\log{p}\right)=1.
\]

\end{enumerate}
\end{lemma}

\subsection*{Proof of Lemma \ref{lemma1}}
The proof of (ii) and (iii) is a trivial modification of Lemma A.1 in \cite{SC11}. Next we only show (i).
For any $\boldsymbol{\nu}_{\boldsymbol{\gamma}}\neq0$, $\frac{\boldsymbol{\nu}_{\boldsymbol{\gamma}}^T\boldsymbol{\epsilon}}{\sigma_0\|\boldsymbol{\nu}_{\boldsymbol{\gamma}}\|}\sim N(0,1)$. By (9.3) of \cite{D05}, if $\xi\sim N(0,1)$,
then $\Pr(|\xi|\ge t)\le C_0\exp(-t^2/2)$ for some positive constant $C_0$. Therefore,
\begin{eqnarray*}
\Pr\left(\max\limits_{\boldsymbol{\gamma}\in \mathcal{S}} \frac{|\boldsymbol{\nu}_{\boldsymbol{\gamma}}^T\boldsymbol{\epsilon}|}{\|\boldsymbol{\nu}_{\boldsymbol{\gamma}}\|}> \sigma_0C\sqrt{\log(\#\mathcal{S})}\right)
\le\sum\limits_{\boldsymbol{\gamma}\in \mathcal{S}}\Pr\left(\frac{|\boldsymbol{\nu}_{\boldsymbol{\gamma}}^T\boldsymbol{\epsilon}|}{\|\boldsymbol{\nu}_{\boldsymbol{\gamma}}\|}> \sigma_0C\sqrt{\log(\#\mathcal{S})}\right)
\le C_0\# \mathcal{S} \cdot (\#\mathcal{S})^{-C^2/2}=C_0 (\#\mathcal{S})^{1-C^2/2},
\end{eqnarray*}
which is small when $C>0$ is chosen as sufficiently large. This shows (\ref{lemma1:eq1}).
To show (\ref{lemma1:eq2}), consider $\mathcal{S}=\bigcup\limits_{s_n\le t_n\le r_n}S_2(t_n)$. Clearly $\mathcal{S}\subset S_2(r_n)$.
Note $\# S_2(r_n)\le {p \choose 1}+\ldots {p\choose r_n}\le \sum\limits_{l\le r_n} p^l/l!\le p^{r_n}$.  Thus we have $\#\mathcal{S}\le p^{r_n}$.
Thus, plugging this into (\ref{lemma1:eq1}), we get (\ref{lemma1:eq2}).

\subsection*{Proof of Theorem \ref{main:thm1}}
The idea of the proof is to derive explicit upper bounds (uniform for the variance-control parameters $c_j$s) for the ratio
$\frac{p(\boldsymbol{\gamma}|\textbf{Z})}{p(\boldsymbol{\gamma}^0|\textbf{Z})}$, where $\gamma\neq\gamma^0$.
By showing that the sum of these upper bounds converges to zero, and using the trivial fact
$p(\boldsymbol{\gamma}^0|\textbf{Z})=\frac{1}{1+\sum\limits_{\boldsymbol{\gamma}\neq\boldsymbol{\gamma}^0}\frac{p(\boldsymbol{\gamma}|\textbf{Z})}{p(\boldsymbol{\gamma}^0|\textbf{Z})}}$,
we will conclude $p(\boldsymbol{\gamma}^0|\textbf{Z})\rightarrow 1$. Throughout the proofs of our theoretical results, we use the shortcut ``w.l.p." to denote the terminology
``with large probability". For any $s_n\le t_n\le r_n$,
We consider the following decomposition for $\boldsymbol{\gamma}\in S_1(t_n)\bigcup S_2(t_n)$,
\begin{eqnarray*}
-\log\left(\frac{p(\boldsymbol{\gamma}|\textbf{Z})}{p(\boldsymbol{\gamma}^0|\textbf{Z})}\right)
&=&-\log\left(\frac{p(\boldsymbol{\gamma})}{p(\boldsymbol{\gamma}^0)}\right)
+\frac{1}{2}\log\left(\frac{\det(\textbf{W}_{\boldsymbol{\gamma}})}{\det(\textbf{W}_{\boldsymbol{\gamma}^0})}\right)+\frac{n+\nu}{2}\log\left(\frac{1+\textbf{Y}^T(\textbf{I}_n-\textbf{X}_{\boldsymbol{\gamma}} \textbf{U}_{\boldsymbol{\gamma}}^{-1} \textbf{X}_{\boldsymbol{\gamma}}^T)\textbf{Y}}{1+\textbf{Y}^T(\textbf{I}_n-\textbf{P}_{\boldsymbol{\gamma}})\textbf{Y}}\right)\\
&&-\frac{n+\nu}{2}\log\left(\frac{1+\textbf{Y}^T(\textbf{I}_n-\textbf{X}_{\boldsymbol{\gamma}^0}\textbf{U}_{\boldsymbol{\gamma}^0}^{-1}\textbf{X}_{\boldsymbol{\gamma}^0}^T)\textbf{Y}}
{1+\textbf{Y}^T(\textbf{I}_n-\textbf{P}_{\boldsymbol{\gamma}^0})\textbf{Y}}\right)+\frac{n+\nu}{2}\log\left(\frac{1+\textbf{Y}^T(\textbf{I}_n-\textbf{P}_{\boldsymbol{\gamma}})\textbf{Y}}
{1+\textbf{Y}^T(\textbf{I}_n-\textbf{P}_{\boldsymbol{\gamma}^0})\textbf{Y}}\right).
\end{eqnarray*}
Denote the above five terms by $I_1, I_2, I_3, I_4, I_5$. Next we approximate these terms respectively.

By Assumption \ref{A1}, $I_1$ is bounded from below. Since $\textbf{U}_{\boldsymbol{\gamma}}\ge \textbf{P}_{\boldsymbol{\gamma}}$, $I_3\ge 0$. By assumption $k_n=O(\underline{\phi}_n)$, $n\psi_n^2\rightarrow\infty$,
and the proof of Theorem 2.2 in \cite{SC11}, $0\le -I_4=O_\Pr(1)$. Next we approximate $I_5$. For $\boldsymbol{\gamma}\in S_2(t_n)$,
let $\boldsymbol{\nu}_{\boldsymbol{\gamma}}=
(\textbf{I}_n-\textbf{P}_{\boldsymbol{\gamma}})\textbf{X}_{\boldsymbol{\gamma}^0\backslash\boldsymbol{\gamma}}\boldsymbol{\beta}^0_{\boldsymbol{\gamma}^0\backslash\boldsymbol{\gamma}}$.
Note Assumption \ref{A3} (iii) implies $r_n\log{p}=o(n\psi_n^2)$.
By Assumption \ref{A1}, it can be shown that
\begin{eqnarray*}
\|\boldsymbol{\nu}_{\boldsymbol{\gamma}}\|^2=
(\boldsymbol{\beta}^0_{\boldsymbol{\gamma}^0\backslash\boldsymbol{\gamma}})^T\textbf{X}_{\boldsymbol{\gamma}^0\backslash\boldsymbol{\gamma}}^T
(\textbf{I}_n-\textbf{P}_{\boldsymbol{\gamma}})\textbf{X}_{\boldsymbol{\gamma}^0\backslash\boldsymbol{\gamma}}\boldsymbol{\beta}^0_{\boldsymbol{\gamma}^0\backslash\boldsymbol{\gamma}}
\ge nc_0^{-1}\|\boldsymbol{\beta}^0_{\boldsymbol{\gamma}^0\backslash\boldsymbol{\gamma}}\|^2\ge nc_0^{-1}\psi_n^2.
\end{eqnarray*}
Note $(\textbf{I}_n-\textbf{P}_{\boldsymbol{\gamma}})\textbf{X}_{\boldsymbol{\gamma}^0}=(\textbf{0},(\textbf{I}_n-\textbf{P}_{\boldsymbol{\gamma}})\textbf{X}_{\boldsymbol{\gamma}^0\backslash\boldsymbol{\gamma}})$.
Then by Lemma \ref{lemma1} (i) and (iii) we have for some fixed $\alpha>2$, w.l.p, for $s_n\le t_n\le r_n$ and uniformly over $\boldsymbol{\gamma}\in S_2(t_n)$,
\begin{eqnarray*}
\textbf{Y}^T(\textbf{I}_n-\textbf{P}_{\boldsymbol{\gamma}})\textbf{Y}&=&\|\boldsymbol{\nu}_{\boldsymbol{\gamma}}\|_2^2+2\boldsymbol{\nu}_{\boldsymbol{\gamma}}^T\boldsymbol{\epsilon}+\boldsymbol{\epsilon}^T
(\textbf{I}_n-\textbf{P}_{\boldsymbol{\gamma}})\boldsymbol{\epsilon}\nonumber\\
&\ge&\|\boldsymbol{\nu}_{\boldsymbol{\gamma}}\|_2^2-4\sigma_0\|\boldsymbol{\nu}_{\boldsymbol{\gamma}}\|\sqrt{r_n\log{p}}+\boldsymbol{\epsilon}^T\boldsymbol{\epsilon}-\alpha\sigma_0^2|\boldsymbol{\gamma}|\log{p}\\
&\ge& \|\boldsymbol{\nu}_{\boldsymbol{\gamma}}\|_2^2\left(1-4\sigma_0\frac{\sqrt{r_n\log{p}}}{\|\boldsymbol{\nu}_{\boldsymbol{\gamma}}\|}
-\alpha\sigma_0^2\frac{t_n\log{p}}{\|\boldsymbol{\nu}_{\boldsymbol{\gamma}}\|^2}\right)+\boldsymbol{\epsilon}^T\boldsymbol{\epsilon}\nonumber\\
&=&\|\boldsymbol{\nu}_{\boldsymbol{\gamma}}\|_2^2(1+o(1))+\boldsymbol{\epsilon}^T\boldsymbol{\epsilon}\nonumber\\
&\ge& nc_0^{-1}\psi_n^2(1+o(1))+\boldsymbol{\epsilon}^T\boldsymbol{\epsilon}.
\end{eqnarray*}
Since $s_n=o(n)$, $\boldsymbol{\epsilon}^T(\textbf{I}_n-\textbf{P}_{\boldsymbol{\gamma}^0})\boldsymbol{\epsilon}=n\sigma_0^2(1+o_\Pr(1))$.
Therefore, there exists a constant $C'>0$ (not depending on $\gamma$) such that w.l.p., for $s_n\le t_n\le r_n$ and for any $\boldsymbol{\gamma}\in S_2(t_n)$,
\begin{eqnarray}\label{thm1:eq1}
I_5\ge\frac{n+\nu}{2}\log\left(\frac{1+nc_0^{-1}\psi_n^2(1+o(1))
+\boldsymbol{\epsilon}^T\boldsymbol{\epsilon}}{1+\boldsymbol{\epsilon}^T(\textbf{I}_n-\textbf{P}_{\boldsymbol{\gamma}^0})\boldsymbol{\epsilon}}\right)\ge\frac{n+\nu}{2}\log\left(1+C'\psi_n^2\right).
\end{eqnarray}

On the other hand, for any fixed $\alpha'>\alpha$, by properties of projection matrices and Lemma \ref{lemma1} (ii), we have,
w.l.p, for $s_n\le t_n\le r_n$ and uniformly for $\boldsymbol{\gamma}\in S_1(t_n)$,
\begin{eqnarray*}
\frac{1+\textbf{Y}^T(\textbf{I}_n-\textbf{P}_{\boldsymbol{\gamma}})\textbf{Y}}{1+\textbf{Y}^T(\textbf{I}_n-\textbf{P}_{\boldsymbol{\gamma}^0})\textbf{Y}}
&=&1-\frac{\textbf{Y}^T(\textbf{P}_{\boldsymbol{\gamma}}-\textbf{P}_{\boldsymbol{\gamma}^0})\textbf{Y}}{1+\textbf{Y}^T(\textbf{I}_n-\textbf{P}_{\boldsymbol{\gamma}^0})\textbf{Y}}\nonumber\\
&=&1-\frac{(\boldsymbol{\beta}_{\boldsymbol{\gamma}^0}^0)^T\textbf{X}_{\boldsymbol{\gamma}^0}^T(\textbf{P}_{\boldsymbol{\gamma}}-\textbf{P}_{\boldsymbol{\gamma}^0})
\textbf{X}_{\boldsymbol{\gamma}^0}\boldsymbol{\beta}_{\boldsymbol{\gamma}^0}
+2(\boldsymbol{\beta}_{\boldsymbol{\gamma}^0}^0)^T\textbf{X}_{\boldsymbol{\gamma}^0}^T(\textbf{P}_{\boldsymbol{\gamma}}-\textbf{P}_{\boldsymbol{\gamma}^0})\boldsymbol{\epsilon}+
\boldsymbol{\epsilon}^T(\textbf{P}_{\boldsymbol{\gamma}}-\textbf{P}_{\boldsymbol{\gamma}^0})\boldsymbol{\epsilon}}
{1+\textbf{Y}^T(\textbf{I}_n-\textbf{P}_{\boldsymbol{\gamma}^0})\textbf{Y}}\nonumber\\
&=&1-\frac{\boldsymbol{\epsilon}^T(\textbf{P}_{\boldsymbol{\gamma}}-\textbf{P}_{\boldsymbol{\gamma}^0})\boldsymbol{\epsilon}}
{1+\boldsymbol{\epsilon}^T(\textbf{I}_n-\textbf{P}_{\boldsymbol{\gamma}^0})\boldsymbol{\epsilon}}\\
&\ge & 1-\frac{\alpha\sigma_0^2 (|\boldsymbol{\gamma}|-s_n)}{n\sigma_0^2(1+o_\Pr(1))}\\
&\ge & 1-\frac{\alpha'(|\boldsymbol{\gamma}|-s_n)\log{p}}{n}.
\end{eqnarray*}
It follows by the inequality that $\log(1-x)\ge -2x$ when $x\in (0,1/2)$,
and by Assumption \ref{A3} (iii) which implies that $(|\boldsymbol{\gamma}|-s_n)\log{p}/n$
approaches zero uniformly for $\boldsymbol{\gamma}\in S_1(t_n)$ with $s_n\le t_n\le r_n$. Therefore, for large $n$, w.l.p, for $s_n\le t_n\le r_n$ and uniformly for $\boldsymbol{\gamma}\in S_1(t_n)$,
\begin{equation}\label{thm1:eq2}
I_5\ge\frac{n+\nu}{2}\log\left(1-\frac{\alpha'(|\boldsymbol{\gamma}|-s_n)\log{p}}{n}\right)\ge -\alpha_0 (|\boldsymbol{\gamma}|-s_n)\log{p},
\end{equation}
where $\alpha_0=2\alpha'$. It follows by Lemma A.2 in \cite{SC11} that
\begin{equation}\label{thm1:eq3}
I_2\ge 2^{-1}(|\boldsymbol{\gamma}|-s_n)\log(1+c_0^{-1}n\underline{\phi}_n),\,\,\textrm{for $s_n\le t_n\le r_n$ and uniformly for $\boldsymbol{\gamma}\in S_1(t_n)$, and}
\end{equation}
\begin{equation}\label{thm1:eq4}
I_2\ge -2^{-1}s_n\log(1+c_0n\bar{\phi}_n),\,\,\textrm{for $s_n\le t_n\le r_n$ and uniformly for $\boldsymbol{\gamma}\in S_2(t_n)$}.
\end{equation}

By Assumption \ref{A3} (v), $\log{p}=o(\log(1+c_0^{-1}n\underline{\phi}_n))$
Using (\ref{thm1:eq1})--(\ref{thm1:eq4}), we have, w.l.p, for $s_n\le t_n\le r_n$ and uniformly for $\underline{\phi}_n\le c_j\le \bar{\phi}_n$,
\begin{equation}\label{thm1:eq5}
\frac{p(\boldsymbol{\gamma}|\textbf{Z})}{p(\boldsymbol{\gamma}^0|\textbf{Z})}\le \widetilde{C}\,\,\left(\frac{1+c_0^{-1}n\underline{\phi}_n}{p^{2\alpha_0}}\right)^{-2^{-1}(|\boldsymbol{\gamma}|-s_n)},\,\,\boldsymbol{\gamma}\in S_1(t_n), s_n\le t_n\le r_n,
\end{equation}
and
\begin{eqnarray}\label{thm1:eq6}
\frac{p(\boldsymbol{\gamma}|\textbf{Z})}{p(\boldsymbol{\gamma}^0|\textbf{Z})}&\le& \widetilde{C}\,\,\exp\left(2^{-1} s_n\log(1+c_0n\bar{\phi}_n)-\frac{n+\nu}{2}\log(1+C'\psi_n^2)\right)\nonumber\\
&\le&\widetilde{C}\,\,(1+C'\psi_n^2)^{-\frac{n+\nu}{4}},\,\,\boldsymbol{\gamma}\in S_2(t_n), s_n\le t_n\le r_n.
\end{eqnarray}
It follows by (\ref{thm1:eq5}), (\ref{thm1:eq6}), and Assumption \ref{A3} (iii) and (v), as $n\rightarrow\infty$,
\begin{eqnarray*}
\sum\limits_{\boldsymbol{\gamma}\in S_1(t_n)}\frac{p(\boldsymbol{\gamma}|\textbf{Z})}{p(\boldsymbol{\gamma}^0|\textbf{Z})}&\le& \widetilde{C}\sum\limits_{\boldsymbol{\gamma}\in S_1(t_n)} \left(\frac{1+c_0^{-1}n\underline{\phi}_n}{p^{2\alpha_0}}\right)^{-2^{-1}(|\boldsymbol{\gamma}|-s_n)}\\
&=&\widetilde{C}\sum\limits_{r=s_n+1}^{t_n} {p-s_n \choose r-s_n}\left(\frac{1+c_0^{-1}n\underline{\phi}_n}{p^{2\alpha_0}}\right)^{-2^{-1}(r-s_n)}\\
&\le&\widetilde{C}\sum\limits_{r=1}^{t_n-s_n}\frac{(p-s_n)^r}{r!}\left(\frac{1+c_0^{-1}n\underline{\phi}_n}{p^{2\alpha_0}}\right)^{-2^{-1}r}\\
&=&\widetilde{C}\sum\limits_{r=1}^{t_n-s_n}\frac{1}{r!}\left[(p-s_n)\left(\frac{1+c_0^{-1}n\underline{\phi}_n}{p^{2\alpha_0}}\right)^{-2^{-1}}\right]^r\\
&\le&\widetilde{C}\left(\exp\left(\sqrt{p^{2\alpha_0+2}/(1+c_0^{-1}n\underline{\phi}_n)}\right)-1\right)\rightarrow0,
\end{eqnarray*}
and
\begin{eqnarray*}
\sum\limits_{\boldsymbol{\gamma}\in S_2(t_n)}\frac{p(\boldsymbol{\gamma}|\textbf{Z})}{p(\boldsymbol{\gamma}^0|\textbf{Z})}&\le& \widetilde{C}\,\,\#S_2(t_n)\cdot (1+C'\psi_n^2)^{-\frac{n+\nu}{4}}
\le \widetilde{C}\,\,p^{t_n}(1+C'\psi_n^2)^{-\frac{n+\nu}{4}}\rightarrow0.
\end{eqnarray*}
Note the above convergence holds in probability and is uniform for $\underline{\phi}_n\le c_j\le \bar{\phi}_n$ and $s_n\le t_n\le r_n$.
As a consequence, $\min\limits_{s_n\le t_n\le r_n}\inf\limits_{\underline{\phi}_n\le c_1,\ldots,c_p\le \bar{\phi}_n}p(\boldsymbol{\gamma}^0|\textbf{Z})\rightarrow1$ in probability.

\subsection*{Proof of Proposition \ref{valid:B1}}
(\ref{B1:eq}) follows immediately from Proposition 2 in \cite{ZH08}.
Next we verify (\ref{B1:eq2}).
Fix $2<\alpha'<\alpha/2$. If $\xi=\chi_\mu^2$, then by Chebyshev's inequality,
\begin{eqnarray}\label{lemma2:eq1}
\Pr(\xi\ge \alpha\mu a_n)&=&\Pr\left(\exp(\xi/\alpha')\ge \exp((\alpha/\alpha')\mu a_n)\right)\nonumber\\
&\le& \exp(-(\alpha/\alpha')\mu a_n) E\left\{\exp(\xi/\alpha')\right\}\nonumber\\
&=& (1-2/\alpha')^{-\mu/2} \exp(-(\alpha/\alpha')\mu a_n).
\end{eqnarray}
Clearly, given $\textbf{P}_{\boldsymbol{\gamma}}$, $\textbf{X}_j^T \textbf{P}_{\boldsymbol{\gamma}} \textbf{X}_j$ follows $\chi_{|\boldsymbol{\gamma}|}^2$.
Then it follows by (\ref{lemma2:eq1}) and the fact ${p\choose r}\le p^r/r!$ that
\begin{eqnarray*}
&&\Pr\left(\max\limits_{\substack{\boldsymbol{\gamma}\in T(t_n)\\ 0<t_n<s_n}}\max\limits_{j\in\boldsymbol{\gamma}^0\backslash\boldsymbol{\gamma}}
\textbf{X}_j^T\textbf{P}_{\boldsymbol{\gamma}} \textbf{X}_j\ge \alpha s_n\log{p}\right)\\
&\le&\Pr\left(\max\limits_{\boldsymbol{\gamma}\in T(s_n-1)}\max\limits_{j\in\boldsymbol{\gamma}^0\backslash\boldsymbol{\gamma}}\textbf{X}_j^T\textbf{P}_{\boldsymbol{\gamma}} \textbf{X}_j\ge \alpha s_n\log{p}\right)\\
&\le&\sum\limits_{\boldsymbol{\gamma}\in T(s_n-1)\backslash\{\emptyset\}}\sum\limits_{j\in\boldsymbol{\gamma}^0\backslash\boldsymbol{\gamma}}\Pr\left(\textbf{X}_j^T\textbf{P}_{\boldsymbol{\gamma}} \textbf{X}_j\ge \alpha s_n\log{p}\right)\\
&=&\sum\limits_{\boldsymbol{\gamma}\in T(s_n-1)\backslash\{\emptyset\}}\sum\limits_{j\in\boldsymbol{\gamma}^0\backslash\boldsymbol{\gamma}}E\left\{\Pr\left(\textbf{X}_j^T\textbf{P}_{\boldsymbol{\gamma}} \textbf{X}_j\ge \alpha s_n\log{p}| \textbf{P}_{\boldsymbol{\gamma}}\right)\right\}\\
&\le&\sum\limits_{\boldsymbol{\gamma}\in T(s_n-1)\backslash\{\emptyset\}}\sum\limits_{j\in\boldsymbol{\gamma}^0\backslash\boldsymbol{\gamma}}\Pr\left(\chi^2_{|\boldsymbol{\gamma}|}\ge \alpha |\boldsymbol{\gamma}|\log{p}\right)\\
&\le& s_n \sum\limits_{\boldsymbol{\gamma}\in T(s_n-1)\backslash\{\emptyset\}}\left[(1-2/\alpha')^{-1/2}\exp\left(-(\alpha/\alpha')\log{p}\right)\right]^{|\boldsymbol{\gamma}|}\\
&\le& s_n \sum\limits_{r=1}^{s_n-1}{p\choose r}\left[(1-2/\alpha')^{-1/2}p^{-\alpha/\alpha'}\right]^r\\
&\le& s_n \sum\limits_{r=1}^{s_n-1}\frac{1}{r!}\left[(1-2/\alpha')^{-1/2}p^{1-\alpha/\alpha'}\right]^r\\
&\le& s_n \left[\exp\left((1-2/\alpha')^{-1/2}p^{-(\alpha/\alpha'-1)}\right)-1\right]=O(s_n/p)=o(1).
\end{eqnarray*}
Thus, with probability approaching one, for any $\boldsymbol{\gamma}\in T(t_n)$ with $0<t_n<s_n$,
\begin{eqnarray*}
\lambda_+\left(\textbf{X}_{\boldsymbol{\gamma}^0\backslash\boldsymbol{\gamma}}^T\textbf{P}_{\boldsymbol{\gamma}} \textbf{X}_{\boldsymbol{\gamma}^0\backslash\boldsymbol{\gamma}}\right)\le \textrm{trace}\left(\textbf{X}_{\boldsymbol{\gamma}^0\backslash\boldsymbol{\gamma}}^T\textbf{P}_{\boldsymbol{\gamma}} \textbf{X}_{\boldsymbol{\gamma}^0\backslash\boldsymbol{\gamma}}\right)
\le s_n \max\limits_{\substack{\boldsymbol{\gamma}\in T(t_n)\\ 0<t_n<s_n}}\max\limits_{j\in\boldsymbol{\gamma}^0\backslash\boldsymbol{\gamma}} \textbf{X}_j^T\textbf{P}_{\boldsymbol{\gamma}} \textbf{X}_j
\le \alpha s_n^2 \log{p}.
\end{eqnarray*}

To prove Theorem \ref{main:thm2}, we need to establish the following preliminary lemma.

\begin{lemma}\label{lemma2} Suppose $\boldsymbol{\epsilon}\sim N(\textbf{0},\sigma_0^2  \textbf{I}_n)$. Adopt the convention that $\boldsymbol{\nu}_{\boldsymbol{\gamma}}^T\boldsymbol{\epsilon}/\|\boldsymbol{\nu}_{\boldsymbol{\gamma}}\|=0$ when $\boldsymbol{\nu}_{\boldsymbol{\gamma}}=0$,
and $\boldsymbol{\epsilon}^T \textbf{P}_{\boldsymbol{\gamma}}\boldsymbol{\epsilon}/|\boldsymbol{\gamma}|=0$ when $\boldsymbol{\gamma}$ is
null.

\begin{enumerate}[(i).]

\item For $\boldsymbol{\gamma}\in T_0(t_n)$, define $\boldsymbol{\nu}_{\boldsymbol{\gamma}}=(\textbf{I}_n-\textbf{P}_{\boldsymbol{\gamma}})\textbf{X}_{\boldsymbol{\gamma}^0\backslash\boldsymbol{\gamma}}
    \boldsymbol{\beta}_{\boldsymbol{\gamma}^0\backslash\boldsymbol{\gamma}}^0$.
Then $\max\limits_{0<t_n<s_n}\max\limits_{\boldsymbol{\gamma}\in T_0(t_n)} \frac{|\boldsymbol{\nu}_{\boldsymbol{\gamma}}^T\boldsymbol{\epsilon}|}{\|\boldsymbol{\nu}_{\boldsymbol{\gamma}}\|}= O_\Pr(\sqrt{s_n})$.

\item For $\boldsymbol{\gamma}\in T_0(t_n)$, define $\boldsymbol{\nu}_{\boldsymbol{\gamma}}=\textbf{P}_{\boldsymbol{\gamma}} \textbf{X}_{\boldsymbol{\gamma}^0}\boldsymbol{\beta}^0_{\gamma^0}$. Then
$\max\limits_{0<t_n<s_n}\max\limits_{\boldsymbol{\gamma}\in T_0(t_n)} \frac{|\boldsymbol{\nu}_{\boldsymbol{\gamma}}^T\boldsymbol{\epsilon}|}{\|\boldsymbol{\nu}_{\boldsymbol{\gamma}}\|}= O_\Pr(\sqrt{s_n})$.

\item For $\boldsymbol{\gamma}\in T_1(t_n)$, denote $\boldsymbol{\gamma}^*=\boldsymbol{\gamma}\cap\boldsymbol{\gamma}^0$ which is nonnull. For any fixed $\alpha>4$,
\[
\lim\limits_{n\rightarrow\infty}\Pr\left(\max\limits_{0<t_n<s_n}\max\limits_{\boldsymbol{\gamma}\in T_1(t_n)}
\frac{\boldsymbol{\epsilon}^T(\textbf{P}_{\boldsymbol{\gamma}}-\textbf{P}_{\boldsymbol{\gamma}^*})\boldsymbol{\epsilon}}{|\boldsymbol{\gamma}|-|\boldsymbol{\gamma}^*|}\le \alpha\sigma_0^2s_n\log{p}\right)=1.
\]

\item Then for any fixed $\alpha>2$,
\[
\lim\limits_{n\rightarrow\infty}\Pr\left(\max\limits_{0<t_n<s_n}\max\limits_{\boldsymbol{\gamma}\in T_2(t_n)}\boldsymbol{\epsilon}^T \textbf{P}_{\boldsymbol{\gamma}}\boldsymbol{\epsilon}/|\boldsymbol{\gamma}| \le \alpha\sigma_0^2\log{p}\right)=1.
\]


\end{enumerate}
\end{lemma}

\subsection*{Proof of Lemma \ref{lemma2}}
The idea of the proof is similar to that of Lemma \ref{lemma1} and Proposition \ref{valid:B1}. But there is some technical difference so we still present some of the details.
We note the trivial fact $\bigcup\limits_{0<t_n<s_n} T_l(t_n)\subset T_l(s_n-1)$ for $l=0,1,2$. Thus, $\# \left(\bigcup\limits_{0<t_n<s_n} T_0(t_n)\right)\le \# T_0(s_n-1)\le 2^{s_n}$.
The proof of parts (i)--(ii) follow immediately by (\ref{lemma1:eq1}).

For part (iii), fix $\alpha'$ such that $2<\alpha'<\alpha/2$.
Then the desired conclusion follows by (\ref{lemma2:eq1}) and the the below argument
\begin{eqnarray*}
&&\Pr\left(\max\limits_{0<t_n<s_n}\max\limits_{\boldsymbol{\gamma}\in T_1(t_n)}\frac{\boldsymbol{\epsilon}^T(\textbf{P}_{\boldsymbol{\gamma}}-\textbf{P}_{\boldsymbol{\gamma}^*})\boldsymbol{\epsilon}}{|\boldsymbol{\gamma}|-|\boldsymbol{\gamma}^*|}\ge \alpha\sigma_0^2s_n\log{p}\right)\\
&\le&\left(\max\limits_{\boldsymbol{\gamma}\in T_1(s_n-1)}\frac{\boldsymbol{\epsilon}^T(\textbf{P}_{\boldsymbol{\gamma}}-\textbf{P}_{\boldsymbol{\gamma}^*})\boldsymbol{\epsilon}}{|\boldsymbol{\gamma}|-|\boldsymbol{\gamma}^*|}\ge \alpha\sigma_0^2s_n\log{p}\right)\\
&\le&\sum\limits_{\boldsymbol{\gamma}\in T_1(s_n-1)} \Pr\left(\boldsymbol{\epsilon}^T(\textbf{P}_{\boldsymbol{\gamma}}-\textbf{P}_{\boldsymbol{\gamma}^*})\boldsymbol{\epsilon}\ge\alpha\sigma_0^2(|\boldsymbol{\gamma}|-|\boldsymbol{\gamma}^*|)s_n\log{p}\right)\\
&\le&\sum\limits_{r=1}^{s_n-2}\sum\limits_{|\boldsymbol{\gamma}|-|\boldsymbol{\gamma}^*|=r}(1-2/\alpha')^{-r/2}p^{-(\alpha/\alpha')rs_n}\\
&\le&\sum\limits_{r=1}^{s_n-2}\left({s_n\choose 1}+\ldots+{s_n \choose s_n-1-r}\right)\cdot {p-s_n\choose r} (1-2/\alpha')^{-r/2}p^{-(\alpha/\alpha')rs_n}\\
&\le&\sum\limits_{r=1}^{s_n-2}(s_n-1-r)\frac{s_n^{s_n-1-r}}{(s_n-1-r)!} \cdot\frac{(p-s_n)^r}{r!}(1-2/\alpha')^{-r/2}p^{-(\alpha/\alpha')rs_n}\\
&\le&s_n^{s_n}\sum\limits_{r=1}^{s_n-1}\frac{1}{r!}\left[(1-2/\alpha')^{-1/2}p^{1-(\alpha/\alpha')s_n}\right]^r\\
&\le&s_n^{s_n}\left[\exp\left((1-2/\alpha')^{-1/2}p^{1-(\alpha/\alpha')s_n}\right)-1\right]\\
&=&O(s_n^{s_n}p^{1-(\alpha/\alpha')s_n})=O(p^{1-(\alpha/\alpha')s_n+s_n})=o(1).
\end{eqnarray*}

For part (iv), fix $\alpha'$ such that $2<\alpha'<\alpha$. Then by (\ref{lemma2:eq1}) with $a_n=\log{p}$ therein, we have
\begin{eqnarray*}
&&\Pr\left(\max\limits_{0<t_n<s_n}\max\limits_{\boldsymbol{\gamma}\in T_2(t_n)}\boldsymbol{\epsilon}^T \textbf{P}_{\boldsymbol{\gamma}}\boldsymbol{\epsilon}/|\boldsymbol{\gamma}| \ge \alpha\sigma_0^2\log{p}\right)\\
&\le&\Pr\left(\max\limits_{\boldsymbol{\gamma}\in T_2(s_n-1)}\boldsymbol{\epsilon}^T \textbf{P}_{\boldsymbol{\gamma}}\boldsymbol{\epsilon}/|\boldsymbol{\gamma}| \ge \alpha\sigma_0^2\log{p}\right)\\
&\le&\sum\limits_{\boldsymbol{\gamma}\in T_2(s_n-1)} \Pr\left(\boldsymbol{\epsilon}^T \textbf{P}_{\boldsymbol{\gamma}}\boldsymbol{\epsilon}\ge \alpha\sigma_0^2|\boldsymbol{\gamma}|\log{p}\right)\\
&\le&\sum\limits_{r=1}^{s_n-1}\sum\limits_{\substack{|\boldsymbol{\gamma}|=r\\ \boldsymbol{\gamma}\in T_2(s_n-1)}}(1-2/\alpha')^{-r/2}\exp(-(\alpha/\alpha')r\log{p})\\
&=&\sum\limits_{r=1}^{s_n-1}{p-s_n\choose r}[(1-2/\alpha')^{-1/2}p^{-(\alpha/\alpha')}]^r\\
&\le&\sum\limits_{r=1}^{s_n-1}\frac{1}{r!}[(1-2/\alpha')^{-1/2}p^{1-(\alpha/\alpha')}]^r\\
&\le&\exp\left((1-2/\alpha')^{-1/2}p^{1-(\alpha/\alpha')}\right)-1=o(1),
\end{eqnarray*}
which shows part (iv).

To show part (v), fix $\alpha>\alpha'>2$. By (\ref{lemma2:eq1}) with $a_n=C\log(2s_n)$ therein, we have
\begin{eqnarray*}
&&\Pr\left(\max\limits_{0<t_n<s_n}\max\limits_{\boldsymbol{\gamma}\in T_0(t_n)}\boldsymbol{\epsilon}^T \textbf{P}_{\boldsymbol{\gamma}}\boldsymbol{\epsilon}/|\boldsymbol{\gamma}|\ge \alpha C\sigma_0^2\log(2s_n)\right)\\
&\le&\Pr\left(\max\limits_{\boldsymbol{\gamma}\in T_0(s_n-1)}\boldsymbol{\epsilon}^T \textbf{P}_{\boldsymbol{\gamma}}\boldsymbol{\epsilon}/|\boldsymbol{\gamma}|\ge \alpha C\sigma_0^2\log(2s_n)\right)\\
&=&\Pr\left(\max\limits_{\substack{\boldsymbol{\gamma}\in T_0(s_n-1)\\ \boldsymbol{\gamma}\neq\emptyset}}\boldsymbol{\epsilon}^T \textbf{P}_{\boldsymbol{\gamma}}\boldsymbol{\epsilon}/|\boldsymbol{\gamma}|\ge\alpha C\sigma_0^2\log(2s_n)\right)\\
&\le&\sum\limits_{\substack{\boldsymbol{\gamma}\in T_0(s_n-1)\\ \boldsymbol{\gamma}\neq\emptyset}} \Pr\left(\boldsymbol{\epsilon}^T \textbf{P}_{\boldsymbol{\gamma}}\boldsymbol{\epsilon}/|\boldsymbol{\gamma}|\ge \alpha C\sigma_0^2\log(2s_n)\right)\\
&\le&\sum\limits_{r=1}^{s_n}\sum\limits_{\substack{\boldsymbol{\gamma}\subset\boldsymbol{\gamma}^0\\ |\boldsymbol{\gamma}|=r}}(1-2/\alpha')^{-r/2}\exp\left(-(\alpha/\alpha')rC\log(s_n)\right)\\
&=&\sum\limits_{r=1}^{s_n}{s_n\choose r} \left[(1-2/\alpha')^{-1/2}\exp\left(-(\alpha/\alpha')C\log(2s_n)\right)\right]^r\\
&=&\left[1+(1-2/\alpha')^{-1/2}(2s_n)^{-(\alpha/\alpha')C}\right]^{s_n}-1,
\end{eqnarray*}
which is small when $C>0$ is chosen to be sufficiently large. This proves part (v).

\subsection*{Proof of Theorem \ref{main:thm2}}
To make it more readable, we sketch the idea of the proof. We will first show that for $\boldsymbol{\gamma}\in T_1(t_n)$ with $0<t_n<s_n$ and $\boldsymbol{\gamma}\cap\boldsymbol{\gamma}^0\neq\emptyset$,
$\max\limits_{\boldsymbol{\gamma}\in T_1(t_n)}p(\boldsymbol{\gamma}|\textbf{Z})/p\boldsymbol{(\gamma}\cap\boldsymbol{\gamma}^0|\textbf{Z})$ converges to zero in probability.
Note the denominator is bounded by $\max\limits_{\boldsymbol{\gamma}\in T_0(t_n)}p(\boldsymbol{\gamma}|\textbf{Z})$,
and thus $\frac{\max\limits_{\boldsymbol{\gamma}\in T_1(t_n)}p(\boldsymbol{\gamma}|\textbf{Z})}{\max\limits_{\boldsymbol{\gamma}\in T_0(t_n)}p(\boldsymbol{\gamma}|\textbf{Z})}\rightarrow0$ in probability.
Secondly, we show that $\frac{\max\limits_{\boldsymbol{\gamma}\in T_2(t_n)}p(\boldsymbol{\gamma}|\textbf{Z})}{p(\emptyset|\textbf{Z})}\rightarrow0$ in probability, i.e., any $\boldsymbol{\gamma}\in T_2(t_n)$ is even worse than the null model.
This will complete the proof. For simplicity, all the arguments in this proof section are built upon (\ref{B1:eq}) and (\ref{B1:eq2}),
which by Assumption \ref{B1} have overwhelming probability when $n$ is large. Next we finish these two steps.

\textbf{Step I}: For $\boldsymbol{\gamma}\in T_1(t_n)$, define $\boldsymbol{\gamma}^*=\boldsymbol{\gamma}\cap\boldsymbol{\gamma}^0$, which by our definition of $T_1(t_n)$, is nonnull.
We will approximate the log-ratio of $p(\boldsymbol{\gamma}|\textbf{Z})$ to $p(\boldsymbol{\gamma}^*|\textbf{Z})$, which can be decomposed as follows
\begin{eqnarray*}
-\log\left(\frac{p(\boldsymbol{\gamma}|\textbf{Z})}{p(\boldsymbol{\gamma}^*|\textbf{Z})}\right)
&=&-\log\left(\frac{p(\boldsymbol{\gamma})}{p(\boldsymbol{\gamma}^*)}\right)
+\frac{1}{2}\log\left(\frac{\det(\textbf{W}_{\boldsymbol{\gamma}})}{\det(\textbf{W}_{\boldsymbol{\gamma}^*})}\right)+\frac{n+\nu}{2}\log\left(\frac{1+\textbf{Y}^T(\textbf{I}_n-\textbf{X}_{\boldsymbol{\gamma}} \textbf{U}_{\boldsymbol{\gamma}}^{-1} \textbf{X}_{\boldsymbol{\gamma}}^T)\textbf{Y}}{1+\textbf{Y}^T(\textbf{I}_n-\textbf{P}_{\boldsymbol{\gamma}})\textbf{Y}}\right)\\
&&-\frac{n+\nu}{2}\log\left(\frac{1+\textbf{Y}^T(\textbf{I}_n-\textbf{X}_{\boldsymbol{\gamma}^*}\textbf{U}_{\boldsymbol{\gamma}^*}^{-1}\textbf{X}_{\boldsymbol{\gamma}^*}^T)\textbf{Y}}
{1+\textbf{Y}^T(\textbf{I}_n-\textbf{P}_{\boldsymbol{\gamma}^*})\textbf{Y}}\right)+\frac{n+\nu}{2}\log\left(\frac{1+\textbf{Y}^T(\textbf{I}_n-\textbf{P}_{\boldsymbol{\gamma}})\textbf{Y}}
{1+\textbf{Y}^T(\textbf{I}_n-\textbf{P}_{\boldsymbol{\gamma}^*})\textbf{Y}}\right).
\end{eqnarray*}
Denote the above five terms by $I_1,I_2,I_3,I_4,I_5$. Clearly, $I_1$ is bounded from below and $I_3\ge0$.
To approximate $I_4$, we use the following Sherman-Morrison-Woodbury matrix identity (pp. 467, \cite{SL03}),
\[
\textbf{U}_{\boldsymbol{\gamma}^*}^{-1}-(\textbf{X}_{\boldsymbol{\gamma}^{*}}^T\textbf{X}_{\boldsymbol{\gamma}^*})^{-1}=
-\left(\textbf{X}_{\boldsymbol{\gamma}^*}^T \textbf{X}_{\boldsymbol{\gamma}^*}\right)^{-1}\left(\boldsymbol{\Sigma}_{\boldsymbol{\gamma}^*}+\left(\textbf{X}_{\boldsymbol{\gamma}^*}^T \textbf{X}_{\boldsymbol{\gamma}^*}\right)^{-1}\right)^{-1}\left(\textbf{X}_{\boldsymbol{\gamma}^*}^T \textbf{X}_{\boldsymbol{\gamma}^*}\right)^{-1}.
\]
Then by $\textbf{Y}=\textbf{X}_{\boldsymbol{\gamma}^0}\boldsymbol{\beta}^0_{\boldsymbol{\gamma}^0}+\boldsymbol{\epsilon}$,
\begin{eqnarray*}
&&\frac{1+\textbf{Y}^T(\textbf{I}_n-\textbf{X}_{\boldsymbol{\gamma}^*}\textbf{U}_{\boldsymbol{\gamma}^*}^{-1}\textbf{X}_{\boldsymbol{\gamma}^*}^T)\textbf{Y}}
{1+\textbf{Y}^T(\textbf{I}_n-\textbf{P}_{\boldsymbol{\gamma}^*})\textbf{Y}}\\
&=&1+\frac{\textbf{Y}^T\textbf{X}_{\boldsymbol{\gamma}^*}((\textbf{X}_{\boldsymbol{\gamma}^{*}}^T\textbf{X}_{\boldsymbol{\gamma}^*})^{-1}-\textbf{U}_{\boldsymbol{\gamma}^*}^{-1})
\textbf{X}_{\boldsymbol{\gamma}^*}^T\textbf{Y}}
{1+\textbf{Y}^T(\textbf{I}_n-\textbf{P}_{\boldsymbol{\gamma}^*})\textbf{Y}}\\
&=&1+\frac{\textbf{Y}^T\textbf{X}_{\boldsymbol{\gamma}^*}(\textbf{X}_{\boldsymbol{\gamma}^{*}}^T\textbf{X}_{\boldsymbol{\gamma}^*})^{-1}
(\boldsymbol{\Sigma}_{\boldsymbol{\gamma}^*}+(\textbf{X}_{\boldsymbol{\gamma}^{*}}^T\textbf{X}_{\boldsymbol{\gamma}^*})^{-1})^{-1}
(\textbf{X}_{\boldsymbol{\gamma}^{*}}^T\textbf{X}_{\boldsymbol{\gamma}^*})^{-1}\textbf{X}_{\boldsymbol{\gamma}^*}^T\textbf{Y}}
{1+\textbf{Y}^T(\textbf{I}_n-\textbf{P}_{\boldsymbol{\gamma}^*})\textbf{X}_{\boldsymbol{\gamma}^{*}}^T\textbf{Y}}\\
&\le&1+\underline{\phi}_n^{-1}\frac{\textbf{Y}^T\textbf{X}_{\boldsymbol{\gamma}^*}(\textbf{X}_{\boldsymbol{\gamma}^{*}}^T\textbf{X}_{\boldsymbol{\gamma}^*})^{-2}\textbf{X}_{\boldsymbol{\gamma}^*}^T\textbf{Y}}
{1+\textbf{Y}^T(\textbf{I}_n-\textbf{P}_{\boldsymbol{\gamma}^*})\textbf{Y}}\\
&\le&1+2\underline{\phi}_n^{-1}\frac{(\boldsymbol{\beta}^0_{\boldsymbol{\gamma}^0})^T\textbf{X}_{\boldsymbol{\gamma}^0}^T\textbf{X}_{\boldsymbol{\gamma}^*}(\textbf{X}_{\boldsymbol{\gamma}^{*}}^T
\textbf{X}_{\boldsymbol{\gamma}^*})^{-2}\textbf{X}_{\boldsymbol{\gamma}^*}^T\textbf{X}_{\boldsymbol{\gamma}^0}\boldsymbol{\beta}^0_{\boldsymbol{\gamma}^0}+
\boldsymbol{\epsilon}^T\textbf{X}_{\boldsymbol{\gamma}^*}(\textbf{X}_{\boldsymbol{\gamma}^{*}}^T\textbf{X}_{\boldsymbol{\gamma}^*})^{-2}\textbf{X}_{\boldsymbol{\gamma}^*}^T\boldsymbol{\epsilon}}
{1+\textbf{Y}^T(\textbf{I}_n-\textbf{P}_{\boldsymbol{\gamma}^*})\textbf{Y}}.
\end{eqnarray*}

Without loss of generality, assume $\textbf{X}_{\boldsymbol{\gamma}^0}=(\textbf{X}_{\boldsymbol{\gamma}^*},\textbf{X}_{\boldsymbol{\gamma}^0\backslash\boldsymbol{\gamma}^*})$ and $\boldsymbol{\beta}^0_{\boldsymbol{\gamma}^0}=((\boldsymbol{\beta}^0_{\boldsymbol{\gamma}^*})^T,(\boldsymbol{\beta}^0_{\boldsymbol{\gamma}^0\backslash\boldsymbol{\gamma}^*})^T)^T$.
By a direct calculation it can be examined that
\begin{eqnarray*}
\textbf{X}_{\boldsymbol{\gamma}^0}^T\textbf{X}_{\boldsymbol{\gamma}^*}(\textbf{X}_{\boldsymbol{\gamma}^{*}}^T\textbf{X}_{\boldsymbol{\gamma}^*})^{-2}
\textbf{X}_{\boldsymbol{\gamma}^*}^T\textbf{X}_{\boldsymbol{\gamma}^0}=
\left(\begin{array}{cc}\textbf{I}_{|\boldsymbol{\gamma}^*|} &
(\textbf{X}_{\boldsymbol{\gamma}^*}^T\textbf{X}_{\boldsymbol{\gamma}^*})^{-1}\textbf{X}_{\boldsymbol{\gamma}^*}^T\textbf{X}_{\boldsymbol{\gamma}^0\backslash\boldsymbol{\gamma}^*}\\
\textbf{X}_{\boldsymbol{\gamma}^0\backslash\boldsymbol{\gamma}^*}^T\textbf{X}_{\boldsymbol{\gamma}^*}(\textbf{X}_{\boldsymbol{\gamma}^*}^T\textbf{X}_{\boldsymbol{\gamma}^*})^{-1} & \textbf{X}_{\boldsymbol{\gamma}^0\backslash\boldsymbol{\gamma}^*}^T\textbf{X}_{\boldsymbol{\gamma}^*}(\textbf{X}_{\boldsymbol{\gamma}^*}^T\textbf{X}_{\boldsymbol{\gamma}^*})^{-2}
\textbf{X}_{\boldsymbol{\gamma}^*}^T\textbf{X}_{\boldsymbol{\gamma}^0\backslash\boldsymbol{\gamma}^*}
\end{array}\right).
\end{eqnarray*}
By Assumption \ref{B1}, w.l.p.,
\[
\lambda_+\left(\textbf{X}_{\boldsymbol{\gamma}^0\backslash\boldsymbol{\gamma}^*}^T\textbf{X}_{\boldsymbol{\gamma}^*}(\textbf{X}_{\boldsymbol{\gamma}^*}^T\textbf{X}_{\boldsymbol{\gamma}^*})^{-2}
\textbf{X}_{\boldsymbol{\gamma}^*}^T\textbf{X}_{\boldsymbol{\gamma}^0\backslash\boldsymbol{\gamma}^*}\right)
\le \frac{d_0}{n}\lambda_+\left(\textbf{X}_{\boldsymbol{\gamma}^0\backslash\boldsymbol{\gamma}^*}\textbf{P}_{\boldsymbol{\gamma}^*}\textbf{X}_{\boldsymbol{\gamma}^0\backslash\boldsymbol{\gamma}^*}\right)\le \frac{d_0\rho_n}{n},
\]
which implies, w.l.p., $\lambda_+\left(\textbf{X}_{\boldsymbol{\gamma}^0}^T\textbf{X}_{\boldsymbol{\gamma}^*}(\textbf{X}_{\boldsymbol{\gamma}^{*}}^T\textbf{X}_{\boldsymbol{\gamma}^*})^{-2}\textbf{X}_{\boldsymbol{\gamma}^*}^T
\textbf{X}_{\boldsymbol{\gamma}^0}\right)\le 1+\frac{d_0\rho_n}{n}=O(1)$.
Thus,
\begin{equation}\label{thm2:eq1}
(\boldsymbol{\beta}^0_{\boldsymbol{\gamma}^0})^T\textbf{X}_{\boldsymbol{\gamma}^0}^T
\textbf{X}_{\boldsymbol{\gamma}^*}(\textbf{X}_{\boldsymbol{\gamma}^{*}}^T\textbf{X}_{\boldsymbol{\gamma}^*})^{-2}\textbf{X}_{\boldsymbol{\gamma}^*}^T
\textbf{X}_{\boldsymbol{\gamma}^0}\boldsymbol{\beta}^0_{\boldsymbol{\gamma}^0}\le (1+\frac{d_0\rho_n}{n})k_n.
\end{equation}
By $\textbf{P}_{\boldsymbol{\gamma}^*}\le \textbf{P}_{\boldsymbol{\gamma}^0}$, $E\{\boldsymbol{\epsilon}^T \textbf{P}_{\boldsymbol{\gamma}^0}\boldsymbol{\epsilon}\}=\sigma_0^2 s_n$ implying $\boldsymbol{\epsilon}^T \textbf{P}_{\boldsymbol{\gamma}^0} \boldsymbol{\epsilon}=O_\Pr(s_n)$,
and (\ref{B1:eq}) of Assumption \ref{B1}, we have, w.l.p.,
\begin{equation}\label{thm2:eq2}
\boldsymbol{\epsilon}^T\textbf{X}_{\boldsymbol{\gamma}^*}(\textbf{X}_{\boldsymbol{\gamma}^{*}}^T\textbf{X}_{\boldsymbol{\gamma}^*})^{-2}\textbf{X}_{\boldsymbol{\gamma}^*}^T\boldsymbol{\epsilon}\le \frac{d_0}{n}\boldsymbol{\epsilon}^T \textbf{P}_{\boldsymbol{\gamma}^0}\boldsymbol{\epsilon}=O_\Pr(s_n/n).
\end{equation}
On the other hand, by Assumption \ref{B3} (i)
\begin{equation}\label{thm2:eq3}
\textbf{Y}^T(\textbf{I}_n-\textbf{P}_{\boldsymbol{\gamma}^*})\textbf{Y}\ge \textbf{Y}^T(\textbf{I}_n-\textbf{P}_{\boldsymbol{\gamma}^0})\textbf{Y}=\boldsymbol{\epsilon}^T(\textbf{I}_n-\textbf{P}_{\boldsymbol{\gamma}^0})\boldsymbol{\epsilon}=n\sigma_0^2(1+o_\Pr(s_n/n))=n\sigma_0^2(1+o_\Pr(1)).
\end{equation}
Combining (\ref{thm2:eq1})--(\ref{thm2:eq3}), and using the fact $k_n\ge s_n\psi_n^2\gg s_n/n$, we have for $0<t_n<s_n$ and uniformly for $\boldsymbol{\gamma}\in T_1(t_n)$,
\begin{eqnarray*}
\frac{1+\textbf{Y}^T(\textbf{I}_n-\textbf{X}_{\boldsymbol{\gamma}^*}\textbf{U}_{\boldsymbol{\gamma}^*}^{-1}\textbf{X}_{\boldsymbol{\gamma}^*}^T)\textbf{Y}}
{1+\textbf{Y}^T(\textbf{I}_n-\textbf{P}_{\boldsymbol{\gamma}^*})\textbf{Y}}\le 1+2\underline{\phi}_n^{-1}\frac{(1+d_0\rho_n/n)k_n+O_\Pr(s_n/n)}{n\sigma_0^2(1+o_\Pr(1))}
&=&1+\frac{2(1+d_0\rho_n/n)k_n}{n\underline{\phi}_n\sigma_0^2}(1+o_\Pr(1)).
\end{eqnarray*}
It follows by $k_n=O(\underline{\phi}_n)$ and $\rho_n=o(n)$ (Assumption \ref{B3} (iv) and (v)) that for $0<t_n<s_n$, uniformly for $c_j$s $\in [\underline{\phi}_n,\bar{\phi}_n]$ and uniformly for $\boldsymbol{\gamma}\in T_1(t_n)$,
$0\le -I_4=O_\Pr(1)$.

Next we present lower bounds for $I_5$. Assume, without loss of generality, that $\textbf{X}_{\boldsymbol{\gamma}^0}=(\textbf{X}_{\boldsymbol{\gamma}^*},\textbf{X}_{\boldsymbol{\gamma}^0\backslash\boldsymbol{\gamma}^*})$ and $\boldsymbol{\beta}^0_{\boldsymbol{\gamma}^0}=((\boldsymbol{\beta}^0_{\boldsymbol{\gamma}^*})^T,(\boldsymbol{\beta}^0_{\boldsymbol{\gamma}^0\backslash\boldsymbol{\gamma}^*})^T)^T$.
Then it follows by $\textbf{Y}=\textbf{X}_{\boldsymbol{\gamma}^0}\boldsymbol{\beta}^0_{\boldsymbol{\gamma}^0}+\boldsymbol{\epsilon}$ and $(\textbf{P}_{\boldsymbol{\gamma}}-\textbf{P}_{\boldsymbol{\gamma}^*})\textbf{X}_{\boldsymbol{\gamma}^*}=0$ that
\begin{eqnarray*}
\textbf{Y}^T(\textbf{P}_{\boldsymbol{\gamma}}-\textbf{P}_{\boldsymbol{\gamma}^*})\textbf{Y}
&=&((\boldsymbol{\beta}^0_{\boldsymbol{\gamma}^*})^T\textbf{X}_{\boldsymbol{\gamma}^*}^T+
(\boldsymbol{\beta}^0_{\boldsymbol{\gamma}^0\backslash\boldsymbol{\gamma}^*})^T\textbf{X}_{\boldsymbol{\gamma}^0\backslash\boldsymbol{\gamma}^*}^T+\boldsymbol{\epsilon}^T)
(\textbf{P}_{\boldsymbol{\gamma}}-\textbf{P}_{\boldsymbol{\gamma}^*})\textbf{X}_{\boldsymbol{\gamma}^0}\boldsymbol{\beta}^0_{\boldsymbol{\gamma}^0}
(\textbf{X}_{\boldsymbol{\gamma}^*}\beta^0_{\gamma^*}+\textbf{X}_{\boldsymbol{\gamma}^0\backslash\boldsymbol{\gamma}^*}
\boldsymbol{\beta}^0_{\boldsymbol{\gamma}^0\backslash\boldsymbol{\gamma}^*}+\boldsymbol{\epsilon})\nonumber\\
&=&(\boldsymbol{\beta}^0_{\boldsymbol{\gamma}^0\backslash\boldsymbol{\gamma}^*})^T\textbf{X}_{\boldsymbol{\gamma}^0\backslash\boldsymbol{\gamma}^*}^T
(\textbf{P}_{\boldsymbol{\gamma}}-\textbf{P}_{\boldsymbol{\gamma}^*})\textbf{X}_{\boldsymbol{\gamma}^0\backslash\boldsymbol{\gamma}^*}\boldsymbol{\beta}^0_{\boldsymbol{\gamma}^0\backslash\boldsymbol{\gamma}^*}
+2(\boldsymbol{\beta}^0_{\boldsymbol{\gamma}^0\backslash\boldsymbol{\gamma}^*})^T\textbf{X}_{\boldsymbol{\gamma}^0\backslash\boldsymbol{\gamma}^*}^T(\textbf{P}_{\boldsymbol{\gamma}}-\textbf{P}_{\boldsymbol{\gamma}^*})
\boldsymbol{\epsilon}+\boldsymbol{\epsilon}^T(\textbf{P}_{\boldsymbol{\gamma}}-\textbf{P}_{\boldsymbol{\gamma}^*})\boldsymbol{\epsilon}\nonumber\\
&\le&2(\boldsymbol{\beta}^0_{\boldsymbol{\gamma}^0\backslash\boldsymbol{\gamma}^*})^T\textbf{X}_{\boldsymbol{\gamma}^0\backslash\boldsymbol{\gamma}^*}^T
(\textbf{P}_{\boldsymbol{\gamma}}-\textbf{P}_{\boldsymbol{\gamma}^*})\textbf{X}_{\boldsymbol{\gamma}^0\backslash\boldsymbol{\gamma}^*}\boldsymbol{\beta}^0_{\boldsymbol{\gamma}^0\backslash\boldsymbol{\gamma}^*}+
2\boldsymbol{\epsilon}^T(\textbf{P}_{\boldsymbol{\gamma}}-\textbf{P}_{\boldsymbol{\gamma}^*})\boldsymbol{\epsilon}.
\end{eqnarray*}
By (\ref{B1:eq2}) of Assumption \ref{B1},
\begin{eqnarray*}
(\boldsymbol{\beta}^0_{\boldsymbol{\gamma}^0\backslash\boldsymbol{\gamma}^*})^T
\textbf{X}_{\boldsymbol{\gamma}^0\backslash\boldsymbol{\gamma}^*}^T(\textbf{P}_{\boldsymbol{\gamma}}-\textbf{P}_{\boldsymbol{\gamma}^*})\textbf{X}_{\boldsymbol{\gamma}^0\backslash\boldsymbol{\gamma}^*}
\boldsymbol{\beta}^0_{\boldsymbol{\gamma}^0\backslash\boldsymbol{\gamma}^*}
&\le&(\boldsymbol{\beta}^0_{\boldsymbol{\gamma}^0\backslash\boldsymbol{\gamma}^*})^T \textbf{X}_{\boldsymbol{\gamma}^0\backslash\boldsymbol{\gamma}^*}^T\textbf{P}_{\boldsymbol{\gamma}} \textbf{X}_{\boldsymbol{\gamma}^0\backslash\boldsymbol{\gamma}^*}\boldsymbol{\beta}^0_{\boldsymbol{\gamma}^0\backslash\boldsymbol{\gamma}^*}\\
&=&(\boldsymbol{\beta}^0_{\boldsymbol{\gamma}^0\backslash\boldsymbol{\gamma}})^T \textbf{X}_{\boldsymbol{\gamma}^0\backslash\boldsymbol{\gamma}}^T\textbf{P}_{\boldsymbol{\gamma}} \textbf{X}_{\boldsymbol{\gamma}^0\backslash\boldsymbol{\gamma}}\boldsymbol{\beta}^0_{\boldsymbol{\gamma}^0\backslash\boldsymbol{\gamma}}
\le\rho_n\|\boldsymbol{\beta}^0_{\boldsymbol{\gamma}^0\backslash\boldsymbol{\gamma}}\|^2.
\end{eqnarray*}
By Lemma \ref{lemma2} (iii), w.l.p., $\boldsymbol{\epsilon}^T(\textbf{P}_{\boldsymbol{\gamma}}-\textbf{P}_{\boldsymbol{\gamma}^*})\boldsymbol{\epsilon}\le \alpha\sigma_0^2 s_n(|\boldsymbol{\gamma}|-|\boldsymbol{\gamma}^*|)\log{p}\le \alpha\sigma_0^2 s_n^2\log{p}$, where $\alpha>4$ is prefixed.
Therefore, w.l.p, for any $0<t_n<s_n$ and $\boldsymbol{\gamma}\in T_1(t_n)$,
\begin{equation}\label{thm2:eq4}
\textbf{Y}^T(\textbf{P}_{\boldsymbol{\gamma}}-\textbf{P}_{\boldsymbol{\gamma}^*})\textbf{Y}\le 2(\rho_n\|\boldsymbol{\beta}^0_{\boldsymbol{\gamma}^0\backslash\boldsymbol{\gamma}}\|^2+\alpha\sigma_0^2 s_n^2\log{p})=2\max\{\rho_n,s_n^2\log{p}\}(\|\boldsymbol{\beta}^0_{\boldsymbol{\gamma}^0\backslash\boldsymbol{\gamma}}\|^2+O(1)).
\end{equation}

We approximate the term $\textbf{Y}^T(\textbf{I}_n-\textbf{P}_{\boldsymbol{\gamma}^*})\textbf{Y}$. Denote $\boldsymbol{\nu}_{\boldsymbol{\gamma}^*}=(\textbf{I}_n-\textbf{P}_{\boldsymbol{\gamma}^*})\textbf{X}_{\boldsymbol{\gamma}^0\backslash\boldsymbol{\gamma}^*}
\boldsymbol{\beta}_{\boldsymbol{\gamma}^0\backslash\boldsymbol{\gamma}^*}^0$.
A direct examination verifies that
$(\textbf{I}_n-\textbf{P}_{\boldsymbol{\gamma}^*})\textbf{X}_{\boldsymbol{\gamma}^0}=(\textbf{0},(\textbf{I}_n-\textbf{P}_{\boldsymbol{\gamma}^*})
\textbf{X}_{\boldsymbol{\gamma}^0\backslash\boldsymbol{\gamma}^*})$ which leads to
\begin{eqnarray}\label{thm2:eq5}
\textbf{Y}^T(\textbf{I}_n-\textbf{P}_{\boldsymbol{\gamma}^*})\textbf{Y}&=&
(\boldsymbol{\beta}^0_{\boldsymbol{\gamma}^0\backslash\boldsymbol{\gamma}^*})^T\textbf{X}_{\boldsymbol{\gamma}^0\backslash\boldsymbol{\gamma}^*}^T
(\textbf{I}_n-\textbf{P}_{\boldsymbol{\gamma}^*})\textbf{X}_{\boldsymbol{\gamma}^0\backslash\boldsymbol{\gamma}^*}\boldsymbol{\beta}^0_{\boldsymbol{\gamma}^0\backslash\boldsymbol{\gamma}^*}
+2(\boldsymbol{\beta}^0_{\boldsymbol{\gamma}^0\backslash\boldsymbol{\gamma}^*})^T\textbf{X}_{\boldsymbol{\gamma}^0\backslash\boldsymbol{\gamma}^*}^T
(\textbf{I}_n-\textbf{P}_{\boldsymbol{\gamma}^*})\boldsymbol{\epsilon}+\boldsymbol{\epsilon}^T(\textbf{I}_n-\textbf{P}_{\boldsymbol{\gamma}^*})\boldsymbol{\epsilon}\nonumber\\
&=&\|\boldsymbol{\nu}_{\boldsymbol{\gamma}^*}\|^2+2\boldsymbol{\nu}_{\boldsymbol{\gamma}^*}^T\boldsymbol{\epsilon}+
\boldsymbol{\epsilon}^T(\textbf{I}_n-\textbf{P}_{\boldsymbol{\gamma}^*})\boldsymbol{\epsilon}\nonumber\\
&\ge&\|\boldsymbol{\nu}_{\boldsymbol{\gamma}^*}\|^2\left(1-\frac{2|\boldsymbol{\nu}_{\boldsymbol{\gamma}^*}^T
\boldsymbol{\epsilon}|}{\|\boldsymbol{\nu}_{\boldsymbol{\gamma}^*}\|}\cdot\frac{1}{\|\boldsymbol{\nu}_{\boldsymbol{\gamma}^*}\|}\right)
+\boldsymbol{\epsilon}^T(\textbf{I}_n-\textbf{P}_{\boldsymbol{\gamma}^0})\boldsymbol{\epsilon}.
\end{eqnarray}
By Lemma \ref{lemma2} (i), uniformly for $\boldsymbol{\gamma}^*$, $\frac{|\boldsymbol{\nu}_{\boldsymbol{\gamma}^*}^T\boldsymbol{\epsilon}|}{\|\boldsymbol{\nu}_{\boldsymbol{\gamma}^*}\|}=O_\Pr(\sqrt{s_n})$.
Since $\boldsymbol{\gamma}^0\backslash\boldsymbol{\gamma}^*\neq\emptyset$, by Assumption \ref{B1},
\begin{equation}\label{thm2:eq7}
\lambda_{-}\left(\frac{1}{n}\textbf{X}_{\boldsymbol{\gamma}^0\backslash\boldsymbol{\gamma}^*}^T(\textbf{I}_n-\textbf{P}_{\boldsymbol{\gamma}^*})\textbf{X}_{\boldsymbol{\gamma}^0\backslash\boldsymbol{\gamma}^*}\right)\ge d_0^{-1}-\frac{\rho_n}{n},
\end{equation}
which implies
\[
\|\boldsymbol{\nu}_{\boldsymbol{\gamma}^*}\|^2\ge n(d_0^{-1}-\frac{\rho_n}{n})\|\boldsymbol{\beta}^0_{\boldsymbol{\gamma}^0\backslash\boldsymbol{\gamma}}\|^2\ge (d_0^{-1}-\frac{\rho_n}{n})n\psi_n^2.
\]
By Assumption \ref{B3} (iii), i.e., $s_n=o(n\psi_n^2)$, we have (\ref{thm2:eq5}) is greater than
\begin{equation}\label{thm2:eq6}
\|\boldsymbol{\nu}_{\boldsymbol{\gamma}^*}\|^2(1+o_\Pr(1))+n\sigma_0^2 (1+o_\Pr(1))\ge \left(n(d_0^{-1}-\frac{\rho_n}{n})\|\boldsymbol{\beta}^0_{\boldsymbol{\gamma}^0\backslash\boldsymbol{\gamma}}\|^2+n\sigma_0^2\right)\cdot(1+o_\Pr(1)).
\end{equation}
Now combined with (\ref{thm2:eq4})--(\ref{thm2:eq6}) we obtain w.l.p.
\begin{eqnarray*}
\frac{1+\textbf{Y}^T(\textbf{I}_n-\textbf{P}_{\boldsymbol{\gamma}})\textbf{Y}}{1+\textbf{Y}^T(\textbf{I}_n-\textbf{P}_{\boldsymbol{\gamma}^*})\textbf{Y}}
=1-\frac{\textbf{Y}^T(\textbf{P}_{\boldsymbol{\gamma}}-\textbf{P}_{\boldsymbol{\gamma}^*})\textbf{Y}}{1+\textbf{Y}^T(\textbf{I}_n-\textbf{P}_{\boldsymbol{\gamma}^*})\textbf{Y}}
\ge1-\frac{C'\max\{\rho_n,s_n^2\log{p}\}}{n},
\end{eqnarray*}
where $C'>0$ is constant unrelated to $\boldsymbol{\gamma}$ and $n$. This shows that w.l.p., for any $0<t_n<s_n$ and uniformly for $\boldsymbol{\gamma}\in T_1(t_n)$,
\[
I_5=\frac{n+\nu}{2}\log\left(\frac{1+\textbf{Y}^T(\textbf{I}_n-\textbf{P}_{\boldsymbol{\gamma}})\textbf{Y}}{1+\textbf{Y}^T(\textbf{I}_n-\textbf{P}_{\boldsymbol{\gamma}^*})\textbf{Y}}\right)\ge -C'' \max\{\rho_n,s_n^2\log{p}\},
\]
where $C''>0$ is constant unrelated to $\gamma$ and $n$.

To conclude Step I, we still need to approximate $I_2$ given as follows. Since $\textbf{U}_{\boldsymbol{\gamma}^*}$ is a submatrix of $\textbf{U}_{\boldsymbol{\gamma}}$,
it follows from the determinant formula for block matrices
(pp. 468, \cite{SL03}), and (\ref{thm2:eq7}) that
\begin{eqnarray*}
\det(\textbf{U}_{\boldsymbol{\gamma}})&=&\det(\textbf{U}_{\boldsymbol{\gamma}^*})
\det\left(\boldsymbol{\Sigma}^{-1}_{\boldsymbol{\gamma}\backslash\boldsymbol{\gamma}^*}+\textbf{X}_{\boldsymbol{\gamma}\backslash\boldsymbol{\gamma}^*}^T
(\textbf{I}_n-\textbf{X}_{\boldsymbol{\gamma}^*}\textbf{U}_{\boldsymbol{\gamma}^*}^{-1}\textbf{X}_{\boldsymbol{\gamma}^*}^T)\textbf{X}_{\boldsymbol{\gamma}\backslash\boldsymbol{\gamma}^*}\right)\\
&\ge&\det(\textbf{U}_{\boldsymbol{\gamma}^*})
\det\left(\boldsymbol{\Sigma}^{-1}_{\boldsymbol{\gamma}\backslash\boldsymbol{\gamma}^*}+\textbf{X}_{\boldsymbol{\gamma}\backslash\boldsymbol{\gamma}^*}^T
(\textbf{I}_n-\textbf{P}_{\boldsymbol{\gamma}^*})\textbf{X}_{\boldsymbol{\gamma}\backslash\boldsymbol{\gamma}^*}\right)\\
&\ge&\det(\textbf{U}_{\boldsymbol{\gamma}^*})\det\left(\boldsymbol{\Sigma}_{\boldsymbol{\gamma}\backslash\boldsymbol{\gamma}^*}^{-1}+(nd_0^{-1}-\rho_n)\textbf{I}_{|\boldsymbol{\gamma}\backslash\boldsymbol{\gamma}^*|}\right).
\end{eqnarray*}
Therefore,
\begin{eqnarray}
\frac{\det(\textbf{W}_{\boldsymbol{\gamma}})}{\det(\textbf{W}_{\boldsymbol{\gamma}^*})}&=&\frac{\det(\boldsymbol{\Sigma}_{\boldsymbol{\gamma}})}{\det(\boldsymbol{\Sigma}_{\boldsymbol{\gamma}^*})}
\frac{\det(\textbf{U}_{\boldsymbol{\gamma}})}{\det(\textbf{U}_{\boldsymbol{\gamma}^*})}\nonumber\\
&\ge&\det(\boldsymbol{\Sigma}_{\boldsymbol{\gamma}\backslash\boldsymbol{\gamma}^*})\det\left(\boldsymbol{\Sigma}_{\boldsymbol{\gamma}\backslash\boldsymbol{\gamma}^*}^{-1}+
(nd_0^{-1}-\rho_n)\textbf{I}_{|\boldsymbol{\gamma}\backslash\boldsymbol{\gamma}^*|}\right)\nonumber\\
&=&\det\left(\textbf{I}_{|\boldsymbol{\gamma}\backslash\boldsymbol{\gamma}^*|}+(nd_0^{-1}-\rho_n)\boldsymbol{\Sigma}_{\boldsymbol{\gamma}\backslash\boldsymbol{\gamma}^0}\right)\nonumber\\
&\ge&\det\left((1+(nd_0^{-1}-\rho_n)\underline{\phi}_n)\textbf{I}_{|\boldsymbol{\gamma}\backslash\boldsymbol{\gamma}^0|}\right)\nonumber\\
&=& (1+(nd_0^{-1}-\rho_n)\underline{\phi}_n)^{|\boldsymbol{\gamma}|-|\boldsymbol{\gamma}^*|}\ge 1+(nd_0^{-1}-\rho_n)\underline{\phi}_n,
\end{eqnarray}
which shows that $I_2\ge 2^{-1}\log(1+(nd_0^{-1}-\rho_n)\underline{\phi}_n)$. By Assumption \ref{B3} (v) we have, w.l.p.,
for some constant $\widetilde{C}>0$, for any $0<t_n<s_n$, uniformly for $c_j$s $\in[\underline{\phi}_n,\bar{\phi}_n]$ and uniformly for $\boldsymbol{\gamma}\in T_1(t_n)$,
\begin{equation}\label{thm2:imeq1}
\frac{p(\boldsymbol{\gamma}|\textbf{Z})}{\max\limits_{\boldsymbol{\gamma}\in T_0(t_n)}p(\boldsymbol{\gamma}|\textbf{Z})}\le\frac{p(\boldsymbol{\gamma}|\textbf{Z})}{p(\boldsymbol{\gamma}^*|\textbf{Z})}
\le\widetilde{C}\cdot\exp\left(-2^{-1}\log(1+(nd_0^{-1}-\rho_n)\underline{\phi}_n)+C''\max\{\rho_n,s_n^2\log{p}\}\right)=o(1).
\end{equation}
This proves $\frac{\max\limits_{\boldsymbol{\gamma}\in T_1(t_n)}p(\boldsymbol{\gamma}|\textbf{Z})}{\max\limits_{\boldsymbol{\gamma}\in T_0(t_n)}p(\boldsymbol{\gamma}|\textbf{Z})}=o_\Pr(1)$.

\textbf{Step II}: To accomplish the second step, we consider the following decomposition for any $0<t_n<s_n$ and $\boldsymbol{\gamma}\in T_2(t_n)$,
\begin{eqnarray*}
-\log\left(\frac{p(\boldsymbol{\gamma}|\textbf{Z})}{p(\emptyset|\textbf{Z})}\right)
&=&-\log\left(\frac{p(\boldsymbol{\gamma})}{p(\emptyset)}\right)+\frac{1}{2}\log\left(\frac{\det(\textbf{W}_{\boldsymbol{\gamma}})}{\det(\textbf{W}_\emptyset)}\right)\\
&&+\frac{n+\nu}{2}\log\left(\frac{1+\textbf{Y}^T(\textbf{I}_n-\textbf{X}_{\boldsymbol{\gamma}} \textbf{U}_{\boldsymbol{\gamma}}^{-1}\textbf{X}_{\boldsymbol{\gamma}}^T)\textbf{Y}}{1+\textbf{Y}^T(\textbf{I}_n-\textbf{P}_{\boldsymbol{\gamma}})\textbf{Y}}\right)
+\frac{n+\nu}{2}\log\left(\frac{1+\textbf{Y}^T(\textbf{I}_n-\textbf{P}_{\boldsymbol{\gamma}})\textbf{Y}}{1+\textbf{Y}^T\textbf{Y}}\right).
\end{eqnarray*}
Denote the above four terms by $I_1,I_2,I_3,I_4$. Similar to the arguments in Step I, $I_1$ is bounded from below, $I_3\ge 0$. So we only approximate $I_2$ and $I_4$.
First we approximate $I_4$.
By (\ref{B1:eq2}) of Assumption \ref{B1}, $\textbf{X}_{\boldsymbol{\gamma}^0}^T \textbf{P}_{\boldsymbol{\gamma}} \textbf{X}_{\boldsymbol{\gamma}^0}\le \rho_n \textbf{I}_{s_n}$. Let $\boldsymbol{\nu}_{\boldsymbol{\gamma}}=\textbf{P}_{\boldsymbol{\gamma}} \textbf{X}_{\boldsymbol{\gamma}^0}\boldsymbol{\beta}_{\boldsymbol{\gamma}^0}^0$, immediately we have
$\|\boldsymbol{\nu}_{\boldsymbol{\gamma}}\|^2\le \rho_n \|\boldsymbol{\beta}_{\boldsymbol{\gamma}^0}^0\|^2=\rho_n k_n$.
By Lemma \ref{lemma2} (iv), we have w.l.p.,
$\boldsymbol{\epsilon}^T \textbf{P}_{\boldsymbol{\gamma}}\boldsymbol{\epsilon}\le \alpha\sigma_0^2 |\boldsymbol{\gamma}| \log{p}$, where $\alpha>2$ is prefixed.
Therefore, w.l.p., for any $0<t_n<s_n$ and uniformly for $\boldsymbol{\gamma}\in T_2(t_n)$,
\begin{eqnarray*}
\textbf{Y}^T \textbf{P}_{\boldsymbol{\gamma}} \textbf{Y}&=&(\boldsymbol{\beta}^0_{\boldsymbol{\gamma}^0})^T\textbf{X}_{\boldsymbol{\gamma}^0}^T \textbf{P}_{\boldsymbol{\gamma}} \textbf{X}_{\boldsymbol{\gamma}^0} \boldsymbol{\beta}^0_{\boldsymbol{\gamma}^0}+2(\boldsymbol{\beta}^0_{\boldsymbol{\gamma}^0})^T\textbf{X}_{\boldsymbol{\gamma}^0}^T \textbf{P}_{\boldsymbol{\gamma}}\boldsymbol{\epsilon}+\boldsymbol{\epsilon}^T \textbf{P}_{\boldsymbol{\gamma}} \boldsymbol{\epsilon}\\
&=&\|\boldsymbol{\nu}_{\boldsymbol{\gamma}}\|^2+2\boldsymbol{\nu}_{\boldsymbol{\gamma}}^T\boldsymbol{\epsilon}+\boldsymbol{\epsilon}^T \textbf{P}_{\boldsymbol{\gamma}}\boldsymbol{\epsilon}\\
&\le&2\|\boldsymbol{\nu}_{\boldsymbol{\gamma}}\|^2+2\boldsymbol{\epsilon}^T \textbf{P}_{\boldsymbol{\gamma}}\boldsymbol{\epsilon}\\
&\le&2\rho_n k_n+2\alpha\sigma_0^2 t_n \log{p}.
\end{eqnarray*}
On the other hand, from $E\{|(\textbf{X}_{\boldsymbol{\gamma}^0}\boldsymbol{\beta}^0_{\boldsymbol{\gamma}^0})^T\boldsymbol{\epsilon}|^2/\|\textbf{X}_{\boldsymbol{\gamma}^0}\boldsymbol{\beta}^0_{\boldsymbol{\gamma}^0}\|^2\}=\sigma_0^2$
we have $|(\textbf{X}_{\boldsymbol{\gamma}^0}\boldsymbol{\beta}^0_{\boldsymbol{\gamma}^0})^T\boldsymbol{\epsilon}|/\|\textbf{X}_{\boldsymbol{\gamma}^0}\boldsymbol{\beta}^0_{\boldsymbol{\gamma}^0}\|=O_\Pr(1)$.
By (\ref{B1:eq}) of Assumption \ref{B1}, $\|\textbf{X}_{\boldsymbol{\gamma}^0}\boldsymbol{\beta}^0_{\boldsymbol{\gamma}^0}\|^2\ge nd_0^{-1}k_n$.
Therefore, we have
\begin{eqnarray*}
\textbf{Y}^T\textbf{Y}&=&\|\textbf{X}_{\boldsymbol{\gamma}^0}\boldsymbol{\beta}^0_{\boldsymbol{\gamma}^0}\|^2+2(\textbf{X}_{\boldsymbol{\gamma}^0}\boldsymbol{\beta}^0_{\boldsymbol{\gamma}^0})^T\boldsymbol{\epsilon}+\boldsymbol{\epsilon}^T\boldsymbol{\epsilon}\\
&=&\|\textbf{X}_{\boldsymbol{\gamma}^0}\boldsymbol{\beta}^0_{\boldsymbol{\gamma}^0}\|^2\left(1+O_\Pr\left(\sqrt{\frac{1}{n k_n}}\right)\right)+\boldsymbol{\epsilon}^T\boldsymbol{\epsilon}\\
&=&\|\textbf{X}_{\boldsymbol{\gamma}^0}\boldsymbol{\beta}^0_{\boldsymbol{\gamma}^0}\|^2\left(1+o_\Pr(1)\right)+n\sigma_0^2 (1+o_\Pr(1))\\
&\ge&(d_0^{-1} n k_n+n\sigma_0^2)\cdot(1+o_\Pr(1)).
\end{eqnarray*}
Then by $t_n\log{p}\le s_n^2\log{p}$, for any $0<t_n<s_n$ and uniformly for $\boldsymbol{\gamma}\in T_2(t_n)$,
\begin{eqnarray*}
\frac{1+\textbf{Y}^T(\textbf{I}_n-\textbf{P}_{\boldsymbol{\gamma}})\textbf{Y}}{1+\textbf{Y}^T\textbf{Y}}=1-\frac{\textbf{Y}^T \textbf{P}_{\boldsymbol{\gamma}} \textbf{Y}}{1+\textbf{Y}^T\textbf{Y}}
\ge 1-\frac{2(\rho_n k_n+\alpha\sigma_0^2 t_n\log{p})}{n(d_0^{-1}k_n+\sigma_0^2)}\cdot (1+o_\Pr(1))
\ge 1-\frac{C'\max\{\rho_n,s_n^2\log{p}\}}{n},
\end{eqnarray*}
where $C'>0$ is constant unrelated to $\gamma$ and $n$. Consequently,
$I_4=\frac{n+\nu}{2}\log\left(\frac{1+\textbf{Y}^T(\textbf{I}_n-\textbf{P}_{\boldsymbol{\gamma}})\textbf{Y}}{1+\textbf{Y}^T\textbf{Y}}\right)\ge -C''\max\{\rho_n,s_n^2\log{p}\}$, where $C''>0$ is unrelated to $\gamma$ and $n$.

Finally we approximate $I_2$ for $\boldsymbol{\gamma}\in T_2(t_n)$. Since $|\boldsymbol{\gamma}|\ge 1$, we have
\begin{eqnarray*}
\det\left(\textbf{W}_{\boldsymbol{\gamma}}\right)=\det\left(\textbf{I}_{|\boldsymbol{\gamma}|}+\boldsymbol{\Sigma}^{1/2}_{\boldsymbol{\gamma}}
\textbf{X}_{\boldsymbol{\gamma}}^T \textbf{X}_{\boldsymbol{\gamma}} \boldsymbol{\Sigma}_{\boldsymbol{\gamma}}^{1/2}\right)
\ge\det\left((1+nd_0^{-1}\underline{\phi}_n)\textbf{I}_{|\boldsymbol{\gamma}|}\right)
\ge1+nd_0^{-1}\underline{\phi}_n.
\end{eqnarray*}
Therefore, $I_2\ge 2^{-1}\log\left(1+nd_0^{-1}\underline{\phi}_n\right)\gg \max\{\rho_n,s_n^2\log{p}\}$ (Assumption \ref{B3} (v)).
As a consequence, we have, w.l.p., for some constant $\widetilde{C}>0$, for any $0<t_n<s_n$, uniformly for $c_j$s $\in[\underline{\phi}_n,\bar{\phi}_n]$ and uniformly for $\boldsymbol{\gamma}\in T_2(t_n)$,
\begin{equation}\label{thm2:imeq2}
\frac{p(\boldsymbol{\gamma}|\textbf{Z})}{p(\emptyset|\textbf{Z})}\le\widetilde{C}
\cdot\exp\left(-2^{-1}\log(1+nd_0^{-1}\underline{\phi}_n)+C''\max\{\rho_n,s_n^2\log{p}\}\right)=o(1).
\end{equation}
This completes Step II, and thus completes the proof of Theorem \ref{main:thm2}.

\subsection*{Proof of Theorem \ref{main:thm3}}
We begin with the following decomposition
\begin{eqnarray*}
-\log\left(\frac{p(\emptyset|\textbf{Z})}{p(\boldsymbol{\gamma}|\textbf{Z})}\right)
&=&-\log\left(\frac{p(\emptyset)}{p(\boldsymbol{\gamma})}\right)+\frac{1}{2}\log\left(\frac{1}{\det(\textbf{W}_{\boldsymbol{\gamma}})}\right)
-\frac{n+\nu}{2}\log\left(\frac{1+\textbf{Y}^T(\textbf{I}_n-\textbf{X}_{\boldsymbol{\gamma}} \textbf{U}_{\boldsymbol{\gamma}}^{-1} \textbf{X}_{\boldsymbol{\gamma}}^T)\textbf{Y}}{1+\textbf{Y}^T(\textbf{I}_n-\textbf{P}_{\boldsymbol{\gamma}})\textbf{Y}}\right)\\
&&+\frac{n+\nu}{2}\log\left(\frac{1+\textbf{Y}^T\textbf{Y}}{1+\textbf{Y}^T(\textbf{I}_n-\textbf{P}_{\boldsymbol{\gamma}})\textbf{Y}}\right).
\end{eqnarray*}
Denote the above four terms by $J_1,J_2,J_3,J_4$. Clearly, $J_1$ is bounded below. The approximation of $J_3$
is exactly the same as the approximation of $I_4$ in Step I of the proof of Theorem \ref{main:thm2}. By replacing
$\boldsymbol{\gamma}^*$ therein with $\boldsymbol{\gamma}$, one can show by going through the same procedure that $0\le -J_3=O_\Pr(1)$, uniformly for $c_j$s $\in[\underline{\phi}_n,\bar{\phi}_n]$.
So we only need to approximate $J_2$ and $J_4$.

To approximate $J_4$, note
$\frac{1+\textbf{Y}^T\textbf{Y}}{1+\textbf{Y}^T(\textbf{I}_n-\textbf{P}_{\boldsymbol{\gamma}})\textbf{Y}}
=1+\frac{\textbf{Y}^T\textbf{P}_{\boldsymbol{\gamma}} \textbf{Y}}{1+\textbf{Y}^T(\textbf{I}_n-\textbf{P}_{\boldsymbol{\gamma}})\textbf{Y}}$. So we only approximate the numerator
and denominator respectively. Let $\boldsymbol{\nu}_{\boldsymbol{\gamma}}=\textbf{P}_{\boldsymbol{\gamma}} \textbf{X}_{\boldsymbol{\gamma}^0}\boldsymbol{\beta}^0_{\boldsymbol{\gamma}^0}$.
Immediately we have
\[
\boldsymbol{\nu}_{\boldsymbol{\gamma}}=\textbf{P}_{\boldsymbol{\gamma}} (\textbf{X}_{\boldsymbol{\gamma}},\textbf{X}_{\boldsymbol{\gamma}^0\backslash\boldsymbol{\gamma}})
{\boldsymbol{\beta}^0_{\boldsymbol{\gamma}}\choose \boldsymbol{\beta}^0_{\boldsymbol{\gamma}^0\backslash\boldsymbol{\gamma}}}
=\textbf{X}_{\boldsymbol{\gamma}}\boldsymbol{\beta}^0_{\boldsymbol{\gamma}}+\textbf{P}_{\boldsymbol{\gamma}}
\textbf{X}_{\boldsymbol{\gamma}^0\backslash\boldsymbol{\gamma}}\boldsymbol{\beta}^0_{\boldsymbol{\gamma}^0\backslash\boldsymbol{\gamma}}.
\]
It follows by (\ref{B1:eq2}) of Assumption \ref{B1} and $\|\boldsymbol{\beta}^0_{\boldsymbol{\gamma}^0\backslash\boldsymbol{\gamma}}\|^2\le f_0\|\boldsymbol{\beta}^0_{\boldsymbol{\gamma}}\|^2$ that
\begin{eqnarray*}
|(\boldsymbol{\beta}^0_{\boldsymbol{\gamma}})^T\textbf{X}_{\boldsymbol{\gamma}}^T \textbf{P}_{\boldsymbol{\gamma}} \textbf{X}_{\boldsymbol{\gamma}^0\backslash\boldsymbol{\gamma}}\boldsymbol{\beta}^0_{\boldsymbol{\gamma}^0\backslash\boldsymbol{\gamma}}|&\le&
\|\textbf{X}_{\boldsymbol{\gamma}} \boldsymbol{\beta}^0_{\boldsymbol{\gamma}}\|\cdot \|\textbf{P}_{\boldsymbol{\gamma}} \textbf{X}_{\boldsymbol{\gamma}^0\backslash\boldsymbol{\gamma}}\boldsymbol{\beta}^0_{\boldsymbol{\gamma}^0\backslash\boldsymbol{\gamma}}\|
\le \|\textbf{X}_{\boldsymbol{\gamma}} \boldsymbol{\beta}^0_{\boldsymbol{\gamma}}\|\cdot \sqrt{\rho_n \|\boldsymbol{\beta}^0_{\boldsymbol{\gamma}^0\backslash\boldsymbol{\gamma}}\|^2}
\le \|\textbf{X}_{\boldsymbol{\gamma}} \boldsymbol{\beta}^0_{\boldsymbol{\gamma}}\|\cdot \sqrt{f_0\rho_n}\|\boldsymbol{\beta}^0_{\boldsymbol{\gamma}}\|.
\end{eqnarray*}
It follows by (\ref{B1:eq}) of Assumption \ref{B1} that $\|\textbf{X}_{\boldsymbol{\gamma}} \boldsymbol{\beta}^0_{\boldsymbol{\gamma}}\|
\ge\sqrt{nd_0^{-1}}\|\boldsymbol{\beta}^0_{\boldsymbol{\gamma}}\|\ge \sqrt{nd_0^{-1}\psi_n^2}$.
Thus, by $\rho_n=o(n)$ (Assumption \ref{B3} (v))
\[
\frac{|(\boldsymbol{\beta}^0_{\boldsymbol{\gamma}})^T\textbf{X}_{\boldsymbol{\gamma}}^T \textbf{P}_{\boldsymbol{\gamma}}
\textbf{X}_{\boldsymbol{\gamma}^0\backslash\boldsymbol{\gamma}}\boldsymbol{\beta}^0_{\boldsymbol{\gamma}^0\backslash\boldsymbol{\gamma}}|}
{\|\textbf{X}_{\boldsymbol{\gamma}}\boldsymbol{\beta}^0_{\boldsymbol{\gamma}}\|^2}
\le \sqrt{\frac{f_0\rho_n}{nd_0^{-1}}}=o(1).
\]
Similarly, one can show $\frac{\|\textbf{P}_{\boldsymbol{\gamma}}
\textbf{X}_{\boldsymbol{\gamma}^0\backslash\boldsymbol{\gamma}}\boldsymbol{\beta}^0_{\boldsymbol{\gamma}^0\backslash\boldsymbol{\gamma}}\|^2}
{\|\textbf{X}_{\boldsymbol{\gamma}}\boldsymbol{\beta}^0_{\boldsymbol{\gamma}}\|^2}=O\left(\frac{f_0\rho_n}{nd_0^{-1}}\right)=o(1)$.
Then
\begin{eqnarray*}
\|\boldsymbol{\nu}_{\boldsymbol{\gamma}}\|^2=\|\textbf{X}_{\boldsymbol{\gamma}}\boldsymbol{\beta}^0_{\boldsymbol{\gamma}}\|^2
\left(1+\frac{(\boldsymbol{\beta}^0_{\boldsymbol{\gamma}})^T\textbf{X}_{\boldsymbol{\gamma}}^T \textbf{P}_{\boldsymbol{\gamma}}
\textbf{X}_{\boldsymbol{\gamma}^0\backslash\boldsymbol{\gamma}}\boldsymbol{\beta}^0_{\boldsymbol{\gamma}^0\backslash\boldsymbol{\gamma}}}
{\|\textbf{X}_{\boldsymbol{\gamma}}\boldsymbol{\beta}^0_{\boldsymbol{\gamma}}\|^2}+
\frac{\|\textbf{P}_{\boldsymbol{\gamma}}
\textbf{X}_{\boldsymbol{\gamma}^0\backslash\boldsymbol{\gamma}}\boldsymbol{\beta}^0_{\boldsymbol{\gamma}^0\backslash\boldsymbol{\gamma}}\|^2}
{\|\textbf{X}_{\boldsymbol{\gamma}}\boldsymbol{\beta}^0_{\boldsymbol{\gamma}}\|^2}\right)
=\|\textbf{X}_{\boldsymbol{\gamma}}\boldsymbol{\beta}^0_{\boldsymbol{\gamma}}\|^2(1+o(1)).
\end{eqnarray*}
Therefore, by Assumption \ref{B3} (iii), and Lemma \ref{lemma2} (ii),
\begin{eqnarray}\label{thm3:eq1}
\textbf{Y}^T\textbf{P}_{\boldsymbol{\gamma}} \textbf{Y}&=&\|\boldsymbol{\nu}_{\boldsymbol{\gamma}}\|^2+2\boldsymbol{\nu}_{\boldsymbol{\gamma}}^T\boldsymbol{\epsilon}+\boldsymbol{\epsilon}^T \textbf{P}_{\boldsymbol{\gamma}}\boldsymbol{\epsilon}\nonumber\\
&\ge&\|\boldsymbol{\nu}_{\boldsymbol{\gamma}}\|^2\left(1+O_\Pr\left(\sqrt{\frac{s_n}{n\psi_n^2}}\right)\right)\nonumber\\
&=&\|\textbf{X}_{\boldsymbol{\gamma}}\boldsymbol{\beta}^0_{\boldsymbol{\gamma}}\|^2(1+o_\Pr(1))\nonumber\\
&\ge&\|\textbf{X}_{\boldsymbol{\gamma}}\boldsymbol{\beta}^0_{\boldsymbol{\gamma}}\|^2/2,\,\,\textrm{w.l.p.}\nonumber\\
&\ge&nd_0^{-1}\|\boldsymbol{\beta}^0_{\boldsymbol{\gamma}}\|^2/2.
\end{eqnarray}

On the other hand, if we let $\widetilde{\boldsymbol{\nu}}_{\boldsymbol{\gamma}}=
(\textbf{I}_n-\textbf{P}_{\boldsymbol{\gamma}})\textbf{X}_{\boldsymbol{\gamma}^0\backslash\boldsymbol{\gamma}}\boldsymbol{\beta}^0_{\boldsymbol{\gamma}^0\backslash\boldsymbol{\gamma}}$,
then by (\ref{B1:eq}) of Assumption \ref{B1}
\[
\|\widetilde{\boldsymbol{\nu}}_{\boldsymbol{\gamma}}\|^2\le (\boldsymbol{\beta}^0_{\boldsymbol{\gamma}^0\backslash\boldsymbol{\gamma}})^T\textbf{X}_{\boldsymbol{\gamma}^0\backslash\boldsymbol{\gamma}}^T\textbf{X}_{\boldsymbol{\gamma}^0\backslash\boldsymbol{\gamma}}\boldsymbol{\beta}^0_{\boldsymbol{\gamma}^0\backslash\boldsymbol{\gamma}}
\le nd_0\|\boldsymbol{\beta}^0_{\boldsymbol{\gamma}^0\backslash\boldsymbol{\gamma}}\|^2\le nd_0 f_0 \|\boldsymbol{\beta}^0_{\boldsymbol{\gamma}}\|^2.
\]
Therefore, by Lemma \ref{lemma2} (i)
and $\boldsymbol{\epsilon}^T(\textbf{I}_n-\textbf{P}_{\boldsymbol{\gamma}})\boldsymbol{\epsilon}=n\sigma_0^2 (1+o_\Pr(1))$,
\begin{eqnarray}\label{thm3:eq2}
\textbf{Y}^T(\textbf{I}_n-\textbf{P}_{\boldsymbol{\gamma}})\textbf{Y}&=&\|\widetilde{\boldsymbol{\nu}}_{\boldsymbol{\gamma}}\|^2
\left(1+\frac{2\widetilde{\boldsymbol{\nu}}_{\boldsymbol{\gamma}}^T\boldsymbol{\epsilon}}{\|\widetilde{\boldsymbol{\nu}}_{\boldsymbol{\gamma}}\|^2}\right)+
\boldsymbol{\epsilon}^T(\textbf{I}_n-\textbf{P}_{\boldsymbol{\gamma}})\boldsymbol{\epsilon}\nonumber\\
&=&\|\widetilde{\boldsymbol{\nu}}_{\boldsymbol{\gamma}}\|^2(1+o_\Pr(1))+n\sigma_0^2(1+o_\Pr(1))\nonumber\\
&\le& 2(\|\widetilde{\boldsymbol{\nu}}_{\boldsymbol{\gamma}}\|^2+n\sigma_0^2)\,\,\textrm{w.l.p.}\nonumber\\
&\le& 2n(d_0 f_0 \|\boldsymbol{\beta}^0_{\boldsymbol{\gamma}}\|^2+\sigma_0^2).
\end{eqnarray}
Define $\zeta_0=\sigma_0^2/(d_0f_0)$. Consequently, by (\ref{thm3:eq1}) and (\ref{thm3:eq2}), and $\|\boldsymbol{\beta}^0_{\boldsymbol{\gamma}}\|^2\ge \psi_n^2$, w.l.p.,
\begin{eqnarray*}
1+\frac{\textbf{Y}^T \textbf{P}_{\boldsymbol{\gamma}} \textbf{Y}}{1+\textbf{Y}^T(\textbf{I}_n-\textbf{P}_{\boldsymbol{\gamma}})\textbf{Y}}\ge
1+\frac{nd_0^{-1}\|\boldsymbol{\beta}^0_{\boldsymbol{\gamma}}\|^2}{4n(d_0 f_0\|\boldsymbol{\beta}^0_{\boldsymbol{\gamma}}\|^2+\sigma_0^2)}
\ge1+\frac{1}{4d_0^2 f_0}\cdot\frac{\psi_n^2}{\psi_n^2+\zeta_0}
\ge1+\frac{1}{4d_0^2 f_0}\min\left\{\frac{1}{2},\frac{\psi_n^2}{2\zeta_0}\right\}.
\end{eqnarray*}
Thus,
\begin{equation}\label{thm3:eq3}
J_4\ge \frac{n+\nu}{2}\log\left(1+\frac{1}{4d_0^2 f_0}\min\left\{\frac{1}{2},\frac{\psi_n^2}{2\zeta_0}\right\}\right).
\end{equation}

Finally we approximate $J_2$. Since $\det\left(\textbf{W}_{\boldsymbol{\gamma}}\right)=\det\left(\textbf{I}_{|\boldsymbol{\gamma}|}+\boldsymbol{\Sigma}^{1/2}_{\boldsymbol{\gamma}}
\textbf{X}_{\boldsymbol{\gamma}}^T \textbf{X}_{\boldsymbol{\gamma}} \boldsymbol{\Sigma}^{1/2}_{\boldsymbol{\gamma}}\right)\le (1+d_0 n\bar{\phi}_n)^{|\boldsymbol{\gamma}|}$. Then
\begin{eqnarray}\label{thm3:eq4}
J_2=\frac{1}{2}\log\left(\frac{1}{\det\left(\textbf{W}_{\boldsymbol{\gamma}}\right)}\right)\ge -\frac{s_n}{2}\log\left(1+d_0 n\bar{\phi}_n\right).
\end{eqnarray}
Combining (\ref{thm3:eq3}) and (\ref{thm3:eq4}), there exists constant $\widetilde{C}$ such that, w.l.p., uniformly for $c_j$s $\in[\underline{\phi}_n,\bar{\phi}_n]$,
\begin{eqnarray*}
\frac{p(\emptyset|Z)}{p(\boldsymbol{\gamma}|\textbf{Z})}\le\widetilde{C}\cdot\exp\left(\frac{s_n}{2}\log\left(1+d_0 n\bar{\phi}_n\right)-\frac{n+\nu}{2}\log\left(1+\frac{1}{4d_0^2 f_0}\min\left\{\frac{1}{2},\frac{\psi_n^2}{2\zeta_0}\right\}\right)\right),
\end{eqnarray*}
which approaches zero by Assumption \ref{A3} (iv). This completes the proof.

\subsection*{Proof of Theorem \ref{main:gp:thm1}}
We observe that
\begin{eqnarray}\label{gp:thm1:eq1}
&&\min\limits_{s_n\le t_n\le r_n}p_g(\boldsymbol{\gamma}^0|\textbf{Z})\\
&=&\min\limits_{s_n\le t_n\le r_n}\int_0^1 p(\boldsymbol{\gamma}^0|c,\textbf{Z})g(c)dc\nonumber
\ge\min\limits_{s_n\le t_n\le r_n}\int_{\underline{\phi}_n}^{\bar{\phi}_n} p(\boldsymbol{\gamma}^0|c,\textbf{Z})g(c)dc\nonumber
\ge\int_{\underline{\phi}_n}^{\bar{\phi}_n} g(c)dc\cdot \min\limits_{s_n\le t_n\le r_n}\inf\limits_{\underline{\phi}_n\le c\le\bar{\phi}_n} p(\boldsymbol{\gamma}^0|c,\textbf{Z}).
\end{eqnarray}
By Theorem \ref{main:thm1}, $\min\limits_{s_n\le t_n\le r_n}\inf\limits_{\underline{\phi}_n\le c\le\bar{\phi}_n} p(\boldsymbol{\gamma}^0|c,\textbf{Z})=1+o_\Pr(1)$.
By Assumption, $\int_{\underline{\phi}_n}^{\bar{\phi}_n} g(c)dc=1+o(1)$.
Thus, by (\ref{gp:thm1:eq1}), $\min\limits_{s_n\le t_n\le r_n}p_g(\boldsymbol{\gamma}^0|\textbf{Z})\ge (1+o(1))\cdot(1+o_\Pr(1))=1+o_\Pr(1)$,
which proves the desired result.

\subsection*{Proof of Theorem \ref{main:gp:thm2}}
Define
\[
D_{1n}=\max\limits_{0<t_n<s_n}\sup\limits_{\underline{\phi}_n\le c\le\bar{\phi}_n}
\max\limits_{\boldsymbol{\gamma}\in T_1(t_n)}\frac{p(\boldsymbol{\gamma}|c,\textbf{Z})}{p(\boldsymbol{\gamma}\cap\boldsymbol{\gamma}^0|c,Z)},\,\,\textrm{and}\,\,
D_{2n}=\max\limits_{0<t_n<s_n}\sup\limits_{\underline{\phi}_n\le c\le\bar{\phi}_n}
\max\limits_{\boldsymbol{\gamma}\in T_2(t_n)}\frac{p(\boldsymbol{\gamma}|c,\textbf{Z})}{p(\emptyset|c,\textbf{Z})}.
\]
By (\ref{thm2:imeq1}) and (\ref{thm2:imeq2}) in the proof of Theorem \ref{main:thm2}, $D_{1n}=o_\Pr(1)$ and $D_{2n}=o_\Pr(1)$.
For any $\boldsymbol{\gamma}\in T_1(t_n)$, denote $\boldsymbol{\gamma}^*=\boldsymbol{\gamma}\cap\boldsymbol{\gamma}^0$. Then
\begin{eqnarray*}
&&p_g(\boldsymbol{\gamma}|\textbf{Z})\\
&=&\int_0^\infty p(\boldsymbol{\gamma}|c,\textbf{Z})g(c)dc
=\int_{\underline{\phi}_n}^{\bar{\phi}_n} p(\boldsymbol{\gamma}|c,\textbf{Z})g(c)dc
\le D_{1n}\int_{\underline{\phi}_n}^{\bar{\phi}_n} p(\boldsymbol{\gamma}^*|c,\textbf{Z})g(c)dc
= D_{1n}\,\,\, p_g(\boldsymbol{\gamma}^*|\textbf{Z})
\le D_{1n}\max\limits_{\boldsymbol{\gamma}\in T_0(t_n)}p_g(\boldsymbol{\gamma}|\textbf{Z}).
\end{eqnarray*}
Therefore,
\begin{equation}\label{gp:thm2:eq1}
\max\limits_{0<t_n<s_n}\frac{\max\limits_{\boldsymbol{\gamma}\in T_1(t_n)}p_g(\boldsymbol{\gamma}|\textbf{Z})}{\max\limits_{\boldsymbol{\gamma}\in T_0(t_n)}p_g(\boldsymbol{\gamma}|\textbf{Z})}\le D_{1n}=o_\Pr(1).
\end{equation}
Likewise, for any $\boldsymbol{\gamma}\in T_2(t_n)$,
\begin{eqnarray*}
p_g(\boldsymbol{\gamma}|\textbf{Z})=\int_{\underline{\phi}_n}^{\bar{\phi}_n} p(\boldsymbol{\gamma}|c,\textbf{Z})g(c)dc
\le D_{2n}\int_{\underline{\phi}_n}^{\bar{\phi}_n} p(\emptyset|c,\textbf{Z})g(c)dc
=D_{2n}\,\,\, p_g(\emptyset|\textbf{Z})
\le D_{2n}\max\limits_{\boldsymbol{\gamma}\in T_0(t_n)}p_g(\boldsymbol{\gamma}|\textbf{Z}).
\end{eqnarray*}
Therefore,
\begin{equation}\label{gp:thm2:eq2}
\max\limits_{0<t_n<s_n}\frac{\max\limits_{\boldsymbol{\gamma}\in T_2(t_n)}p_g(\boldsymbol{\gamma}|\textbf{Z})}{\max\limits_{\boldsymbol{\gamma}\in T_0(t_n)}p_g(\boldsymbol{\gamma}|\textbf{Z})}\le D_{2n}=o_\Pr(1).
\end{equation}
The desired conclusion follows immediately from (\ref{gp:thm2:eq1}) and (\ref{gp:thm2:eq2}).

\subsection*{Proof of Theorem \ref{main:gp:thm3}}
Define $D_n=\sup\limits_{\underline{\phi}_n\le c\le\bar{\phi}_n}\frac{p(\emptyset|c,\textbf{Z})}{p(\boldsymbol{\gamma}|c,\textbf{Z})}$.
Theorem \ref{main:thm3} implies $D_n=o_\Pr(1)$.
Then
\begin{eqnarray*}
p_g(\emptyset|\textbf{Z})=\int_{\underline{\phi}_n}^{\bar{\phi}_n} p(\emptyset|c,\textbf{Z})g(c)dc
\le D_n\int_{\underline{\phi}_n}^{\bar{\phi}_n} p(\boldsymbol{\gamma}|c,\textbf{Z})g(c)dc
=D_n\,\,\, p_g(\boldsymbol{\gamma}|\textbf{Z}).
\end{eqnarray*}
Thus, $\frac{p_g(\emptyset|\textbf{Z})}{p_g(\boldsymbol{\gamma}|\textbf{Z})}\le D_n=o_\Pr(1)$, which completes the proof.


\begin{thebibliography}{34}

\bibitem{BP96} Berger, J. O. and Pericchi, L. (1996).
The intrinsic {B}ayes factor for model selection and prediction.
{\it Journal of the American Statistical Association}
{\bf 91}, 109--122.

\bibitem{BGM03} Berger, J. O., Ghosh, J. K. and Mukhopadhyay, N. (2003).
Approximations and consistency of Bayes factors as model dimension grows.
{\it Journal of Statistical Planning and Inference}
{\bf 112}, 241--258.


\bibitem{BR12} Bondell, H. D. and Reich, B. J. (2012). Consistent high-dimensional Bayesian variable selection via penalized credible regions.
\emph{Journal of the American Statistical Association}. In Press.


\bibitem{BFV01} Brown, P., Fearn, T. and Vannucci, M. (2001).  Bayesian wavelet regression on curves With application to a spectroscopic
calibration problem. \textit{Journal of the American Statistical Association} \textbf{96}, 398--408.

\bibitem{BVF02} Brown, P., Vannucci, M. and Fearn, T. (2002). Bayes model averaging with selection of regressors. \textit{Journal of the Royal Statistical Society, Series B} \textbf{64}, 519--536.

\bibitem{BKM10} B\"{u}hlmann, P., Kalisch, M. and Maathuis, M. H. (2010).
Variable selection in high-dimensional linear models: partially faithful distributions and the
PC-simple algorithm. \textit{Biometrika} \textbf{97}, 261--278.

\bibitem{CT05} Cand\`{e}s, E. J. and Tao, T. (2005).
The Dantzig selector: statistical estimation when $p$ is much larger than $n$. \emph{Annals of Statistics} \textbf{35}, 2313--2351.

\bibitem{CS09} Carvalho, C. M. and Scott, J. G. (2009). Objective Bayesian model selection in Gaussian graphical models.
\textit{Biometrika} \textbf{96}, 497--512.

\bibitem{CGMM09} Casella, C., Gir{\'o}n, F. J., Mart\'{\i}nez, M.~L. and Moreno, E. (2009).
Consistency of {B}ayesian procedures for variable selection.
{\it Annals of Statistics}
{\bf 37}, 1207--1228.

\bibitem{CG00} Clyde, M. and George, E. I. (2000).
Flexible empirical {B}ayes estimation for wavelets.
{\it Journal of the Royal Statistical Society, Series B} {\bf 62}, 681--698.

\bibitem{CPV98} Clyde, M., Parmigiani, G. and Vidakovic, B. (1998).
Multiple shrinkage and subset selection in wavelets.
{\it Biometrika} {\bf 85}, 391--401.

\bibitem{D05} Durrett, R. (2005).
{\it Probability: Theorey and Examples}. 3rd Ed. Wadsworth-Brooks/Cole, Pacific Grove.

\bibitem{FG94} Foster, D. P. and George, E. I. (1994). The risk in
ation criterion for multiple regression. \emph{Annals of Statistics} \textbf{22}, 1947--1975.


\bibitem{FL08} Fan, J. and Lv, J. (2008). Sure independence screening for ultrahigh dimensional feature space.
\textit{Journal of the Royal Statistical Society, Series B} \textbf{70}, 849--911.

\bibitem{FL10} Fan, J. and Lv, J. (2010).
A selective overview of variable selection in high dimensional feature space.
\textit{Statistica Sinica} \textbf{20}, 101--148.

\bibitem{FS10} Fan, J. and Song, R. (2010). Sure independence screening in generalized linear models with NP-dimensionality.
\textit{Annals of Statistics} \textbf{38}, 3567--3604.

\bibitem{FLS01} Fern{\'a}ndez, C.,  Ley, E. and Steel, M. F. J. (2001).
Benchmark priors for Bayesian model averaging.
{\it Journal of Econometrics}
{\bf 100}, 381--427.

\bibitem{GCSR03} Gelman, A., Carlin, J. B., Stern, H. S. and Rubin, D. B.
(2003). \textit{Bayesian Data Analysis} (2nd ed). Chapman $\&$ Hall/CRC.


\bibitem{GF00} George, E. and Foster, D. (2000).
Calibration and empirical Bayes variable selection.
\textit{Biometrika} \textbf{87}, 731--747.

\bibitem{GM93} George, E. and McCulloch, R. (1993).
Variable selection via Gibbs sampling.
\textit{Journal of the American Statistical Association} \textbf{88}, 881--889.

\bibitem{GM97} George, E. and McCulloch, R. (1997).
Approaches for Bayesian variable selection.
\textit{Statistica Sinica} \textbf{7}, 339--373.


\bibitem{GMCM10} Gir\'{o}n, F. J., Moreno, E., Casella, G. and Mart\'{i}nez, M. L.
(2010). Consistency of objective Bayes factors for nonnested linear
models and increasing model dimension.
\textit{Revista de la Real Academia de Ciencias Exactas, Fisicas y Naturales. Serie A. Matematicas}
\textbf{104}, 57--67.

\bibitem{GR98} Godsill, J.~S. and Rayner, P.~J.~W. (1998).
Robust reconstruction and analysis of autoregressive signals in impulsive noise using the Gibbs sampler.
\textit{IEEE Trans. Speech Audio Process} \textbf{6}, 352--372.

\bibitem{HHM08} Huang, J., Horowitz, J. and Ma, S. (2008). Asymptotic properties of bridge estimators in sparse high-dimensional regression models.
\textit{Annals of Statistics} \textbf{36}, 587--613.

\bibitem{J07} Jiang, W. (2007).
Bayesian variable selection for high dimensional generalized linear models: Convergence rates of the fitted densities.
{\it Annals of Statistics} {\bf 35}, 1487--1511.


\bibitem{JR12} Johnson, V. E. and Rossell, D. (2012). Bayesian model selection in high-dimensional settings.
\emph{Journal of the American Statistical Association} \textbf{107}, 649--660

\bibitem{LZ10} Li, F. and Zhang, N. R. (2010). Bayesian variable selection
in structured high-dimensional covariate spaces with applications in
genomics. \textit{Journal of the American Statistical Association}
\textbf{105}, 1202--1214.

\bibitem{LF09} Lv, J. and Fan, Y. (2009). A unified approach to model selection and sparse recovery using regularized least squares.
\textit{Annals of Statistics} \textbf{37}, 3498--3528.


\bibitem{LPMCB08} Liang, F., Paulo, R., Molina, G., Clyde, M. and Berger, J. O. (2008).
Mixtures of $g$-priors for Bayesian variable selection.
{\it Journal of the American Statistical Association} {\bf 103}, 410--423.

\bibitem{LS09} Ley, E. and Steel, M. F. J. (2009). On the effect of prior assumptions in Bayesian
model averaging with applications to growth regression. \emph{Journal of Applied
Econometrics} \textbf{24}, 651?674.

\bibitem{LS12} Ley, E. and Steel, M. F. J. (2012). Mixtures of g-priors for Bayesian model averaging with economic applications.
\textit{Journal of Econometrics}. In press.

\bibitem{MB06} Meinshausen, N. and B\"{u}hlmann, P. (2006). High dimensional graphs and variable selection with the Lasso. \textit{Annals of Statistics} \textbf{34}, 1436--1462.

\bibitem{MY09} Meinshausen, N. and Yu, B. (2009). Lasso-type recovery of sparse representations for high-dimensional data. \textit{Annals of Statistics} \textbf{37}, 246--270.

\bibitem{MG05} Moreno, E. and Gir{\'o}n, F. J. (2005).
Consistency of {B}ayes factors for intrinsic priors in normal linear models.
{\it C. R. Math. Acad. Sci. Paris}
{\bf 340}, 911--914.

\bibitem{MGC10} Moreno, E., Gir\'{o}n, F. J. and Casella, G. (2010).
Consistency of objective Bayes factors as the model dimension grows.
\textit{Annals of Statistics} \textbf{38}, 1937--1952.

\bibitem{NK05} Nott, D. J. and Kohn, R. (2005). Adaptive sampling for Bayesian variable selection.
\emph{Biometrika} \textbf{92}, 747-?63.

\bibitem{SB10} Scott, J. G., Berger, J. O. (2010). Bayes and empirical Bayes multiplicity adjustment in
the variable-selection problem. \emph{Annals of Statistics} \textbf{38}, 2587?2619.

\bibitem{S11} Shang, Z. (2011). Bayesian variable selection: theory and applications (PhD Dissertation).

\bibitem{SC11} Shang, Z. and Clayton, M. K. (2011). Consistency of Bayesian model selection for linear models
with a growing number of parameters. {\it Journal of Statistical Planning and Inference},  \textbf{11}, 3463--3474.

\bibitem{SC12} Shang, Z. and Clayton, M. K. (2012). An application of Bayesian variable selection to spatial concurrent linear models.
\emph{Environmental and Ecological Statistics} \textbf{19}, 521--544.

\bibitem{SM95} Shun, Z. and McCullagh, P. (1995). Laplace approximation of high dimensional
integrals. \emph{Journal of the Royal Statistical Society, Series B} \textbf{57}, 749--760.

\bibitem{SL03} Seber, G.~A.~F. and Lee, A. J. (2003).
{\it Linear Regression Analysis}, 2nd Ed.
Wiley-Interscience [John Wiley \& Sons], Hoboken, NJ.

\bibitem{SPZ12} Shen, X., Pan, W., Zhu, Y. (2012). Likelihood-based selection and sharp parameter estimation. \textit{Journal of American Statistical Association} \textbf{107}, 223-232.

\bibitem{SK96} Smith, M. S. and Kohn, R. (1996).
Nonparametric regression using Bayesian variable selection. \textit{Journal of Econometrics}
\textbf{75}, 317--344.

\bibitem{VD08} van de Geer, S. A. (2008). High-dimensional generalized linear models and the Lasso. \textit{Annals of Statistics} \textbf{36}, 614--645.

\bibitem{WLL09} Wang, H., Li, B. and Leng, C. (2009). Shrinkage tuning parameter selection with a diverging number of parameters.
\textit{Journal of Royal Statistical Society, Series B} \textbf{71}, 671--683.

\bibitem{WG07} Wang, X. and George, E. (2007). Adaptive Bayesian criteria in variable selection
for generalized linear models. \emph{Statistica Sinica} \textbf{17}, 667--690.

\bibitem{WGN04} Wolfe, P. J., Godsill, S. J. and Ng, W.-J. (2004).
Bayesian variable selection and regularization for time-frequency surface estimation.
{\it Journal of the Royal Statistical Society, Series B}
{\bf 66}, 575--589.

\bibitem{WR09} Wasserman, L. and Roeder, K. (2009). High-dimensional variable selection. \textit{Annals of Statistics}
\textbf{37}, 2178--2201.

\bibitem{ZS80} Zellner, A. and Siow, A. (1980). Posterior odds ratios for selected regression hypotheses. In
\textit{Bayesian analysis in econometrics and statistics: the Zellner view and papers}, (ed. A. Zellner), 389--399.
Edward Elgar Publishing Limited.

\bibitem{ZH08} Zhang, C.-H. and Huang, J. (2008). The sparsity and bias of the Lasso selection in high-dimensional linear regression.
\textit{Annals of Statistics} \textbf{36}, 1567--1594.

\bibitem{ZY06} Zhao, P. and Yu, B. (2006). On model selection consistency of Lasso. \textit{Journal of Machine Learning Research} \textbf{7}, 2541--2567.

\bibitem{ZH12} Zhao, Z. and Hwang, J. T. (2012). Empirical Bayes FCR controlling confidence interval. \emph{Journal of the Royal Statistical Society, Series B}
\textbf{74}, 871--891.

\end{thebibliography}
\end{document}